\documentclass[12pt]{iopart}
\usepackage{graphicx}
\usepackage{iopams}
\usepackage{amssymb}
\usepackage{setstack}
\usepackage{cite}
\usepackage[colorlinks=true, linkcolor=blue,citecolor=blue]{hyperref}	
\hypersetup{breaklinks=true}
\begin{document}

\title[Hatsugai-Kohmoto models]{Topic Review: Hatsugai-Kohmoto models: Exactly solvable playground for Mottness and Non-Fermi Liquid}

\author{Miaomiao Zhao$^{1,2}$}
\address{$^{1}$ Key Laboratory of Quantum Theory and Applications of MoE $\&$ School of Physical Science and Technology, Lanzhou University, Lanzhou 730000, People Republic of China}
\address{$^{2}$ Lanzhou Center for Theoretical Physics, Key Laboratory of Theoretical Physics of Gansu Province, Lanzhou University, Lanzhou 730000, People Republic of China}

\author{Wei-Wei Yang$^{3}$}
\address{$^{3}$Beijing National Laboratory for Condensed Matter Physics and Institute of Physics, Chinese Academy of Sciences, Beijing 100190, China}

\author{Yin Zhong$^{1,2,*}$}
\address{$^{1}$ Key Laboratory of Quantum Theory and Applications of MoE $\&$ School of Physical Science and Technology, Lanzhou University, Lanzhou 730000, People Republic of China}
\address{$^{2}$ Lanzhou Center for Theoretical Physics, Key Laboratory of Theoretical Physics of Gansu Province, Lanzhou University, Lanzhou 730000, People Republic of China}
\ead{zhongy@lzu.edu.cn}
%\vspace{10pt}
%\begin{indented}
%\item[]May 2023
%\end{indented}

\begin{abstract}
This pedagogic review aims to give a gentle introduction to an exactly solvable model, the Hatsugai-Kohmoto (HK) model, which has infinite-ranged interaction but conserves the center of mass. Although this model is invented in 1992, intensive studies on its properties ranging from unconventional superconductivity, topological ordered states to non-Fermi liquid behaviors are made since 2020. We focus on its emergent non-Fermi liquid behavior and provide discussion on its thermodynamics, single-particle and two-particle correlation functions. Perturbation around solvable limit has also been explored with the help of perturbation theory, renormalization group and exact diagonalization calculation. We hope the present review will be helpful for graduate students or researchers interested in HK-like models or more generic strongly correlated electron systems.
\end{abstract}

%\vspace{2pc}

\submitto{\JPCM}
\maketitle
%\ioptwocol

%
% Uncomment for keywords
%\vspace{2pc}
%\noindent{\it Keywords}: XXXXXX, YYYYYYYY, ZZZZZZZZZ
%
% Uncomment for Submitted to journal title message
%\submitto{\JPCM}
%
% Uncomment if a separate title page is required
%\maketitle
%
% For two-column output uncomment the next line and choose [10pt] rather than [12pt] in the \documentclass declaration

%

\section{Introduction}\label{sec:1}
\subsection{Landau's Fermi liquid and non-Fermi liquid}
The Landau's Fermi liquid (FL) theory has been the cornerstone of modern condensed matter physics for several decades.\cite{Nozieres1966,Baym1991,Coleman2015,Sachdev2023} In this paradigm, the low-energy physics of interacting many fermions system is described by weakly interacting quasiparticles, whose properties are similar to the non-interacting electron gas but with modified mass and magnetic moment. The He-$3$ liquid (above its superfluid critical temperature) and simple metals are typical examples of FL and people believe that FL description is valid for all of metallic phases in nature.\cite{Wen2004} However, the discovery of cuprate high-$T_{c}$ superconductivity has challenged this viewpoint, e.g. the linear-in-$T$ resistivity observed in strange metal phase and unclosed Fermi surface (Fermi arc) in the pseudogap phase.\cite{PLee2006,Greene2020} The linear-$T$ resistivity among several temperature regimes is rather different from the classic $T^{2}$ behavior predicted by FL, suggesting the non-quasiparticles-like transport and the breakdown of semiclassical Boltzmann theory.\cite{Varma2020,Hartnoll2022} Furthermore, many heavy fermion compounds are tuned near magnetic quantum critical points with anomalous power-law behaviors in resistivity, specific heat and susceptibility,\cite{Lohneysen2007,Stockert2011,Kirchner2020} implying the destruction of FL quasiparticle due to critical fluctuations. These non-Fermi liquid (NFL) behaviors are now widely observed in $d$/$f$-electron systems and even in the light $s,p$-electron materials such as twisted bilayer graphene.\cite{YCao2020}

\begin{figure}
\begin{center}
\includegraphics[width=0.45\columnwidth]{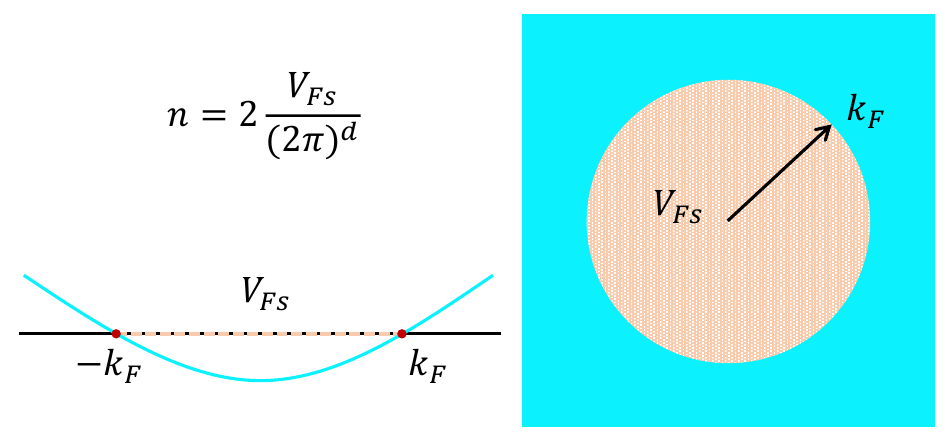}
\caption{\label{fig:Luttinger} The Luttinger theorem for one and two-dimensional fermion systems, which is valid for FL and Luttinger liquid.}
\end{center}
\end{figure}

To understand the mentioned NFL behaviors, we recall an essential feature of FL, the Luttinger theorem/Luttinger sum-rule, which states that the volume closed by Fermi surface is proportional to the density of electron and cannot be changed by interaction if no phase transition appears.\cite{Luttinger1960,LW1960} (Fig.~\ref{fig:Luttinger}) This theorem is the underlying assumption of Landau's phenomenological construction of FL and has been proved nonperturbatively by flux arguments.\cite{Oshikawa2000} Generally, the violation of Luttinger theorem implies the breakdown of FL and the emergence of NFL. Amusingly, as the first NFL state, the Luttinger liquid with power-law behaviors in its correlation function obeys the Luttinger theorem because of its emergent particle-hole symmetry.\cite{Blagoev1997,Seki2017}
In literature, theories of NFL may be classified into four kinds: 1) Fermi surface+$X$, where $X$ denotes bosonic gapless excitation results from critical fluctuation of symmetry-breaking order in spin/charge channel, e.g. metallic nematic and antiferromagnetic spin-density-wave order.\cite{Metlitski2010a,Metlitski2010b} The first theory in this category must be the Hertz-Millis-Moriya (HMM) theory, which integrates fermions out and leads to bosonic theory with Landau damping. When tuned into critical points, those bosons generates NFL-like thermodynamics and render fermions acquiring NFL self-energy correction.\cite{Hertz1976,Millis1993,Mishra1998} Although HMM theory is applicable in three spatial dimension, it seems to break down in $d=2$ and even more sophisticated perturbative calculation with fermions taken into account does not lead to conclusive results.\cite{Metlitski2010a,Metlitski2010b} 2) The second case involves fermions interacting with emergent transverse gauge field and bosonic Higgs degree of freedom (hybridization field in the language of heavy fermion).\cite{Florens2004,Senthil2004,Pepin2007,Kaul2008,Podolsky2009,Senthil2008a,Senthil2008b} In contrast to the first case, no sensible symmetry-breaking is involved and the whole Fermi surface becomes critical instead of hot spots in HMM theory. In fact, those theories are based on certain slave-particle techniques,\cite{Florens2004} which often begin with splitting electrons into partons and binding them with some kinds of compact gauge field. To proceed with slave-particle theory, mean-field approximations are unavoidable, aided with RPA-like correction with the assumption of deconfined gauge fields, thus it is difficult to evaluate the validity of such fractionalized theories,\cite{Mross2010,Lee2018} let alone the possibility of confinement of gauge fields in low-spatial dimension.\cite{Kogut1979}
3) The higher-dimensional bosonization is the generalization of the successful one-dimensional bosonization,\cite{Luther1979,Neto1994,Kwon1994,Chubukov2006,Delacretaz2022,Ding2012} which maps strongly interacting fermions into weakly coupled bosons.\cite{Giamarchi2003} However, it is a major obstacle to treat the umklapp scattering and only RPA-like results are obtained in present version of higher-dimensional bosonization. 4) NFL generated from gauge-gravity duality (holographic duality).\cite{Sachdev2012,Zaanen2015} The holographic duality itself is inspired by superstring theory and relates the field theory with gravity theory in higher dimension.\cite{Maldacena1998} Applications to NFL phenomena exist but it is not sure the extra symmetry (e.g. supersymmetry and conformal symmetry) enforced in holographic duality leads to consistent description on the low-energy dynamics of electrons.\cite{Zaanen2015} In the perspective of quantum criticality, only time-direction is critical in the holographic description but the spatial degree of freedom is intact,\cite{Faulkner2011} contradicting with critical Fermi surface in many NFL models.\cite{Senthil2008a,Senthil2008b} In addition to these analytical studies, NFL behaviors have also been observed in numerical techniques developed in recent three decades such as dynamic mean-field theory (DMFT) and determinant quantum Monte Carlo.\cite{Georges1996,TLee2016,Berg2012,Berg2019}

To be honest, in spite of those mentioned NFL theories, a coherent, robust and unified theory (like the formalism of FL theory) is still awaiting. (It is possible that distinct mechanisms may be responsible for NFL behaviors in mentioned systems, if so a unified theory has to be a dream for theorists.) We mention that the extension of Sachdev-Ye-Kitaev (SYK) model seems to provide useful information on metallic phase without quasiparticle though its solvability requires the randomly-all-to-all interaction and artificial large-$N$ limit.\cite{Sachdev1993,Maldacena2016,Chowdhury2022} ($N$ maybe the number of component of electrons) If the system is translation invariant without disorder or only has finite $N$, the solvability of SYK-like models is lost and controllable approximations are difficult to perform. On the other hand, solvable models without random interaction/disorder or any large-$N$ limit indeed exist, e.g. the Hatsugai-Kohmoto (HK) model.\cite{Hatsugai1992}

\subsection{History of Hatsugai-Kohmoto model}
The HK model is invented by two Japanese theorists Hatsugai and Kohmoto in 1992, (Similar models are independently discovered by others.\cite{Spalek1988,Baskaran1991}) who are inspired by the seminal spin-glass paper of Sherrington and Kirpatrick (SK).\cite{Sherrington1975} Although the interaction range in SK model is infinite, which is unrealistic, it serves an excellent starting point for understanding spin-glass and generic complex systems.\cite{Fischer1991,Parisi2020} It is this point that motivates Hatsugai and Kohmoto to construct their own HK model. HK model can be easily solved by Fourier transformation since its infinite-ranged interaction conserves the motion of center of mass, which leads to decoupling of Hamiltonian into each momentum sector and only electrons with opposite spin flavor in the same momentum interact. Because dimension of Hilbert space in each momentum sector is only $4$, all eigenstates are readily to be constructed as the product-state of each momentum. Importantly, the ground-states with product-state structure are not trivial and exhibit an exact Mott insulating phase for half-filled electron density if interaction is larger than the bandwidth while other parameters admit Luttinger-theorem-violating NFL for any spatial dimension, which is rare in statistical mechanics and condensed matter physics. After the work of Hatsugai and Kohmoto, several extensions of HK model are proposed, including a bosonic version, momentum dependent interaction and multi-component of electrons.\cite{Continentino1994,Lidsky1998,Nogueira1996} The phase diagram of bosonic HK model proposed by Continentino and Coutinho-Filho has superfluid phase and Mott insulator,\cite{Continentino1994} but the former state is not stable to Bose metal with hidden Fermi surface.\cite{WWYang2023} Although thermodynamics and electron distribution function have been given in Ref.~\cite{Hatsugai1992}, explicit formalism for single-particle Green function and dynamic spin susceptibility first appear in the work of Nogueira and Anda.\cite{Nogueira1996} At the same time, it is realized that the quasiparticles in HK model are not the counterpart in FL but the particles satisfy the exclusion statistics of Haldane.\cite{Haldane1991,YSWu1994,Nayak1994,Byczuk1994} This fact uncovers the NFL feature of quasiparticles and has simplified the treatment of thermodynamics.\cite{Hatsugai1996,Vitoriano2001a,Vitoriano2001b,Ramos2022} Another direction in the early time involves the nature of phase transition between Mott insulator and metallic NFL. Using generic scaling arguments, Continentino and Coutinho-Filho find that both the interaction-driven and chemical-potential-driven Mott-metal transition belong to the universality of Lifshitz transition.\cite{Continentino1994}

After those intensive studies, HK model seems to be forgotten by the condensed matter community and only limited works are published.\cite{Tarasewicz2009,Verhulst2011,Phillips2018,Yeo2019} Such disappointing situation has changed after the work of Phillips, Yeo and Huang,\cite{Phillips2020} who cleverly combine HK Hamiltonian with Bardeen-Cooper-Schrieffer (BCS) pairing interaction. (One may call it HK-BCS model.) Due to the intrinsic NFL feature built in HK model, the resultant superconducting phase exhibits non-BCS behaviors where Bogoliubov quasiparticle is formed by doublon and holon but not the mixture of electron and hole in regular BCS superconductor. Another feature of HK-driven superconductor is that its superfluid density is significantly suppressed, particularly near half-filling, consistent with phenomena in cuprate superconductor. This is a direct consequence of the Mott interaction since proximity to Mott insulator reduces the kinetic energy and hence the effective carrier density. We note that adding BCS pairing into HK Hamiltonian has been done in Ref.~\cite{Tarasewicz2009,Verhulst2011}, about ten years ago but their works do not receive enough attention. The Phillips group's work stimulates further exploration on HK-driven superconductivity and revives the study for generic HK-like systems.\cite{Setty2021a,Setty2021b,Nesselrodt2021,Yang2021,Zhu2021,Li2022,JZhao2022,Zhong2022,Zhong2023,Leeb2023,Wang2023,JZhao2023,MZhao2023,Worm2024,Huang2022,Mai2023,Wysokinski2023,Manning-Coe2023,Guerci2024,YMa2024,Skolimowski2024,JZhao2023b,Jablonowski2023,JWang2024,JWang2023,PMai2024,ZSun2024,Setty2023,Flores-Calderon2024,Sinha2024,PLi2023
,ZQi2024,PMai2024b,Tenkila2024,Souza2023,Setty2023b,Setty2023c,PMai2023c,Bacsi2024,YLLi2024}

Interestingly, the HK-BCS model can be topologically nontrivial with fermionic parity of ground-state as the many-body topological invariant.\cite{Zhu2021} In terms of exact diagonalization (ED) on a $10$-site chain, Zhu et al. also find that the bulk-boundary correspondence works and there exists fermion-like zero mode around boundary.\cite{Zhu2021} The thermodynamics of HK-BCS model has been clarified by Li et al. and Zhao et al..\cite{Li2022,JZhao2022} Li et al. find the system exhibits a two-stage superconductivity feature as temperature
decreases: a first-order superconducting transition occurs at a temperature $T_{c}$ that is followed by a
sudden increase of the superconducting order parameter at a lower temperature $T_{c}'<T_{c}$. At the
first stage, the pairing function arises and the entropy is released only in the vicinity of the two Fermi surfaces; while at the second stage, the pairing function becomes significant and the entropy is further released in deep (single-occupied) region in the Fermi sea.
The first-order transition is confirmed by Zhao et al. and their calculation on nuclear magnetic resonance (NMR) relaxation rate shows the absence of Hebel-Slichter peak.\cite{JZhao2022} Until now, the pairing interaction in HK-BCS model has only been treated with mean-field approximation, does more sophisticated calculation change the established results?

In principle, the interaction in HK model admits non-trivial momentum-dependence which enriches the NFL-like metal states. Yang considers an anisotropic momentum-dependent interaction and it gives rise to Fermi arc-like structure widely observed in pseudogap phase of cuprate
superconductor.\cite{Yang2021} When including on-site Hubbard interaction, those Fermi arcs are proven to be stable under the perturbative Feynman diagram calculation.\cite{Wang2023} It is amusing that a Hartree-Fock treatment on HK interaction in fact leads to exact single-particle Green function, which establishes the starting point of the mentioned Feynman diagram technique. Instead of Yang's construction, if electrons with opposite spin flavor have a non-zero momentum transfer, particularly such momentum is just the antiferromagnetic characteristic wavevector, Fermi arc will emerge naturally as shown by Worm et al..\cite{Worm2024}

The original HK model is a single-band model and its extension to multi-band or multi-orbital version has been pursued by several authors.
\cite{Zhong2022,Jablonowski2023,JWang2024,JWang2023,Manning-Coe2023,Souza2023,PMai2024b,Tenkila2024,ZSun2024}
An asymmetric two-band model, the periodic Anderson model, has been studied by Zhong,\cite{Zhong2022} who shows that the ground-states have Kondo insulator (hybridization insulator), Mott insulator, NFL phase and their existence agrees with a Lieb-Schultz-Mattis (LSM) argument.\cite{Yamanaka1997} A Schrieffer-Wolf transformation on the periodic Anderson model has been performed by Wang, Li and Yang, and it leads to a modified Kondo lattice, which simplifies the analysis on NFL and superconducting pairing.\cite{JWang2024} However, Kondo insulator in these models shows no Kondo effect of single magnetic impurity as seen from the temperature evolution of spectral function. As comparison, a poorman's scaling analysis by Setty implies the magnetic impurity in HK model may form Kondo singlet,\cite{Setty2021a} but state-of-art numerical renormalization group (NRG) calculation is not available since the bath electron is now interacting.\cite{Bulla2008} Furthermore, spin-orbit coupling has been introduced into the model of Zhong, and Jab{\l}onowski et al. find topological Mott insulating phases for odd-integer electron filling.\cite{Jablonowski2023} To stabilize the topological Mott phase, HK-like inter-band and intra-band interactions have to be included, otherwise only trivial Mott phase exists. Two-band/orbital model has also been investigated by Manning-Coe and Bradlyn, whose motivation is to reduce the ground-state degeneracy and break the spin conservation.\cite{Manning-Coe2023} The resulting model in momentum space is similar to multi-site Hubbard model, whose ground-states degeneracy is known to be lifted. The reduction of degeneracy removes the tendency towards magnetic ordering, rendering the ground state stable to infinitesimal Zeeman fields. We note that such stability has already been observed in the Kondo insulator phase of periodic Anderson model.\cite{Zhong2022} Along this line, Phillips' group has explored the multi-orbital HK model and they find large-orbital HK system seems to be a good playground for understanding Mott physics in the classic Hubbard model.\cite{PMai2024b,Tenkila2024,Phillips2010}

Physical observable in FL under orbital magnetic field exhibits periodic oscillation and such magnetic quantum oscillation has been the standard experimental tool to extract Fermi surface and effective mass of electron for metallic phases.\cite{Shoenberg1984,Luttinger1961,Wasserman1996,Chakravarty2011} Calculations of Zhong, Leeb and Knolle suggest quantum oscillation appears in the NFL phase of HK model, as well.\cite{Zhong2023,Leeb2023} Importantly, with the Landau level wavefunction as basis, Leeb and Knolle find the famous Onsager relation (connecting oscillation frequencies with the Fermi surface areas) breaks down due to strong Landau level repulsion. In addition, they discover
unconventional temperature dependencies of quantum oscillation amplitudes
and effective mass renormalizations beyond the classic Lifshitz-Kosevich theory.\cite{Leeb2023} We should emphasize that lattice calculation has not been done and it is not clear if the above findings are valid when flux in each plaquette is not negligible small.\cite{Zhong2023,Chakravarty2011}

In 2001, Anderson and Haldane point out that the quasiparticle spectrum of FL
has an extra $Z_{2}$ symmetry, local in momentum space, which is not generic to
the Hamiltonian with interactions.\cite{Anderson2001} Motivated by this observation,
Huang, Nava and Phillips have examined HK model and they conclude that
although the Mott transition from FL is correctly believed to arise without breaking
any continuous symmetry, a discrete $Z_{2}$ symmetry is broken.\cite{Huang2022} In their perspective, the $Z_{2}$ symmetry-breaking serves as an organizing principle for
Mott physics whatever it arises from the tractable HK model or the intractable Hubbard model.
A follow-up work of Zhao, Nava and Phillips performs a perturbative renormalization-group (RG) analysis on HK model perturbed by short-ranged Hubbard interaction,\cite{JZhao2023} just like Shankar's RG treatment on FL.\cite{Shankar1994} They find that the HK fixed point is stable to perturbations in forward or exchange scattering channel while it will flow to superconducting BCS fixed point described in Ref.~\cite{Phillips2020} if interaction is attractive in the BCS scattering channel. It is fair to say that these works have established the stability of HK fixed point under  weak perturbations.

Although single-particle Green function in HK model is known two decades ago, detailed calculation on two-particle correlation, such as the charge susceptibility or current-current correlation function has only been done in recent years.\cite{MZhao2023,Guerci2024,YMa2024} Zhao et al. focus on the impurity-induced Friedel oscillation in the NFL phase. Combined with linear-response theory and $T$-matrix formalism, they show that Friedel oscillation is dominated by the inter-band transition.  As a byproduct, the charge susceptibility is proved to have Fermi gas form as if the Wick theorem is valid in HK models.\cite{MZhao2023} We note that an exact calculation based on equation of motion for retarded Green function gives the same result. Unexpectedly, Guerci et al. find the long-wave limit of charge susceptibility calculated from Kubo formula (linear-response theory) does not reproduce the correct thermodynamic response since it gives a finite value in the insulating Mott phase.\cite{Guerci2024} The current-current correlation function is also singular, with Kohn's trick to differentiate between metals and insulators by threading a flux in a torus geometry, Guerci et al. uncover the striking property that HK models with an interaction-induced gap in the spectrum sustain a current. Therefore, the results of Guerci et al. indicate that the Mott insulator phase seems to be not insulating due to the infinite-ranged HK interaction. However, the conclusion of Guerci et al. has been challenged by Ma et al., who point out the importance of the non-commutativity of the long-wavelength and thermodynamic limits.\cite{YMa2024} They claim that
when the correct limits are taken, the correlation function from Kubo formula can yield physically reasonable results. Whatever which group's results are correct, such controversy warns us that the
lesson from HK model should be taken with care, particularly when confronting with more realistic models.

Most of studies on HK models work in the momentum space and the periodic boundary condition (PBC) is essential. Replacing PBC with open boundary condition (OBC) damages the solvability and one has to solve the model with numerical tools, such as ED. Skolimowski performs a ED calculation for a $8$-site HK model and the results show that hard edges introduced in OBC enhance the ferromagnetic correlations and the
system undergoes a magnetic transition before reaching the strong coupling limit.\cite{Skolimowski2024} The results of Skolimowski remind us that it is a crucial step
to understand the impact of hard edges in OBC before answering the looming question of the existence of edge states and other topological phenomena in systems with HK interaction. ED calculation has also been used to extract chaotic-integrable transition in
disordered HK model, which suggests a limitation in using out-of-time-order correlator plateau values as a diagnostic tool for chaos.\cite{YLLi2024}

We have introduced several interesting works in the issue of HK models. It seems to us that with the help of HK-like model, our understanding on Mott insulator and related NFL behaviors has been sharpened, Fermi arc and non-BCS-like superconducting phenomena grow with our hand without any formidable numerical simulations. The magnetic quantum oscillation and Friedel oscillation in HK model admit a detailed analysis in the background of NFL, which cannot be realized in generic interacting fermion models. The hidden $Z_{2}$ symmetry-breaking in HK system defines Mottness itself and related RG analysis confirms the robustness of HK physics. Although we are not able to predict the future of HK-related issues, we believe the reviewed works in existing literature are interesting enough and more unexpected physics will be uncovered in near future.
\subsection{The issues not covered in this review}
Due to the limited space and our ability, other interesting issues like topological insulator or more generic symmetry-protected topological states in HK model will not be discussed.\cite{Mai2023,PMai2024,PMai2023c,Wysokinski2023,Setty2023,Setty2023b,Setty2023c,Flores-Calderon2024,Sinha2024,JZhao2023b,Wen2017}
Interested readers may consult papers in the reference list. It is noted that topological order (in fractional quantum Hall effect or gapped quantum spin liquids) has not been discovered in the present HK models, whose existence seems to require external orbital magnetic field and non-HK interaction.

The remaining part of the present review is devoted to a pedagogic introduction of HK model, where its phase diagram, thermodynamics, single-particle and two-particle correlation function are discussed in detail. It is beyond our ability to present all important results in literature of HK models, but we hope readers equipped with those working knowledge will be able to derive main results in HK-related papers by themselves.

\section{Hatsugai-Kohmoto model and its solution}\label{sec:2}

\subsection{Hatsugai-Kohmoto model}
\begin{figure}
\begin{center}
\includegraphics[width=0.45\columnwidth]{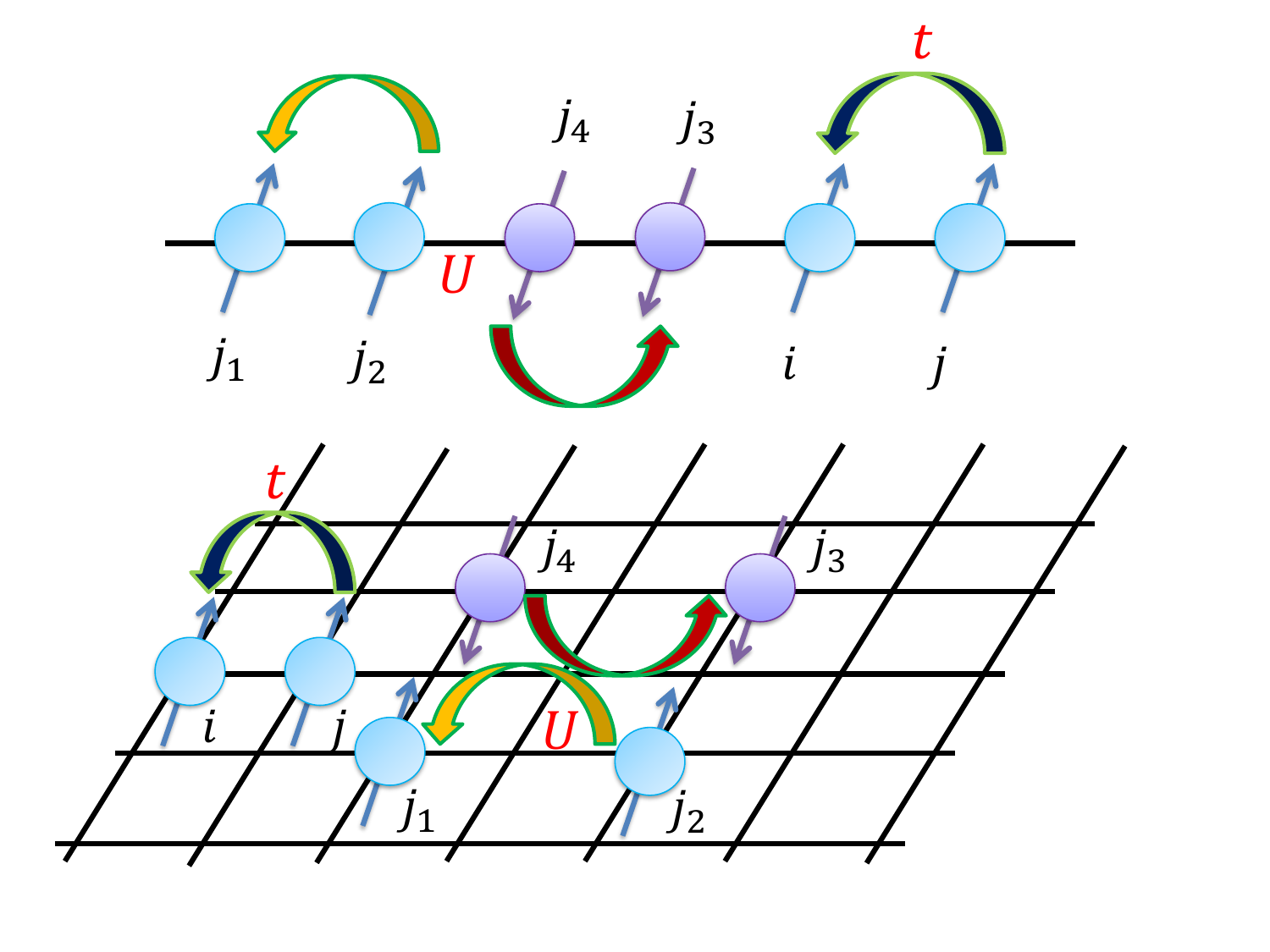}
\includegraphics[width=0.45\columnwidth]{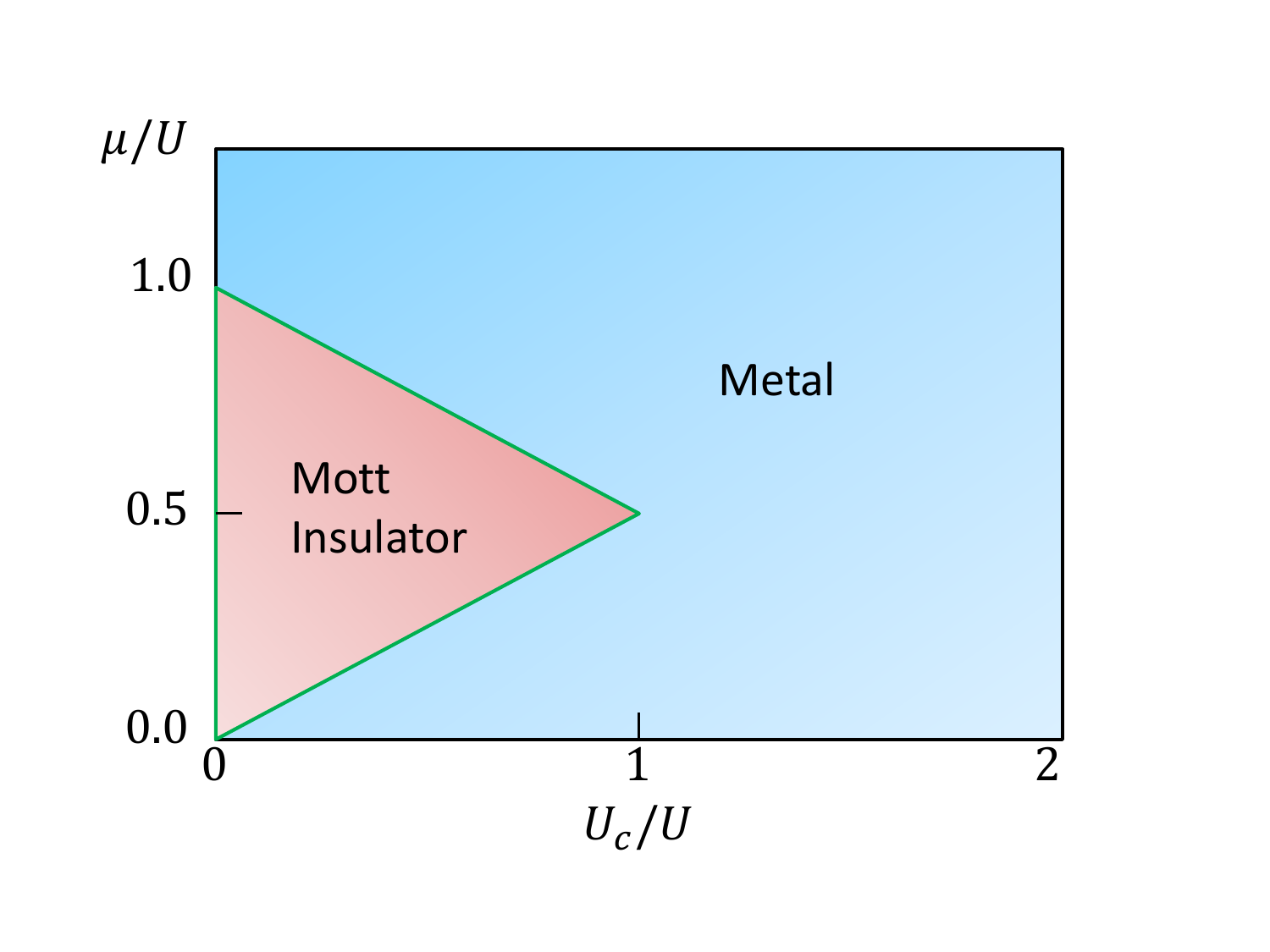}
\caption{\label{fig:HK_int}(Left) The HK model defined on one-dimensional chain and square lattice, noting the conservation of center of mass before and after electrons interact. (Right) Ground-state phase diagram of HK model on hypercubic lattice, $U_{c}=W$ and $W$ denotes the non-interacting electron band-width.}
\end{center}
\end{figure}

Inspired by the randomly-interacting-infinite-ranged spin model invented by Sherrington and Kirpatrick,\cite{Sherrington1975} who desire to understand the elusive phase diagram of spin-glass phase, in 1992 Japanese theorists Hatsugai and Kohmoto propose an exactly solvable interacting many-body model, now named Hatsugai-Kohmoto (HK) model,\cite{Hatsugai1992}
\begin{equation}
\hat{H}_{HK}=-t\sum_{\langle i,j\rangle\sigma}\hat{c}_{i\sigma}^{\dag}\hat{c}_{j\sigma}-\mu\sum_{j\sigma}\hat{c}_{j\sigma}^{\dag}\hat{c}_{j\sigma}+\frac{U}{N_{s}}\sum_{j_{1},j_{2},j_{3},j_{4}}\delta_{j_{1}+j_{3}=j_{2}+j_{4}}\hat{c}_{j_{1}\uparrow}^{\dag}\hat{c}_{j_{2}\uparrow}\hat{c}_{j_{3}\downarrow}^{\dag}\hat{c}_{j_{4}\downarrow},
\end{equation}
where a $d$-dimensional hypercubic lattice with nearest-neighbor-hopping $t$ has been considered though more generic lattice geometry and hopping are easy to include. $\hat{c}_{i\sigma}^{\dag}$ denotes the fermionic creation operator and satisfies standard anti-commutative relation, $\{\hat{c}_{i\sigma},\hat{c}_{j\sigma'}^{\dag}\}=\delta_{ij}\delta_{\sigma\sigma'}$. ($\hat{c}_{i\sigma}^{\dag}$ can also describe bosons but the resulting physics is rather different from fermion systems.\cite{WWYang2023}) $\mu$ denotes the chemical potential, $U$ is the strength of interaction, and $N_{s}$ is the total number of sites. If the lattice constant is setting to be $a$, the volume of hypercube is $V=N_{s}a^{d}=L^{d}$. The key feature of HK model relies on the infinite-ranged interaction term $\hat{H}_{U}=\frac{U}{N_{s}}\sum_{j_{1},j_{2},j_{3},j_{4}}\delta_{j_{1}+j_{3}=j_{2}+j_{4}}\hat{c}_{j_{1}\uparrow}^{\dag}\hat{c}_{j_{2}\uparrow}\hat{c}_{j_{3}\downarrow}^{\dag}\hat{c}_{j_{4}\downarrow}$, which means any electrons feel equal interaction strength if their center of mass is preserved before and after interaction. (denoted by the constraint $\delta_{j_{1}+j_{3}=j_{2}+j_{4}}$). (Fig.~\ref{fig:HK_int})
To give more insight on the above interaction, consider just two-site case,
$\hat{H}_{U}=\frac{U}{2}\left(\hat{c}_{1\uparrow}^{\dag}\hat{c}_{1\uparrow}\hat{c}_{1\downarrow}^{\dag}\hat{c}_{1\downarrow}+
\hat{c}_{1\uparrow}^{\dag}\hat{c}_{1\uparrow}\hat{c}_{2\downarrow}^{\dag}\hat{c}_{2\downarrow}+
\hat{c}_{1\uparrow}^{\dag}\hat{c}_{2\uparrow}\hat{c}_{2\downarrow}^{\dag}\hat{c}_{1\downarrow}+
\hat{c}_{2\uparrow}^{\dag}\hat{c}_{1\uparrow}\hat{c}_{1\downarrow}^{\dag}\hat{c}_{2\downarrow}+
\hat{c}_{2\uparrow}^{\dag}\hat{c}_{2\uparrow}\hat{c}_{1\downarrow}^{\dag}\hat{c}_{1\downarrow}+
\hat{c}_{2\uparrow}^{\dag}\hat{c}_{2\uparrow}\hat{c}_{2\downarrow}^{\dag}\hat{c}_{2\downarrow}\right).$
Except for the on-site Hubbard interaction $\hat{c}_{1\uparrow}^{\dag}\hat{c}_{1\uparrow}\hat{c}_{1\downarrow}^{\dag}\hat{c}_{1\downarrow},~\hat{c}_{2\uparrow}^{\dag}\hat{c}_{2\uparrow}\hat{c}_{2\downarrow}^{\dag}\hat{c}_{2\downarrow}$, there are inter-site interaction term $\hat{c}_{1\uparrow}^{\dag}\hat{c}_{1\uparrow}\hat{c}_{2\downarrow}^{\dag}\hat{c}_{2\downarrow},~\hat{c}_{2\uparrow}^{\dag}\hat{c}_{2\uparrow}\hat{c}_{1\downarrow}^{\dag}\hat{c}_{1\downarrow}$ and inter-site spin exchange interaction $\hat{c}_{1\uparrow}^{\dag}\hat{c}_{2\uparrow}\hat{c}_{2\downarrow}^{\dag}\hat{c}_{1\downarrow},
~\hat{c}_{2\uparrow}^{\dag}\hat{c}_{1\uparrow}\hat{c}_{1\downarrow}^{\dag}\hat{c}_{2\downarrow}$. Thus, the HK interaction term not only includes the standard Hubbard interaction but also the magnetic exchange interaction.

Use the Fourier transformation $\hat{c}_{j\sigma}=\frac{1}{\sqrt{N_{s}}}\sum_{k}\hat{c}_{k\sigma}e^{ikR_{j}}$ with PBC, the interaction term $\hat{H}_{U}$ is transformed into
\begin{eqnarray}
  \hat{H}_{U} &=& \frac{U}{N_{s}^{3}}\sum_{k_{1},k_{2},k_{3},k_{4}}\hat{c}_{k_{1}\uparrow}^{\dag}\hat{c}_{k_{2}\uparrow}\hat{c}_{k_{3}\downarrow}^{\dag}\hat{c}_{k_{4}\downarrow}\sum_{j_{1},j_{2},j_{3},j_{4}}\delta_{j_{1}+j_{3}=j_{2}+j_{4}} e^{-i(k_{1}R_{j_{1}}+k_{3}R_{j_{3}})}e^{i(k_{2}R_{j_{2}}+k_{4}R_{j_{4}})} \nonumber\\
   &=&U\sum_{k_{1},k_{2},k_{3},k_{4}}\hat{c}_{k_{1}\uparrow}^{\dag}\hat{c}_{k_{2}\uparrow}\hat{c}_{k_{3}\downarrow}^{\dag}\hat{c}_{k_{4}\downarrow} \frac{1}{N_{s}}\sum_{j_{1}}e^{i(k_{4}-k_{1})R_{j_{1}}}\frac{1}{N_{s}}\sum_{j_{3}}e^{i(k_{4}-k_{3})R_{j_{3}}} \frac{1}{N_{s}}\sum_{j_{2}}e^{i(-k_{4}+k_{2})R_{j_{2}}}\nonumber\\
   &=& U\sum_{k_{1},k_{2},k_{3},k_{4}}\hat{c}_{k_{1}\uparrow}^{\dag}\hat{c}_{k_{2}\uparrow}\hat{c}_{k_{3}\downarrow}^{\dag}\hat{c}_{k_{4}\downarrow} \delta_{k_{4},k_{1}}\delta_{k_{4},k_{3}}\delta_{k_{4},k_{2}}\nonumber\\
   &=&U\sum_{k_{4}}\hat{c}_{k_{4}\uparrow}^{\dag}\hat{c}_{k_{4}\uparrow}\hat{c}_{k_{4}\downarrow}^{\dag}\hat{c}_{k_{4}\downarrow}=U\sum_{k}\hat{c}_{k\uparrow}^{\dag}\hat{c}_{k\uparrow}\hat{c}_{k\downarrow}^{\dag}\hat{c}_{k\downarrow}=U\sum_{k}\hat{n}_{k\uparrow}\hat{n}_{k\downarrow}
\end{eqnarray}
Here, $\hat{n}_{k\sigma}\equiv\hat{c}_{k\sigma}^{\dag}\hat{c}_{k\sigma}$. Note that the interaction only affects electrons with the same momentum and opposite spin-direction. Now, the Hamiltonian reads
\begin{equation}
\hat{H}_{HK}=\sum_{k}\hat{H}_{k}=\sum_{k}\left[(\varepsilon_{k}-\mu)(\hat{n}_{k\uparrow}+\hat{n}_{k\downarrow})+U\hat{n}_{k\uparrow}\hat{n}_{k\downarrow}\right]
\end{equation}
It is clear that the HK Hamiltonian $\hat{H}_{HK}$ is decoupled into each momentum sector $\hat{H}_{k}$, as a result of infinite-ranged interaction and the conservation of center of mass. If the condition of infinite-ranged interaction is relaxed, the resulting model becomes the dipole-conserved Hubbard model with exotic NFL phase and fracton-like dynamics.\cite{Lake2023,Nandkishore2019} For $d$-dimensional hypercubic lattice with the nearest-neighbor-hopping $t$, the non-interacting electron dispersion is $\varepsilon_{k}=-2t\sum_{a=1}^{d}\cos k_{a}$. Furthermore, the interaction can have momentum-dependence $U\rightarrow U(k)$ or $U(k,k')=(2\pi)^{2}U(|k|)\delta(|k|-|k'|)\frac{g(\phi)}{k}$
with $\cos\phi=\frac{k\cdot k'}{|k||k'|}$. Interestingly, the latter one corresponds to   $\frac{1}{2V}\sum_{k,k'}f_{k,k'}\hat{n}_{k}\hat{n}_{k'}$ just like the Landau's energy functional in FL,\cite{Lidsky1998} although the operator feature of $\hat{n}_{k},\hat{n}_{k'}$ are replaced by quasiparticle distribution $\delta n_{k},\delta n_{k'}$ in FL. It is this replacement that neglects the correlation between electrons and leads to distinct physics in HK and FL. We also notice a similar model due to Baskaran
\begin{eqnarray}
\hat{H}=\sum_{k}(\varepsilon_{k}-\mu)\hat{c}_{k\sigma}^{\dag}\hat{c}_{k\sigma}+J\left(\sum_{k}\hat{S}_{k}\right)^{2}-J\sum_{k}\hat{S}_{k}^{2},~~\hat{S}_{k}=\sum_{\alpha\beta}\hat{c}_{k\alpha}^{\dag}\sigma_{\alpha\beta}\hat{c}_{k\beta}\nonumber
\end{eqnarray}
whose properties are equivalent to HK model if setting $U=3J/2$.\cite{Baskaran1991}
\subsubsection{Eigenstates and eigen-energy of HK model}
Now, the Hamiltonian decouples into independent sectors $\hat{H}_{k}$ and each $\hat{H}_{k}$ works as a Hubbard atom in momentum space. For $\hat{H}_{k}$, it has four eigenstates and eigen-energy,
\begin{eqnarray}
&&  |0\rangle_{k}~~~~~~~~~~~~~~~~~~~~~~~E_{0}(k)=0 \nonumber\\
&&  |\uparrow\rangle_{k}=\hat{c}_{k\uparrow}^{\dag}|0\rangle_{k}~~~~~~~~~E_{u}(k)=\varepsilon_{k}-\mu \nonumber\\
&&  |\downarrow\rangle_{k}=\hat{c}_{k\downarrow}^{\dag}|0\rangle_{k}~~~~~~~~~E_{n}(k)=\varepsilon_{k}-\mu \nonumber\\
&&  |\uparrow\downarrow\rangle_{k}=\hat{c}_{k\uparrow}^{\dag}\hat{c}_{k\downarrow}^{\dag}|0\rangle_{k}~~~~E_{d}(k)=2\varepsilon_{k}-2\mu+U.
\end{eqnarray}
These states correspond to empty occupation, single occupation with spin-up/down electron and double occupation. Thus, any many-body state of HK model can be constructed as
\begin{eqnarray}
  |\Psi\rangle=\prod_{k}|\Phi\rangle_{k},~~|\Phi\rangle_{k}=a|0\rangle_{k}+b|\uparrow\rangle_{k}+c|\downarrow\rangle_{k}+d|\uparrow\downarrow\rangle_{k}=\sum_{\alpha=0,\uparrow,\downarrow,\uparrow\downarrow}c_{\alpha}|\alpha\rangle_{k}\nonumber
\end{eqnarray}
Particularly, the eigenstates are readily to be obtained if $|\Phi\rangle_{k}$ is chosen as one of four eigenstates. Since any eigenstates can be constructed, HK model is solvable. Note that the above construction of eigenstates does not depend on specific lattice geometry or electron filling, so HK model is solvable for any dimension or any electron density. (For lattice with sublattice structure like honeycomb or Lieb lattice,\cite{Manning-Coe2023} $\hat{H}_{U}$ has to be tuned to enforce the solvability.) In fact, just like Kitaev's toric-code and honeycomb lattice model,\cite{Kitaev2003,Kitaev2006} HK model belongs to the frustration-free many-body system since $[\hat{H}_{k},\hat{H}_{k'}]=0,[\hat{H}_{k},\hat{H}]=0$.\cite{Zeng2019} As a result, the solvability of HK model is robust and permits perturbation around such solvable limit.

\begin{figure}
\begin{center}
\includegraphics[width=0.65\columnwidth]{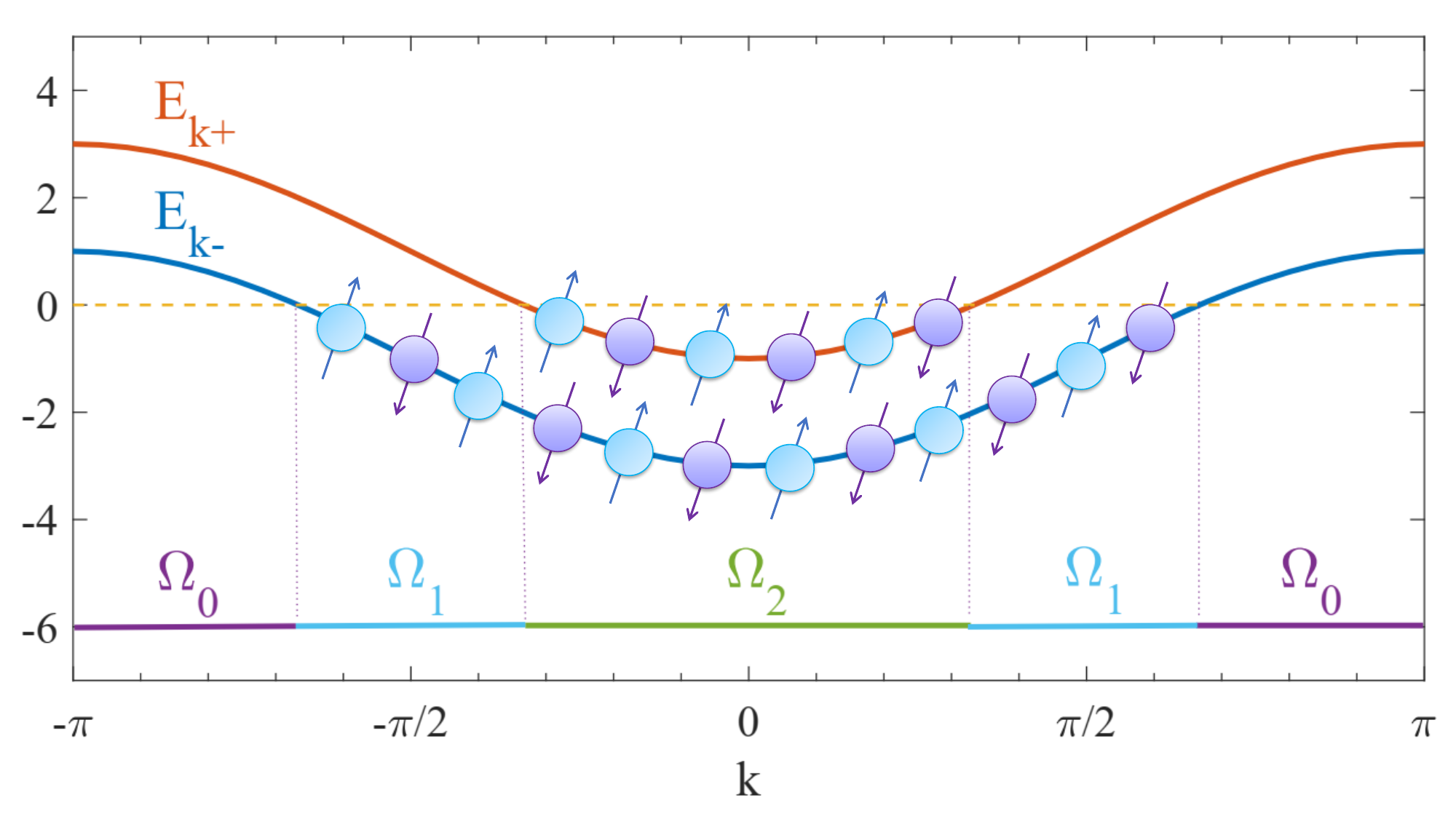}
\caption{\label{fig:HK_bands} The empty ($\Omega_{0}$), single ($\Omega_{1}$) and double ($\Omega_{2}$) occupation regime for HK model in momentum space. Here, a $d=1$ HK system is assumed and $E_{k+}=\varepsilon_{k}-\mu+U,E_{k-}=\varepsilon_{k}-\mu$ are quasi-particle bands.}
\end{center}
\end{figure}

Because each eigenstate has different occupation configuration, alternatively, it can be written as
(Fig.~\ref{fig:HK_bands})
\begin{equation}
  |\Psi\rangle=\prod_{k\in \Omega_{0}}|0\rangle_{k}\prod_{k\in \Omega_{1}}|\alpha=\uparrow,\downarrow\rangle_{k}\prod_{k\in \Omega_{2}}|\uparrow\downarrow\rangle_{k}
\end{equation}
where $\Omega_{0},\Omega_{1},\Omega_{2}$ denote the momentum range of empty occupation, single occupation and double occupation, respectively. Obviously, the single occupation regime is degenerated, which means the eigenstates are highly-degenerated, including the ground-states. The degeneracy can be lifted due to Zeeman energy if an infinitesimal uniform magnetic field is applied, then a unique ferromagnetic ground-state is obtained. ($|\Psi_{FM}\rangle=\prod_{k\in \Omega_{0}}|0\rangle_{k}\prod_{k\in \Omega_{1}}|\uparrow\rangle_{k}\prod_{k\in \Omega_{2}}|\uparrow\downarrow\rangle_{k}$ (magnetic field is along $-z$-direction) or $|\Psi_{FM}\rangle=\prod_{k\in \Omega_{0}}|0\rangle_{k}\prod_{k\in \Omega_{1}}|\downarrow\rangle_{k}\prod_{k\in \Omega_{2}}|\uparrow\downarrow\rangle_{k}$ (magnetic field is along $z$-direction))

Now, the essential issue is what the ground-states are for generic parameters ($t,U,\mu$ or $t,U,n$ with $n$ denoting electron density)
\subsection{The ground-state phase diagram of HK model}
To find the ground-states of HK model, let us consider a special case, namely all energy of single occupation is smaller than any empty or double occupation energy:
\begin{eqnarray}
(E_{u}(k))_{max}<(E_{0}(k))_{min},(E_{d}(k))_{min}\nonumber
\end{eqnarray}
In this case, the ground-states are formed by single occupation states, i.e. $|\Psi_{g}\rangle=\prod_{k\in\Omega_{1}}|\alpha=\uparrow,\downarrow\rangle_{k}$ with $\Omega_{1}$ occupies the first Brillouin zone (BZ). At the same time, the number of electron is equal to the total number of sites, (each momentum is occupied by one electron) thus the system is half-filled. An obvious question is that whether the system is a metal or an insulator. This can be identified by the gap of electron-hole excitation. When the gap is zero, one finds a metal while an insulator has finite gap.

Consider an electron has been injected into the ground-state $|\Psi_{g}\rangle$, its wave-function reads
\begin{eqnarray}
|\Psi_{e}^{k',\sigma'}\rangle=\hat{c}_{k'\sigma'}^{\dag}|\Phi_{g}\rangle=|\uparrow\downarrow\rangle_{k'}\prod_{k\neq k',k\in\Omega_{1}}|\alpha=\uparrow,\downarrow\rangle_{k}\nonumber
\end{eqnarray}
The excitation energy above ground-state is
\begin{eqnarray}
\Delta E_{k'\sigma'}^{e}=\langle \Psi^{k',\sigma'}_{e}|\hat{H}|\Psi_{e}^{k',\sigma'}\rangle-\langle\Phi_{g}|\hat{H}|\Phi_{g}\rangle=E_{d}(k')-E_{u}(k')=\varepsilon_{k'}-\mu+U\nonumber
\end{eqnarray}
Similarly, wave-function with injected hole in ground-state is $|\Psi_{h}^{k',\sigma'}\rangle=\hat{c}_{k'\sigma'}|\Phi_{g}\rangle$, and its excitation energy is $\Delta E_{k'\sigma'}^{h}=E_{0}(k')-E_{u}(k')=\mu-\varepsilon_{k'}$. For insulating states, it is required that for each $k'$, $\Delta E_{k'\sigma'}^{h},\Delta E_{k'\sigma'}^{e}$ should be larger than zero, i.e. $(\Delta E_{k'\sigma'}^{e})_{min}>0, (\Delta E_{k'\sigma'}^{h})_{min}>0$. In contrast, the metallic state means some of excitation energy is smaller than zero.

Obviously, $(\Delta E_{k'\sigma'}^{e})_{min},(\Delta E_{k'\sigma'}^{h})_{min}\rightarrow0^{+}$ give rise to the boundary between metal and insulator:
\begin{eqnarray}
(\varepsilon_{k'})_{min}-\mu+U=0,~~~~\mu-(\varepsilon_{k'})_{max}=0\nonumber
\end{eqnarray}
For $d$-dimensional hypercubic lattice, we assume $\varepsilon_{k}\in[-\frac{W}{2},\frac{W}{2}]$ with $W=4td$ being the non-interacting electron band-width. Inserting $(\varepsilon_{k})_{max}=\frac{W}{2},(\varepsilon_{k})_{min}=-\frac{W}{2}$, we find two solutions
\begin{equation}
\frac{\mu}{U}=1-\frac{W}{2U};~~~~\frac{\mu}{U}=\frac{W}{2U}.
\end{equation}
Combined with $W/U=0$, these three equations determine the triangular regime,\cite{Continentino1994,Hatsugai1996} corresponding to the insulating phase while other regime describes the metallic phase. (Fig.~\ref{fig:HK_int})

Note that if we turn off the interaction, the present half-filled insulating state ($n=1$) must be metallic, which means such insulator is interaction-driven, and we may call it Mott insulator. In Hubbard model, Mott insulator in $d\geqslant2$ can have long-ranged antiferromagnetic order but the $d=1$ case is a paramagnet with strong antiferromagnetic fluctuation. Mott insulating state in HK model is always paramagnetic for any spatial dimension and extra non-HK interaction is required to induce magnetic orders. (To arrive at the paramagnetic solution, one must average over all degenerated ground-states.) When particle-hole symmetry is conserved, $\mu=U/2$ is satisfied. By using $\mu/U=W/(2U)$, the critical strength of metal-Mott insulator transition is $U_{c}=W$. This is the interaction-tuned Mott insulator. For fixed $U$, density-tuned Mott insulator can be obtained by tuning chemical potential $\mu$.
\begin{figure}
\begin{center}
\includegraphics[width=0.65\columnwidth]{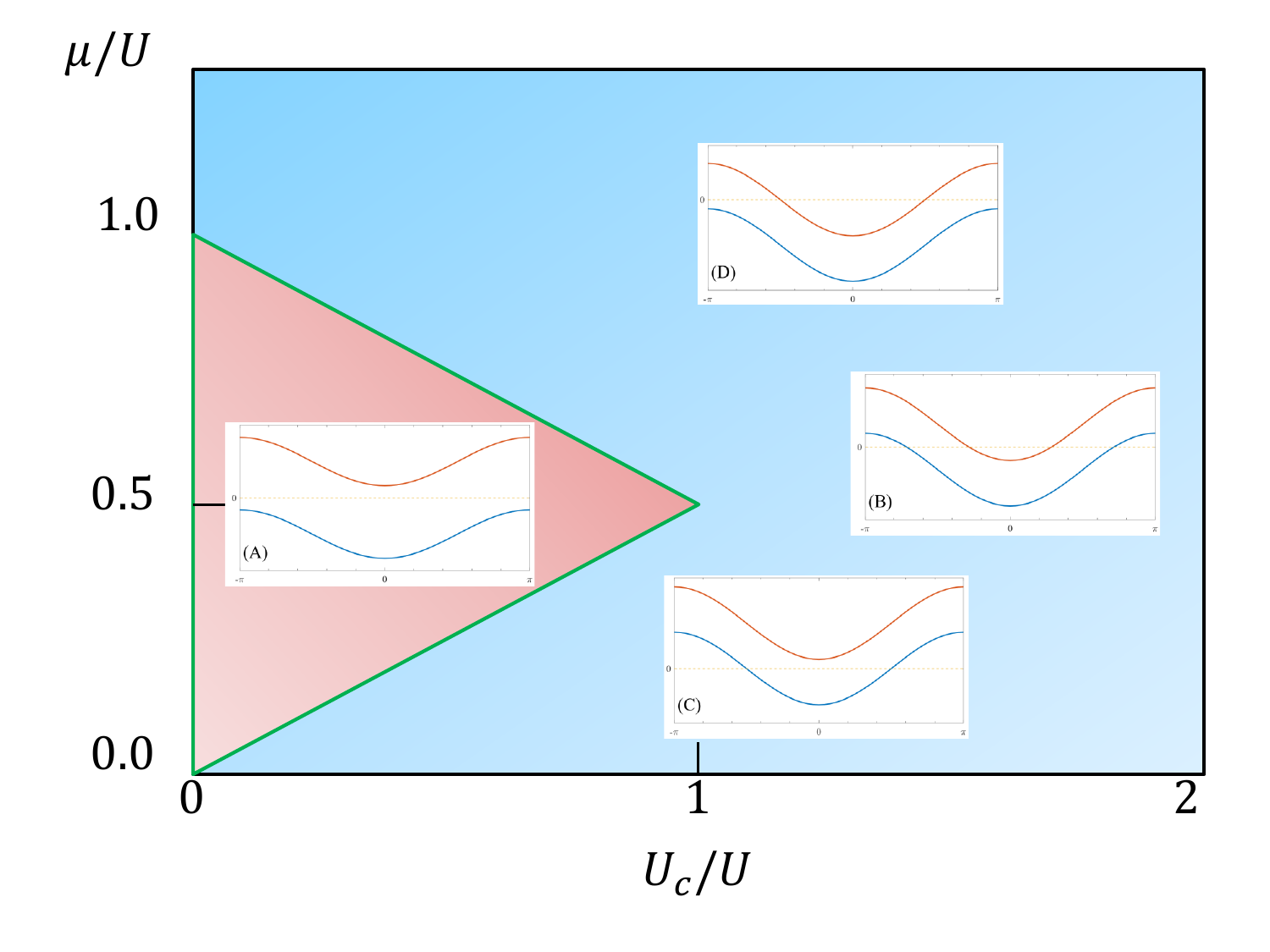}
\caption{\label{fig:HK_bands2} Band structure of HK model for different parameters, (A) $U>W=U_{c},\mu=U/2$,(B) $U<W,\mu=U/2$,(C) $U<W,\mu=0$ and (D) $U<W,\mu=U$. The Mott insulator has definite gap between two bands while the bands in metallic state are partially occupied.}
\end{center}
\end{figure}

\subsubsection{A Lieb-Schultz-Mattis argument}
We note that the Mott insulator appears in the exactly half-filling case and any deviation from this special filling gives rise to metallic states. This fact can be understood in terms of a Lieb-Schultz-Mattis (LSM) argument,\cite{Yamanaka1997} which has been used for the periodic Anderson model with HK interaction.\cite{Zhong2022}

Consider a $d=1$ HK model in real space with PBC,
\begin{eqnarray}
\hat{H}=-t\sum_{j\sigma}(\hat{c}_{j\sigma}^{\dag}\hat{c}_{j+1\sigma}+h.c)-\mu\sum_{j\sigma}\hat{c}_{j\sigma}^{\dag}\hat{c}_{j\sigma}+\frac{U}{N_{s}}
\sum_{j_{1}j_{2}j_{3}j_{4}}\delta_{j_{1}+j_{3}=j_{2}+j_{4}}\hat{c}_{j_{1}\uparrow}^{\dag}\hat{c}_{j_{2}\uparrow}\hat{c}_{j_{3}\downarrow}^{\dag}\hat{c}_{j_{4}\downarrow}.\nonumber
\end{eqnarray}
and define the twist operator $\hat{U}=e^{i\sum_{j=1}^{N_{s}}\frac{2\pi j}{N_{s}}\sum_{\sigma}\hat{c}_{j\sigma}^{\dag}\hat{c}_{j\sigma}}$. If we denote the ground-state as $|\Psi_{0}\rangle$, then a new state is constructed by applying $\hat{U}$, i.e. the twisted state $\hat{U}|\Psi_{0}\rangle$. So, one can calculate the energy difference
\begin{eqnarray}
\Delta E=\langle\Psi_{0}|\hat{U}^{-1}\hat{H}\hat{U}|\Psi_{0}\rangle-\langle \Psi_{0}|\hat{H}|\Psi_{0}\rangle=\sum_{\sigma}\sum_{j=1}^{N_{s}}(2-e^{-i2\pi/N_{s}}-e^{i2\pi/N_{s}})t\langle \hat{c}_{j\sigma}^{\dag}\hat{c}_{j+1\sigma}\rangle,\nonumber
\end{eqnarray}
which does not depend on HK interaction, thanks to its conservation of center of mass.
When $N_{s}>>1$, $\Delta E\sim \mathcal{O}(1/N_{s})$, thus there exists at least one low-energy state near ground-state. Furthermore, for the translation operator $\hat{T}$, we have $\hat{T}\hat{U}\hat{T}^{-1}=\hat{U}e^{-i2\pi \hat{n}}$. ($\hat{n}=\frac{1}{N_{s}}\sum_{j\sigma}\hat{c}_{j\sigma}^{\dag}\hat{c}_{j\sigma}$) Assume the ground-state $|\Psi_{0}\rangle$ has particle density $n$ and momentum $P_{0}$, then
\begin{equation}
\hat{T}\hat{U}|\Psi_{0}\rangle=\hat{U}\hat{T}e^{-i2\pi \hat{n}}|\Psi_{0}\rangle=e^{-i2\pi n}e^{-iP_{0}}\hat{U}|\Psi_{0}\rangle,
\end{equation}
which means the twisted state is the eigenstate of momentum $2\pi n+P_{0}$. If $n$ is not an integer, $\hat{U}|\Psi_{0}\rangle$ and $|\Psi_{0}\rangle$ must be orthogonal, thus the system is gapless in this situation and it corresponds to metallic state. In contrast, when $n$ is an integer, we expect $\hat{U}|\Psi_{0}\rangle$ and $|\Psi_{0}\rangle$ are the same state which suggests that there exists no low-energy state and the system should be an insulator. Therefore, $n=0,1,2$ in HK model correspond to insulating states ($n=0,2$ is band insulator while $n=1$ is Mott insulator) and generic electron's density implies metallic state. Although above argument is only applicable in one spatial dimension, we believe its extension still tells us that the Mott insulator is stable in $n=1$ for $d\geqslant2$ systems.\cite{Oshikawa2000b,Hastings2004}

\subsection{Partition function and thermodynamics}
\subsubsection{Electron occupation and quasi-Fermi wavevector}

To clarify the difference between Mott and metal state, we calculate the electron distribution function in ground-state, i.e. $n_{k\sigma}=\langle \hat{n}_{k\sigma}\rangle$. However, due to large degeneracy in the ground-state, we have to average over all degenerated states. In other words, the considered ground-state is in fact a mixed state.

To do the calculation over all degenerated ground-states, we first calculate quantities in finite temperature, and then take the zero-temperature limit. When finite-temperature formalism is used, the Hamiltonian is decoupled into each sector in momentum space, such that the partition function is just the multiple of each part,
\begin{eqnarray}
\mathcal{Z}&=&\mathrm{Tr} e^{-\beta\hat{H}}=\prod_{k}\mathcal{Z}_{k}=\prod_{k}\mathrm{Tr}e^{-\beta \hat{H}_{k}}
=\prod_{k}\sum_{\alpha=0,\uparrow,\downarrow,\uparrow\downarrow}\langle\alpha|_{k}e^{-\beta \hat{H}_{k}}|\alpha\rangle_{k}\nonumber\\
&=&\prod_{k}(1+2e^{-\beta(\varepsilon_{k}-\mu)}+e^{-\beta(2\varepsilon_{k}-2\mu+U)})=\prod_{k}f_{k}
\end{eqnarray}
Here, we have defined the factor
\begin{equation}
f_{k}=1+2z_{k}+z_{k}^{2}e^{-\beta U};~~~~z_{k}=e^{-\beta(\varepsilon_{k}-\mu)}.
\end{equation}
Using the above partition function, we calculate $n_{k\sigma}$,
\begin{eqnarray}
n_{k\sigma}&=&\frac{1}{\mathcal{Z}}\mathrm{Tr }\hat{n}_{k\sigma}e^{-\beta \hat{H}}=\frac{\prod_{k'\neq k}\mathcal{Z}_{k'}\mathrm{Tr}\hat{n}_{k\sigma}e^{-\beta \hat{H}_{k}}}{\prod_{k'}\mathcal{Z}_{k'}}=\frac{\mathrm{Tr}\hat{n}_{k\sigma}e^{-\beta \hat{H}_{k}}}{\mathcal{Z}_{k}}\nonumber\\
&=&\frac{e^{-\beta(\varepsilon_{k}-\mu)}+e^{-\beta(2\varepsilon_{k}-2\mu+U)}}{1+2e^{-\beta(\varepsilon_{k}-\mu)}+e^{-\beta(2\varepsilon_{k}-2\mu+U)}}=\frac{f_{k}-z_{k}-1}{f_{k}}\nonumber\\
&=&\frac{f_{F}(\varepsilon_{k}-\mu)}{f_{F}(\varepsilon_{k}-\mu)+1-f_{F}(\varepsilon_{k}-\mu+U)}.
\end{eqnarray}
where $f_{F}(x)=1/(e^{x/T}+1)$ is the standard Fermi distribution function. It is clear that the distribution function in HK model is different from the ones in free electron gas. Furthermore, let us consider its zero-temperature limit
\begin{eqnarray}
n_{k\sigma}^{T=0}=\frac{1}{2}\left[\theta(\mu-\varepsilon_{k})+\theta(\mu-\varepsilon_{k}-U)\right],
\end{eqnarray}
where the unitstep function has the property $\theta(x)=1$ for $x>0$ while $\theta(x)=0$ for $x<0$.

\begin{figure}
\begin{center}
\includegraphics[width=0.75\columnwidth]{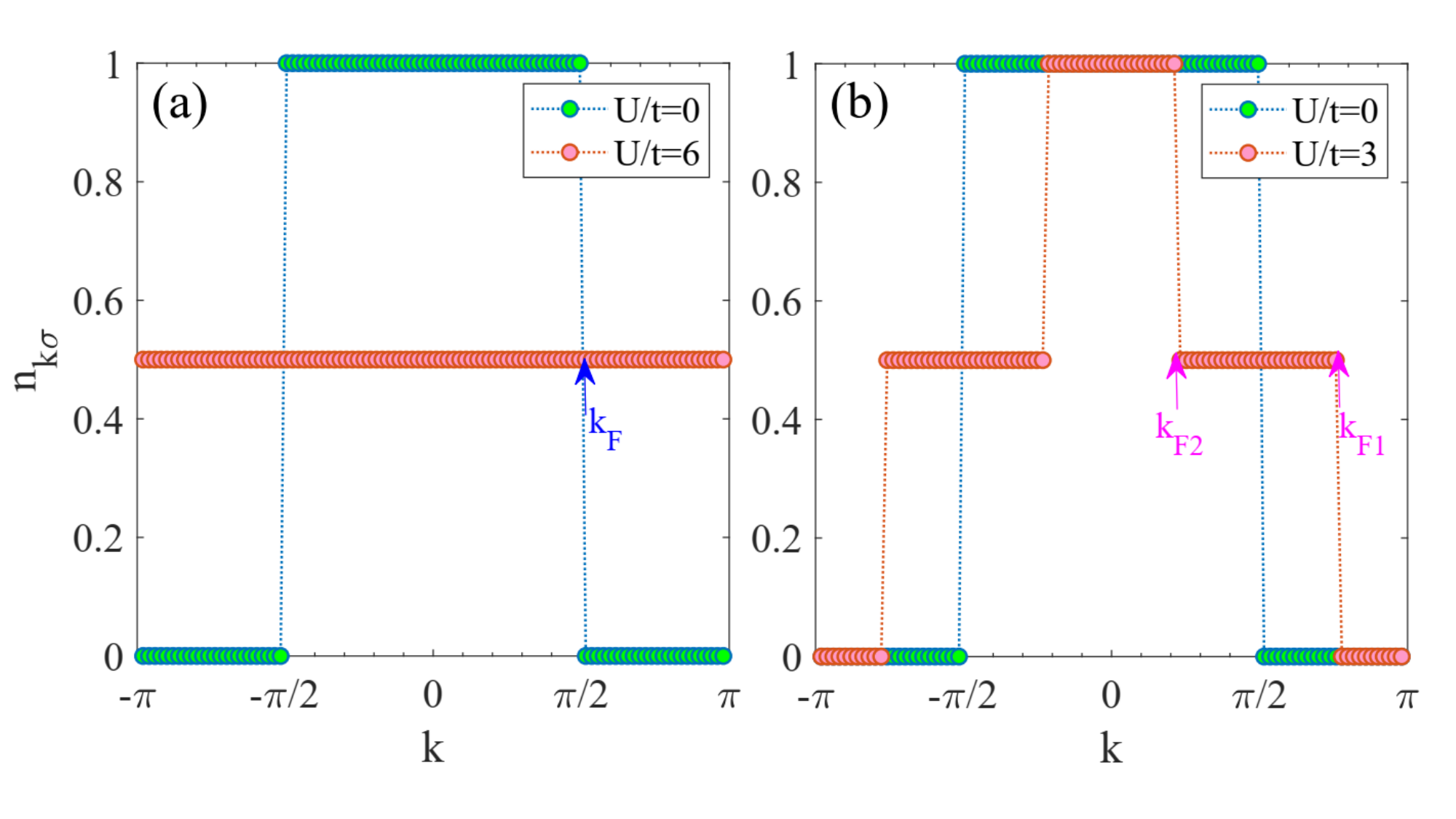}
\caption{\label{fig:HK_1D}The ground-state electron occupation $n_{k\sigma}=\langle \hat{c}_{k\sigma}^{\dag}\hat{c}_{k\sigma}\rangle$. (a) Mott insulator with $U>W=4t$; (b) Metal state for $U<W=4t$. For comparison, Fermi gas with $U/t=0$ is also shown.}
\end{center}
\end{figure}

For one-dimensional HK model, its dispersion is $\varepsilon_{k}=-2t\cos k$ and the band-width is $W=4t$. In this case,
\begin{eqnarray}
n_{k\sigma}^{T=0}=\frac{1}{2}\left[\theta(\mu+2t\cos k)+\theta(\mu+2t\cos k-U)\right].
\end{eqnarray}
For given $\mu,U,t$, we obtain the electron occupation. In the symmetric half-filling case with $\mu=U/2$, $n_{k\sigma}^{T=0}=\frac{1}{2}\left[\theta(2t\cos k+U/2)+\theta(2t\cos k-U/2)\right]$.
When $U>W=4t$, the system is in Mott insulator, which means each momentum is single occupied and the distribution function should be $1/2$ for any momentum in the first BZ. Interestingly, the half-filled $d=1$ Hubbard model in the strong coupling limit is an Mott insulator and has  $n_{k\sigma}^{T=0}=1/2-2.78\frac{t\cos k}{U_{H}}+\mathcal{O}(t^{2}/U_{H}^{3})$,\cite{Takahashi1977} ($U_{H}$ is the on-site Hubbard interaction) which is similar to the present result. In contrast,
the Fermi gas state with $U=0$ has two electrons in each occupied momentum
$k\in[-k_{F},k_{F}],k_{F}=\pi/2$. (Fig.~\ref{fig:HK_1D}(a))

If $U<W$, Mott state transits into a metal state. In this situation, there are single and double occupation regime. In the double occupation regime, $n_{k\sigma}=1$ and it is $1/2$ in the single occupation regime, which is different from Fermi gas, in other words, metallic state in HK model with $U<W$ should be a kind of NFL.

From Fig.~\ref{fig:HK_1D}(b), we see that there are two jumps in electron distribution function, ($k>0$ regime) they correspond to $n_{k\sigma}=1\rightarrow1/2$ and $1/2\rightarrow0$. Now, we label the corresponding momentum as $k_{F2}$ and $k_{F1}$, (as quasi-Fermi wavevector) which are determined by $\varepsilon_{k}=\mu-U$, $\varepsilon_{k}=\mu$ as
\begin{equation}
k_{F2}=\left|\arccos\frac{U-\mu}{2t}\right|,~~~~k_{F1}=\left|\arccos\frac{-\mu}{2t}\right|.
\end{equation}
In terms of $k_{F2},k_{F1}$, the density of electron is found to be
\begin{equation}
n=\overbrace{2k_{F2}}^{\Omega_{2}}\frac{2}{2\pi}+\overbrace{2(k_{F1}-k_{F2})}^{\Omega_{1}}\frac{2}{2\pi}\frac{1}{2}=(k_{F1}+k_{F2})\frac{2}{2\pi}.
\end{equation}
Here $\frac{2}{2\pi}$ is the density of state in momentum space including the spin degeneracy, and the factor $\frac{1}{2}$ considers the fact that $n_{k\sigma}=1/2$ in the single occupation regime. Note that this result is not consistent with the Luttinger theorem satisfied by Fermi gas, which states $n=2k_{F}\frac{2}{2\pi}$.\cite{Luttinger1960} Specifically, Luttinger theorem means the density of electron $n$ is determined by volume closed by Fermi surface $V_{FS}$,\cite{Dzyaloshinskii2003}
\begin{equation}
n=2\int\frac{d^{d}k}{(2\pi)^{d}}\theta(\mathrm{Re}G(k,\omega=0))=
2\int_{\mathrm{Re}G(k,\omega=0)>0}\frac{d^{d}k}{(2\pi)^{d}}
=2\frac{V_{FS}}{(2\pi)^{d}}.\label{eq:Luttinger_theorem}
\end{equation}
where $G(k,\omega)$ is the single-particle Green function and interaction does not change the above formula. For $d=1$ Fermi gas,
$G(k,\omega)=1/(\omega-\varepsilon_{k}+\mu)$ and $G(k=k_{F},\omega=0)=0$ gives Fermi wavevector $k_{F}$. Using Eq.~\ref{eq:Luttinger_theorem}, we find $n=2k_{F}\frac{2}{2\pi}$, which is not equal to $(k_{F1}+k_{F2})\frac{2}{2\pi}$. Thus, this violation of Luttinger theorem suggests the NFL feature of HK model.

\begin{figure}
\begin{center}
\includegraphics[width=0.75\columnwidth]{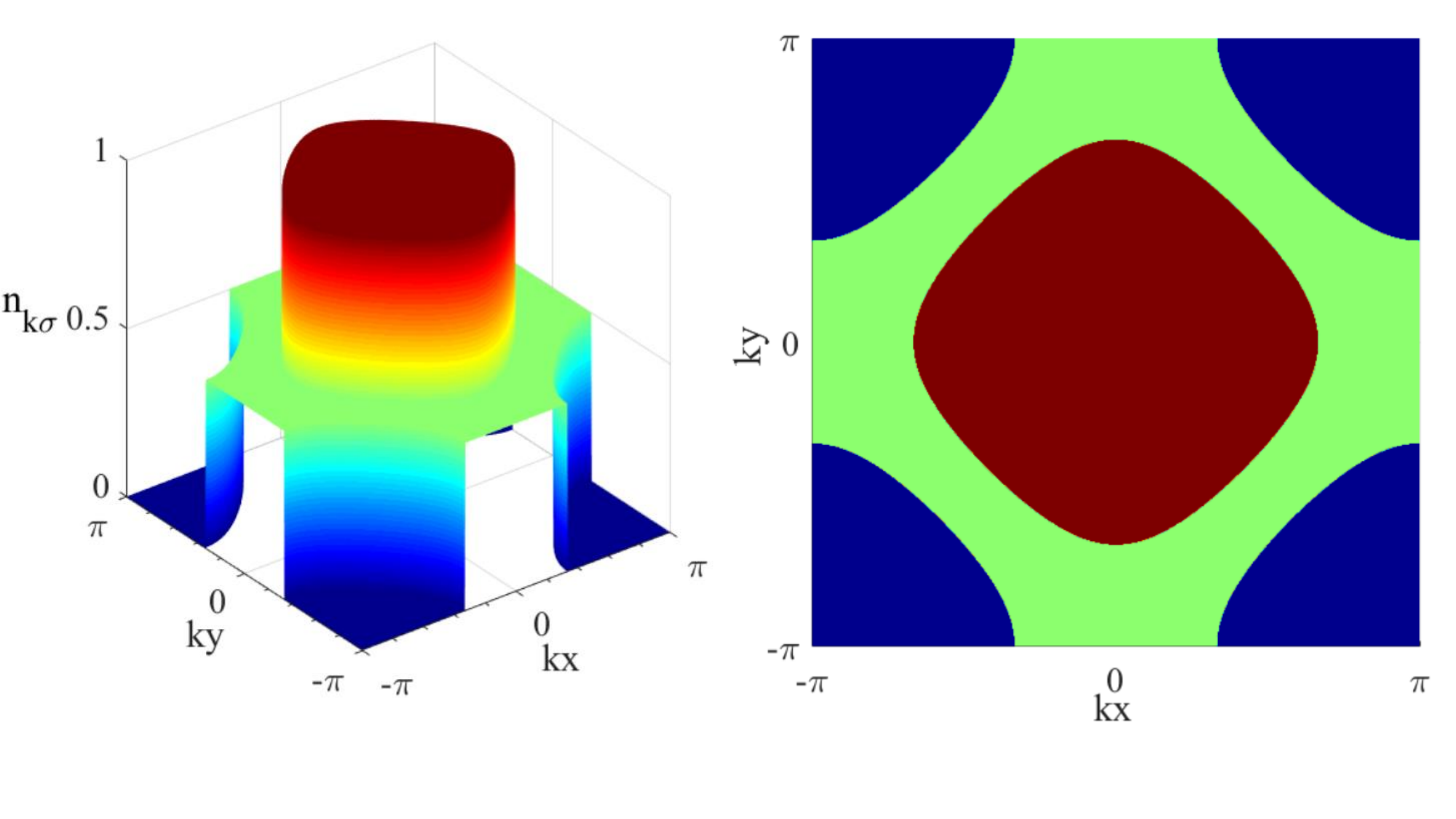}
\caption{\label{fig:HK_2D} (Left) The metallic state's electron distribution $n_{k\sigma}=\langle \hat{c}_{k\sigma}^{\dag}\hat{c}_{k\sigma}\rangle$ of HK model on square lattice with two jumps; (Right) The corresponding quasi-Fermi surface structure. ($U/t=2,\mu=U/2$)}
\end{center}
\end{figure}

Importantly, for free Fermi gas or FL, the electron density uniquely determines the Fermi wavevector $k_{F}$, irrespective on interaction. For metal state in HK model, electron density cannot be uniquely determined by quasi-Fermi wavevector $k_{F2},k_{F1}$, without considering the strength of interaction. From $k_{F2}=\left|\arccos\frac{U-\mu}{2t}\right|$, it is clear that the quasi-Fermi wavevector itself depends on interaction, which means the quasi-Fermi wavevector is a function of interaction and this contradicts with Luttinger theorem. Thus, the metallic phase belongs to a kind of NFL. At last, we emphasize that these results are also valid for $d>1$. For example, on square lattice, the metal phase is NFL and it has two quasi-Fermi surface, in which $n_{k\sigma}$ jumps from $1$ to $0.5$ and $0.5$ to $0$. (Fig.~\ref{fig:HK_2D})

\subsubsection{Thermodynamics}
The thermodynamics of HK model is determined by its free energy,
\begin{equation}
\mathcal{F}=-T\ln \mathcal{Z}=-T\sum_{k}\ln f_{k}=-T\sum_{k}\ln(1+2e^{-\beta(\varepsilon_{k}-\mu)}+e^{-\beta(2\varepsilon_{k}-2\mu+U)})\label{eq:free_energy}
\end{equation}
and $f_{k}$ acts as partition function for each $k$. Thus, the energy of the system is
\begin{eqnarray}
&&E=\mathcal{F}-T\frac{\partial \mathcal{F}}{\partial T}=\sum_{k}\left(\frac{2z_{k}}{f_{k}}(\varepsilon_{k}-\mu)+\frac{2z_{k}^{2}e^{-\beta U}}{f_{k}}\left(\varepsilon_{k}-\mu+\frac{U}{2}\right)\right)\nonumber\\
&&=\sum_{k}\left(\frac{2z_{k}+z_{k}^{2}e^{-\beta U}}{f_{k}}(\varepsilon_{k}-\mu)+\frac{z_{k}^{2}e^{-\beta U}}{f_{k}}(\varepsilon_{k}-\mu+U)\right).\label{eq:energy}
\end{eqnarray}
\begin{figure}
\begin{center}
\includegraphics[width=0.65\columnwidth]{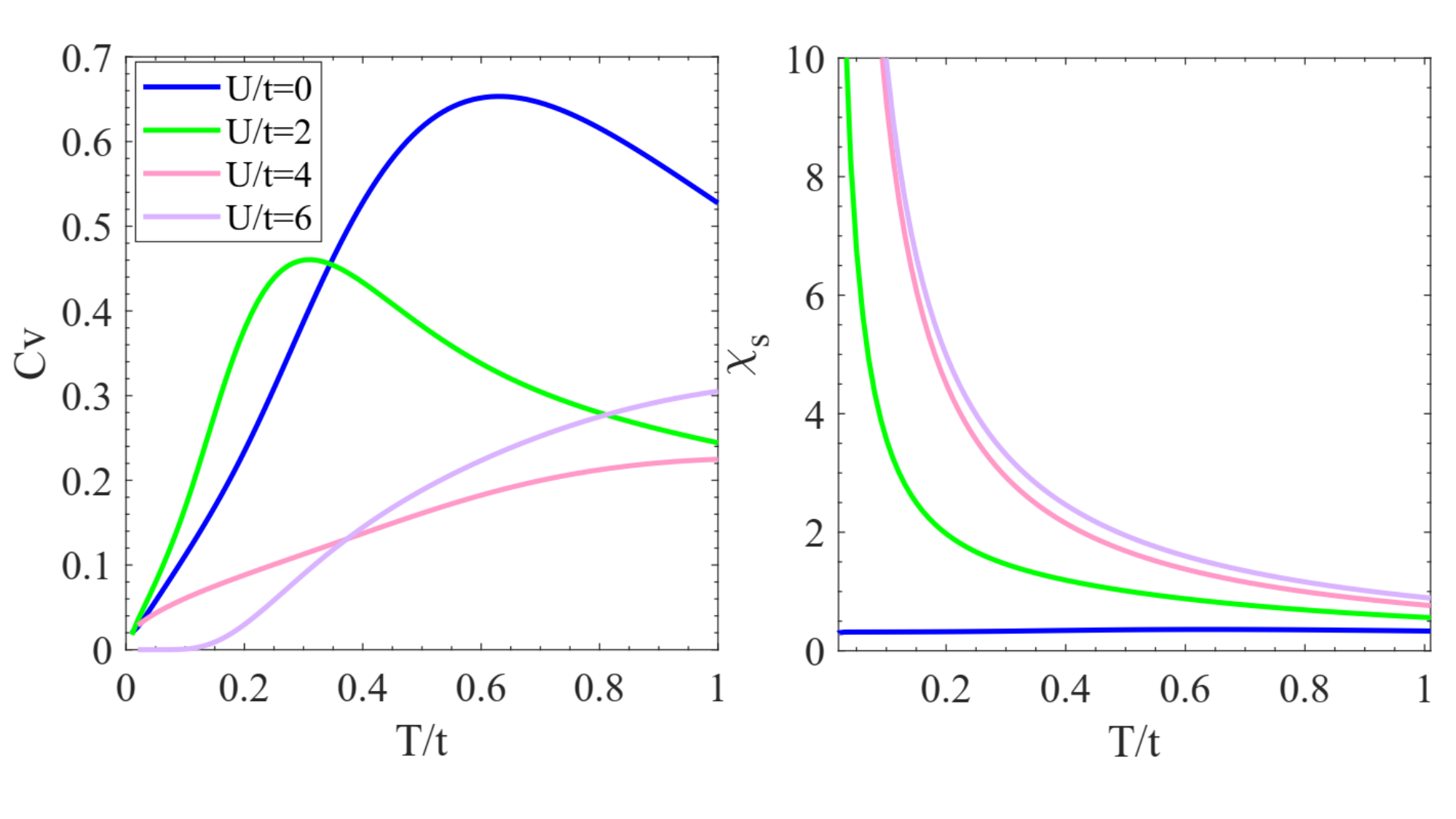}
\caption{\label{fig:HK_1D_Cv_chis} (Left) The specific heat $C_{v}$ and (Right) spin susceptibility $\chi_{s}$ of one-dimensional HK model with $\mu=U/2$.}
\end{center}
\end{figure}
Then, one obtains specific heat $C_{v}=\frac{d E}{dT}$ as shown in Fig.~\ref{fig:HK_1D_Cv_chis} where the specific heat of $d=1$ system is shown. $C_{v}$ in metallic state has linear-$T$ behavior ($U/t<4$) while it shows exponential behavior in Mott insulator for $U/t>4$ due to the interaction-induced gap $\Delta=U-W$.

At $T=0$, the ground-state energy is contributed from two spectrums,
\begin{equation}
E\rightarrow E_{g}=\sum_{k}\left[(\varepsilon_{k}-\mu)\theta(\mu-\varepsilon_{k})+(\varepsilon_{k}-\mu+U)\theta(\mu-\varepsilon_{k}-U)\right].
\end{equation}
Then, the electron density is
\begin{equation}
n=-\frac{1}{N_{s}}\frac{\partial \mathcal{F}}{\partial \mu}=\frac{2}{N_{s}}\sum_{k}\frac{z_{k}}{f_{k}}(1+z_{k}e^{-\beta U}).
\end{equation}
\begin{figure}
\begin{center}
\includegraphics[width=0.75\columnwidth]{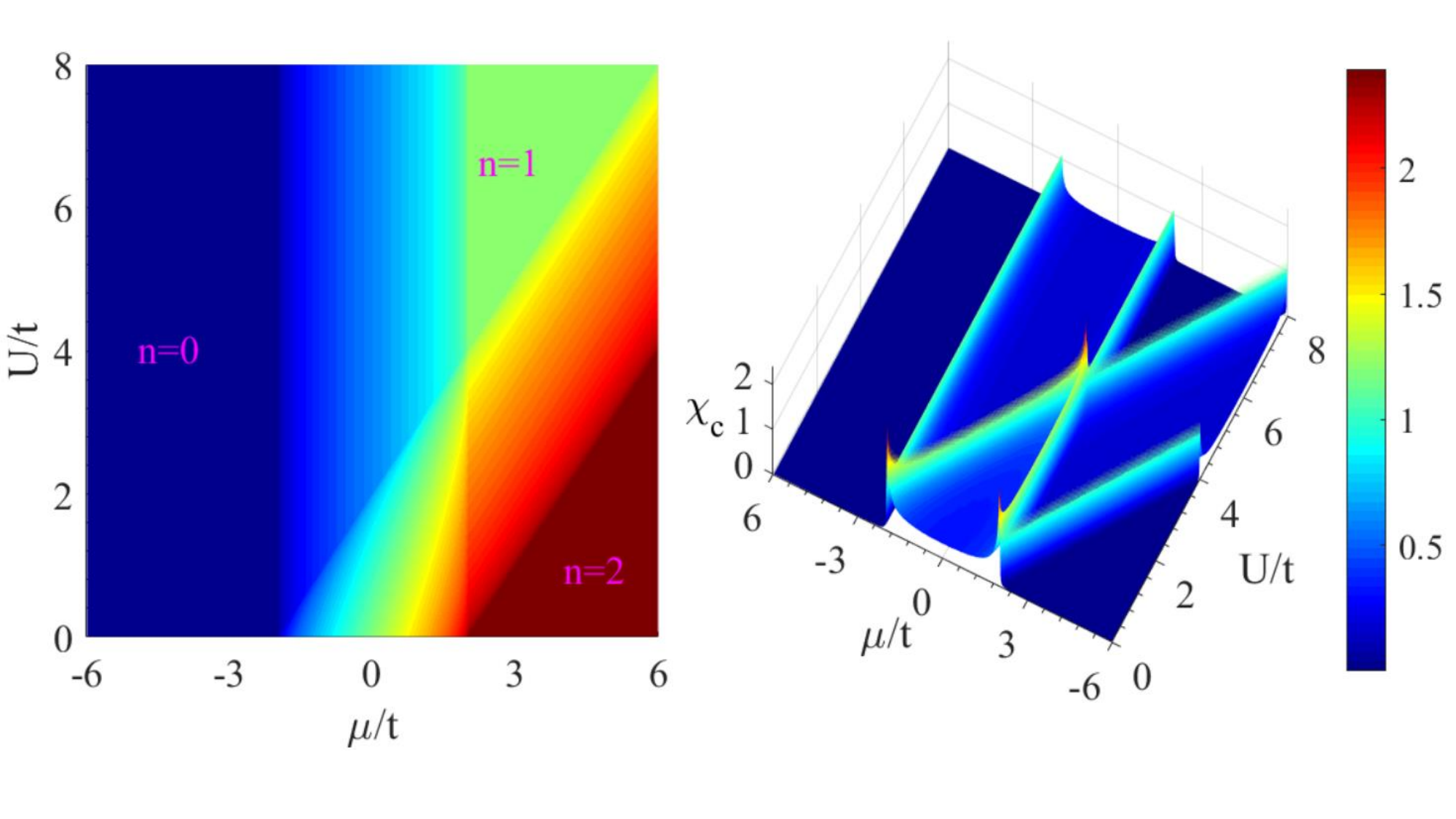}
\caption{\label{fig:HK_1D_n_chi} (Left) The one-dimensional HK model's electron density $n$ and (Right) the charge susceptibility $\chi_{c}$ at $T=0$.}
\end{center}
\end{figure}
At $T=0$,
\begin{equation}
n=-\frac{1}{N_{s}}\frac{\partial E_{g}}{\partial \mu}=\frac{1}{N_{s}}\sum_{k}\left(\theta(\mu-\varepsilon_{k})+\theta(\mu-\varepsilon_{k}-U)\right)
\end{equation}
while the charge susceptibility at $T=0$ is
\begin{equation}
\chi_{c}=\frac{\partial n}{\partial \mu}=\frac{1}{N_{s}}\sum_{k}\left(\delta(\mu-\varepsilon_{k})+\delta(\mu-\varepsilon_{k}-U)\right)=N_{0}(\mu)+N_{0}(\mu-U)
\end{equation}
Here $N_{0}(\omega)$ is the density of state of non-interacting system.

Fig.~\ref{fig:HK_1D_n_chi} shows electron density $n$ and charge susceptibility $\chi_{c}$ versus $U$ and $\mu$. ($\varepsilon_{k}=-2t\cos k$) It is seen that $n=0,2$ correspond to band insulator, while the $n=1$ regime is divided into Mott insulator with large $U$ and metal state with small $U$. The boundary of these regimes is characterized by the peak of $\chi_{c}$, denoting the phase transition lines.

In hypercubic lattice, $N_{0}(\omega)$ is symmetric versus $\omega$ and is nonzero in regime $[-W/2,W/2]$. For symmetric half-filling case, $\mu=U/2$, and increasing $U$ from $U=0$, we find $\chi_{c}$ is finite for $U<W$ and it denotes a metallic state. For $U>W$, $\chi_{c}=0$ characterizing Mott insulator. Thus, $U=U_{c}=W$ is the critical point of interaction-driven Mott transition. On the other hand, starting with half-filled Mott insulator with $U>U_{c}$ and tuning chemical potential $\mu$, we find $\mu_{c}=U-\frac{W}{2}$ is the critical point of chemical-potential-tuned Mott transition.

For $d=1$ system,
\begin{eqnarray}
\chi_{c}=\frac{1}{\pi }\frac{1}{\sqrt{W^{2}-4\mu^{2}}}\theta(W-2|\mu|)+\frac{1}{\pi }\frac{1}{\sqrt{W^{2}-4(\mu-U)^{2}}}\theta(W/2-2|\mu-U|)\nonumber
\end{eqnarray}
Thus, the interaction-driven Mott transition has $\chi_{c}(U\rightarrow U_{c})\sim \frac{1}{\sqrt{U_{c}-U}}$, ($U_{c}=W,\mu=U/2$), while the chemical potential-driven transition has $\chi_{c}(\mu\rightarrow \mu_{c})\sim \frac{1}{\sqrt{\mu-\mu_{c}}}$ ($\mu_{c}=U-W/2$). Note that the singularity comes from the edge behavior of density of state. For hypercubic lattice, the density of state is approximated as the one of free fermion, so $N_{0}(\omega)\sim(W/2-|\omega|)^{d/2-1}$ and the singular part of ground-state energy is from the integral of density of state and energy, namely $e_{g}\sim\int d\omega \omega N_{0}(\omega)\sim (W/2-|\mu|)^{d/2+1}+(W/2-|\mu-U|)^{d/2+1}$. Obviously, it leads to
\begin{eqnarray}
e_{g}(\mu\rightarrow \mu_{c})\propto (\mu-\mu_{c})^{\frac{d}{2}+1},~~~~e_{g}(U\rightarrow U_{c})\propto (U_{c}-U)^{\frac{d}{2}+1}\nonumber
\end{eqnarray}
So, the critical exponent is $\alpha=1-d/2$. Then, $\partial^{2} e_{g}/\partial \mu^{2}$ gives
\begin{eqnarray}
\chi_{c}(\mu\rightarrow \mu_{c})\propto (\mu-\mu_{c})^{\frac{d}{2}-1},~~~~\chi_{c}(U\rightarrow U_{c})\propto (U_{c}-U)^{\frac{d}{2}-1}\nonumber
\end{eqnarray}
We find the critical exponent $\gamma=1-d/2$. Since $\Delta n=1-n\sim(\mu-\mu_{c})^{\frac{d}{2}}$, we know that the two kinds of Mott transition in hypercubic lattice have the same universality class, i.e. the free fermion Lifshitz transition class.\cite{Continentino1994,Continentino2017}

Let us consider the (static) spin susceptibility $\chi_{s}$. To calculate $\chi_{s}$, just add the Zeeman coupling term, $\hat{H}_{Z}=-\mu_{B}h\sum_{j\sigma}\sigma \hat{c}_{j\sigma}^{\dag}\hat{c}_{j\sigma}=-\mu_{B}h\sum_{k\sigma}\sigma \hat{c}_{k\sigma}^{\dag}\hat{c}_{k\sigma}$, which means we can replace $\varepsilon_{k}$ with $\varepsilon_{k\sigma}=\varepsilon_{k}-\mu_{B}h\sigma$. In this case,
\begin{eqnarray}
n_{k\sigma}&=&\frac{1}{\mathcal{Z}}\mathrm{Tr }\hat{n}_{k\sigma}e^{-\beta \hat{H}}=\frac{\mathrm{Tr}\hat{n}_{k\sigma}e^{-\beta \hat{H}_{k}}}{\mathcal{Z}_{k}}\nonumber\\
&=&\frac{e^{-\beta(\varepsilon_{k\sigma}-\mu)}+e^{-\beta(2\varepsilon_{k}-2\mu+U)}}{1+e^{-\beta(\varepsilon_{k\uparrow}-\mu)}+e^{-\beta(\varepsilon_{k\downarrow}-\mu)}+e^{-\beta(2\varepsilon_{k}-2\mu+U)}}.
\end{eqnarray}
So, the magnetization of the system is $M=\frac{1}{N_{s}}\sum_{k\sigma}(n_{k\uparrow}-n_{k\downarrow})\mu_{B}$, and the spin susceptibility in $h\rightarrow0$ is
\begin{equation}
\chi_{s}=\lim_{h\rightarrow0}\frac{\partial M}{\partial h}=\lim_{h\rightarrow0}\frac{1}{N_{s}}\sum_{k\sigma}\left(\frac{\partial n_{k\uparrow}}{\partial h}-\frac{\partial n_{k\downarrow}}{\partial h}\right)\mu_{B}.
\end{equation}
Fig.~\ref{fig:HK_1D_Cv_chis} gives the temperature-dependent spin susceptibility $\chi_{s}$ for one-dimensional HK model. It is seen that except the Fermi gas with $U/t=0$, all other cases including NFL metal ($U/t=2,4$) and Mott insulator ($U/t=6$) show the Curie-like susceptibility $\chi_{s}\sim1/T$ instead of the Pauli-like susceptibility in usual FL.\cite{Vitoriano2001a} This means there are many free local moments, and the system itself has magnetic instability under weak magnetic field. In addition, susceptibility is enhanced when interaction increases, which is a result of increasing probability of single occupation when interaction is enhanced while the number of empty and double occupation is reduced. Finally, in Mott state, all $k$-state is single occupied, and the effective local moment is maximized, which reflects the saturation behavior in Fig.~\ref{fig:HK_1D_Cv_chis} where $\chi_{s}$ with $U/t=6$ is comparable with $U/t=4$.
\subsection{Single-particle excitation}
To study the behavior of single-particle excitation, we calculate the single-particle Green function. Just like the treatment of the Hubbard-I approximation,\cite{Hubbard1963} we define single-particle Green function as $G_{\sigma}(k,\omega)=\langle\langle \hat{c}_{k\sigma}|\hat{c}_{k\sigma}^{\dag}\rangle\rangle_{\omega}$, then its equation of motion reads (use $[\hat{c}_{k\sigma},\hat{H}]=(\varepsilon_{k}-\mu)\hat{c}_{k\sigma}+U\hat{c}_{k\sigma}\hat{n}_{k\bar{\sigma}}$)
\begin{eqnarray}
\omega\langle\langle \hat{c}_{k\sigma}|\hat{c}_{k\sigma}^{\dag}\rangle\rangle_{\omega}=1+(\varepsilon_{k}-\mu)\langle\langle \hat{c}_{k\sigma}|\hat{c}_{k\sigma}^{\dag}\rangle\rangle_{\omega}+U\langle\langle \hat{c}_{k\sigma}\hat{n}_{k\bar{\sigma}}|\hat{c}_{k\sigma}^{\dag}\rangle\rangle_{\omega}\nonumber
\end{eqnarray}
while the equation of motion of $\langle\langle \hat{c}_{k\sigma}\hat{n}_{k\bar{\sigma}}|\hat{c}_{k\sigma}^{\dag}\rangle\rangle_{\omega}$ is closed (use $[\hat{c}_{k\sigma}\hat{n}_{k\bar{\sigma}},\hat{H}]=(\varepsilon_{k}-\mu+U)\hat{c}_{k\sigma}\hat{n}_{k\bar{\sigma}}$)
\begin{eqnarray}
\omega\langle\langle \hat{c}_{k\sigma}\hat{n}_{k\bar{\sigma}}|\hat{c}_{k\sigma}^{\dag}\rangle\rangle_{\omega}=\langle \hat{n}_{k\bar{\sigma}}\rangle+(\varepsilon_{k}-\mu+U)\langle\langle \hat{c}_{k\sigma}\hat{n}_{k\bar{\sigma}}|\hat{c}_{k\sigma}^{\dag}\rangle\rangle_{\omega}\nonumber
\end{eqnarray}
Thus
\begin{equation}
\langle\langle \hat{c}_{k\sigma}\hat{n}_{k\bar{\sigma}}|\hat{c}_{k\sigma}^{\dag}\rangle\rangle_{\omega}=\frac{\langle \hat{n}_{k\bar{\sigma}}\rangle}{\omega-\varepsilon_{k}+\mu-U}
\end{equation}
\begin{equation}
G_{\sigma}(k,\omega)=\frac{1+\frac{U\langle\hat{n}_{k\bar{\sigma}}\rangle}{\omega-(\varepsilon_{k}-\mu+U)}}{\omega-(\varepsilon_{k}-\mu)}=\frac{1-\langle\hat{n}_{k\bar{\sigma}}\rangle}{\omega-(\varepsilon_{k}-\mu)}+\frac{\langle\hat{n}_{k\bar{\sigma}}\rangle}{\omega-(\varepsilon_{k}-\mu+U)}
\end{equation}
Different from Hubbard model, we have exactly solved the single-particle Green function in HK model. This is a result of the smallness of Hilbert space of $\hat{H}_{k}$. Use the spectral theorem of Green function, we find $\langle\hat{n}_{k\sigma}\rangle=(1-\langle\hat{n}_{k\bar{\sigma}}\rangle)f_{F}(\varepsilon_{k}-\mu)+\langle \hat{n}_{k\bar{\sigma}}\rangle f_{F}(\varepsilon_{k}-\mu+U)$. Because the system is assumed to be in the paramagnetic state, $n_{k}=\langle\hat{n}_{k\uparrow}\rangle=\langle\hat{n}_{k\downarrow}\rangle$, we find $n_{k}=\frac{f_{F}(\varepsilon_{k}-\mu)}{f_{F}(\varepsilon_{k}-\mu)+1-f_{F}(\varepsilon_{k}-\mu+U)}$. Similarly, we have expectation value of double occupation as  $\langle\hat{n}_{k\sigma}\hat{n}_{k\bar{\sigma}}\rangle=\frac{f_{F}(\varepsilon_{k}-\mu)f_{F}(\varepsilon_{k}-\mu+U)}{f_{F}(\varepsilon_{k}-\mu)+1-f_{F}(\varepsilon_{k}-\mu+U)}$.

From the poles of $G_{\sigma}(k,\omega)$, the quasi-particle excitation has the dispersion $\omega=\varepsilon_{k}-\mu,\varepsilon_{k}-\mu+U$. This is just the excitation energy of single-hole or electron in the Mott state. Note that the different weight of dispersion is the essential feature of doped Mott insulator. When $U>W$, there is a gap $\Delta=U-W$ between the two dispersions, and it forms the upper/lower Hubbard bands. If $U<W$, the two dispersions have overlap, thus there exists gapless excitation.

Particularly, in symmetric half-filling case, $\langle\hat{n}_{k\bar{\sigma}}\rangle=\frac{1}{2},\mu=\frac{U}{2}$,
\begin{eqnarray}
G_{\sigma}(k,\omega)=\frac{1}{\omega-\varepsilon_{k}-\frac{(U/2)^{2}}{\omega-\varepsilon_{k}}}=\frac{1}{\omega-\varepsilon_{k}-\Sigma(k,\omega)},~~\Sigma(k,\omega)=\frac{(U/2)^{2}}{\omega-\varepsilon_{k}}.\nonumber
\end{eqnarray}
At the non-interacting Fermi surface $\varepsilon_{k}=0$, the self-energy at zero frequency $\Sigma(k_{F},\omega=0)$ diverges, thus the single-particle Green function $G_{\sigma}(k_{F},\omega=0)=0$. It defines the Luttinger surface, namely the set of momentum involved in $\mathrm{Re}G(k,0)$ when it changes sign.\cite{Dzyaloshinskii2003} The momentum at Luttinger surface is called the Luttinger momentum, $k_{L}$, which has the property  $\mathrm{Re}G(k_{L},0)=0$.

\subsubsection{The $d=1$ model}
\begin{figure}
\begin{center}
\includegraphics[width=0.95\columnwidth]{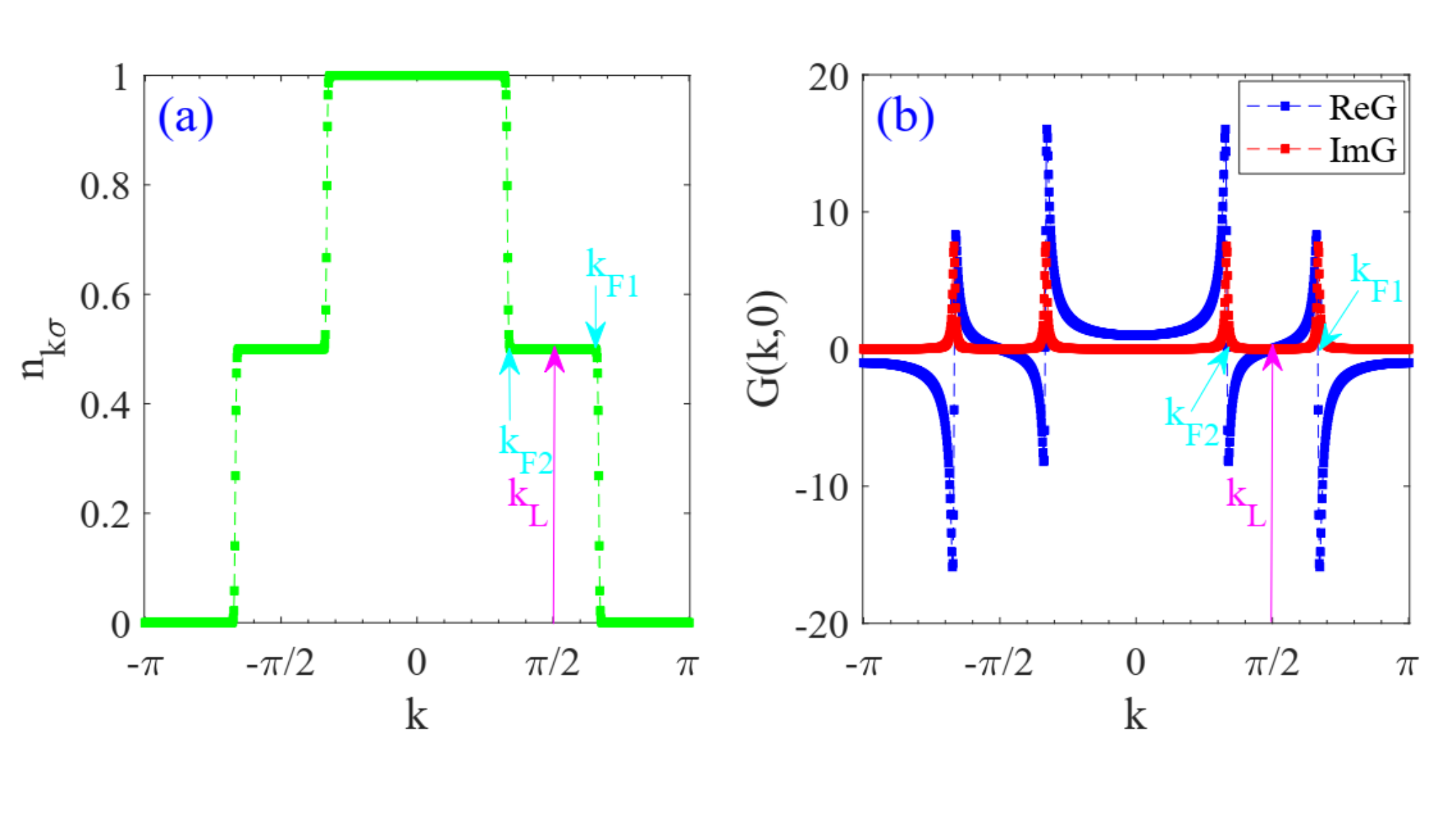}
\caption{\label{fig:HK_1D_kL}(a) Electron distribution function $n_{k\sigma}$ for $d=1$ HK model; (b) The corresponding real and imaginary part of Green function.}
\end{center}
\end{figure}

For $d=1$ model, the symmetric half-filling condition leads to $\varepsilon_{k_{L}}=-2t\cos k_{L}=0$ and it gives $k_{L}=\pi/2$. On the other hand, the quasi-Fermi wavevector $k_{F1},k_{F2}$ is given by $\varepsilon_{k}=U/2,\varepsilon_{k}=-U/2$. In Fig.~\ref{fig:HK_1D_kL}, the jump in electron distribution function is determined by $k_{F2},k_{F1}$, ($n_{k\sigma }$ changes from $1\rightarrow1/2,1/2\rightarrow0$) where the real part of Green function evolves from $\infty$ to $-\infty$ and the imaginary part of Green function shows maximal value. The zero of Green function is determined by $k_{L}$, where the real part of Green function changes from positive to negative value with crossing zero. Under above consideration, Eq.~\ref{eq:Luttinger_theorem} gives
\begin{equation}
LI=\frac{2}{2\pi}\int_{\mathrm{Re}G(k,\omega=0)>0}dk=\frac{2}{2\pi}(2k_{F2}+2(k_{F1}-k_{L})),
\end{equation}
but note that $k_{L}=\frac{k_{F1}-k_{F2}}{2}+k_{F2}=\frac{\pi}{2}$ in the symmetric half-filling case, thus $LI=\frac{2}{2\pi}(k_{F2}+k_{F1})=1=n$, and the Luttinger theorem seems to be valid in this situation because of the particle-hole symmetry.\cite{Seki2017}

The weight of quasiparticle $Z$ at Fermi wavevector $k_{F1}$ is found by expanding the self-energy
\begin{eqnarray}
\Sigma(k_{F1},\omega)=\frac{(U/2)^{2}}{\omega-U/2}\simeq-\frac{U}{2}\left(1+2\frac{\omega}{U}\right)+\mathcal{O}((\omega/U)^{2})\nonumber
\end{eqnarray}
which leads to $Z=\frac{1}{1-\partial_{\omega}\Sigma(\omega)}=\frac{1}{2}>0$. Similarly, $k_{F2}$ has the same formalism, which suggests that although the system is in a NFL state, the self-energy of electron has FL-like form near quasi-Fermi surface and its weight is nonzero. In contrast, the self-energy near Luttinger surface is $\Sigma(k_{L},\omega)=\frac{(U/2)^{2}}{\omega}\sim\frac{1}{\omega}$, which does not have a meaningful series expansion since the self-energy shows pole structure and diverges when $\omega\rightarrow0$.

When deviating from half-filling, $G_{\sigma}(k_{L},\omega=0)=0$ leads to Luttinger wavevector $k_{L}$, which satisfies
\begin{equation}
1+\frac{U\langle\hat{n}_{k_{L}\bar{\sigma}}\rangle}{-(\varepsilon_{k_{L}}-\mu+U)}=0\Rightarrow \varepsilon_{k_{L}}=Un_{k_{L}}-U+\mu.
\end{equation}
Inserting the expression of $n_{k_{L}}$ gives
\begin{equation}
\varepsilon_{k_{L}}=U\frac{f_{F}(\varepsilon_{k_{L}}-\mu)}{f_{F}(\varepsilon_{k_{L}}-\mu)+1-f_{F}(\varepsilon_{k_{L}}-\mu+U)}-U+\mu.
\end{equation}
\subsubsection{holon and doublon}

Because of the divergency of zero-frequency self-energy, FL weight $Z$ loses its meaning, and the system is in NFL, namely the metallic phase is in fact a NFL. Except for the symmetric half-filling case, it is found that for all other parameters, $LI\neq n$ which means the system does not satisfy the Luttinger theorem valid for FL, implying the NFL nature of metal state in HK model.

So, the quasiparticle in metal state is not the FL quasiparticle, which is adiabatically evolved from non-interacting limit. For every momentum $k$ there exist four eigenstates and this indicates its relation to quasiparticles, which suggests the following quasiparticle operators, namely holon $\hat{h}$ and doublon $\hat{d}$
\begin{equation}
\hat{c}_{k\sigma}=\hat{h}_{k\sigma}+\hat{d}_{k\sigma},~~~~  \hat{h}_{k\sigma}=\hat{c}_{k\sigma}(1-\hat{n}_{k\bar{\sigma}}),~~~~\hat{d}_{k\sigma}=\hat{c}_{k\sigma}\hat{n}_{k\bar{\sigma}}
\end{equation}
Using the commutative relation $[\hat{c}_{k\sigma},\hat{H}]=(\varepsilon_{k}-\mu)\hat{c}_{k\sigma}+U\hat{c}_{k\sigma}\hat{n}_{k\bar{\sigma}}$,$[\hat{c}_{k\sigma}\hat{n}_{k\bar{\sigma}},\hat{H}]=(\varepsilon_{k}-\mu+U)\hat{c}_{k\sigma}\hat{n}_{k\bar{\sigma}}$, we obtain the Green function of holon and doublon
\begin{equation}
\langle\langle \hat{h}_{k\sigma}|\hat{h}_{k\sigma}^{\dag}\rangle\rangle_{\omega}=\frac{1-\langle \hat{n}_{k\bar{\sigma}}\rangle}{\omega-\varepsilon_{k}+\mu},
\end{equation}
\begin{equation}
\langle\langle \hat{d}_{k\sigma}|\hat{d}_{k\sigma}^{\dag}\rangle\rangle_{\omega}=\frac{\langle \hat{n}_{k\bar{\sigma}}\rangle}{\omega-\varepsilon_{k}+\mu-U}.
\end{equation}
Note that the electron's Green function is the sum of above two, $\langle\langle \hat{c}_{k\sigma}|\hat{c}_{k\sigma}^{\dag}\rangle\rangle_{\omega}=\langle\langle \hat{h}_{k\sigma}|\hat{h}_{k\sigma}^{\dag}\rangle\rangle_{\omega}+\langle\langle \hat{d}_{k\sigma}|\hat{d}_{k\sigma}^{\dag}\rangle\rangle_{\omega}$,
which means the quasiparticle excitation is in fact the holon and doublon. The quasiparticle operator $\hat{h},\hat{d}$ is not the standard fermion because they do not satisfy the standard anti-commutative relation, $[\hat{d}_{k\sigma},\hat{d}_{k'\sigma'}^{\dag}]_{+}=\delta_{kk'}\delta_{\sigma\sigma'}\hat{n}_{k\bar{\sigma}},[\hat{h}_{k\sigma},\hat{h}_{k'\sigma'}^{\dag}]_{+}=\delta_{kk'}\delta_{\sigma\sigma'}(1-\hat{n}_{k\bar{\sigma}})$.
Note $\hat{d}_{k\uparrow}^{\dag}\hat{d}_{k\uparrow}=\hat{d}_{k\downarrow}^{\dag}\hat{d}_{k\downarrow}$, suggesting the number of double occupation does not depend on the spin-direction.
\subsubsection{Density of state}
\begin{figure}
\begin{center}
\includegraphics[width=0.75\columnwidth]{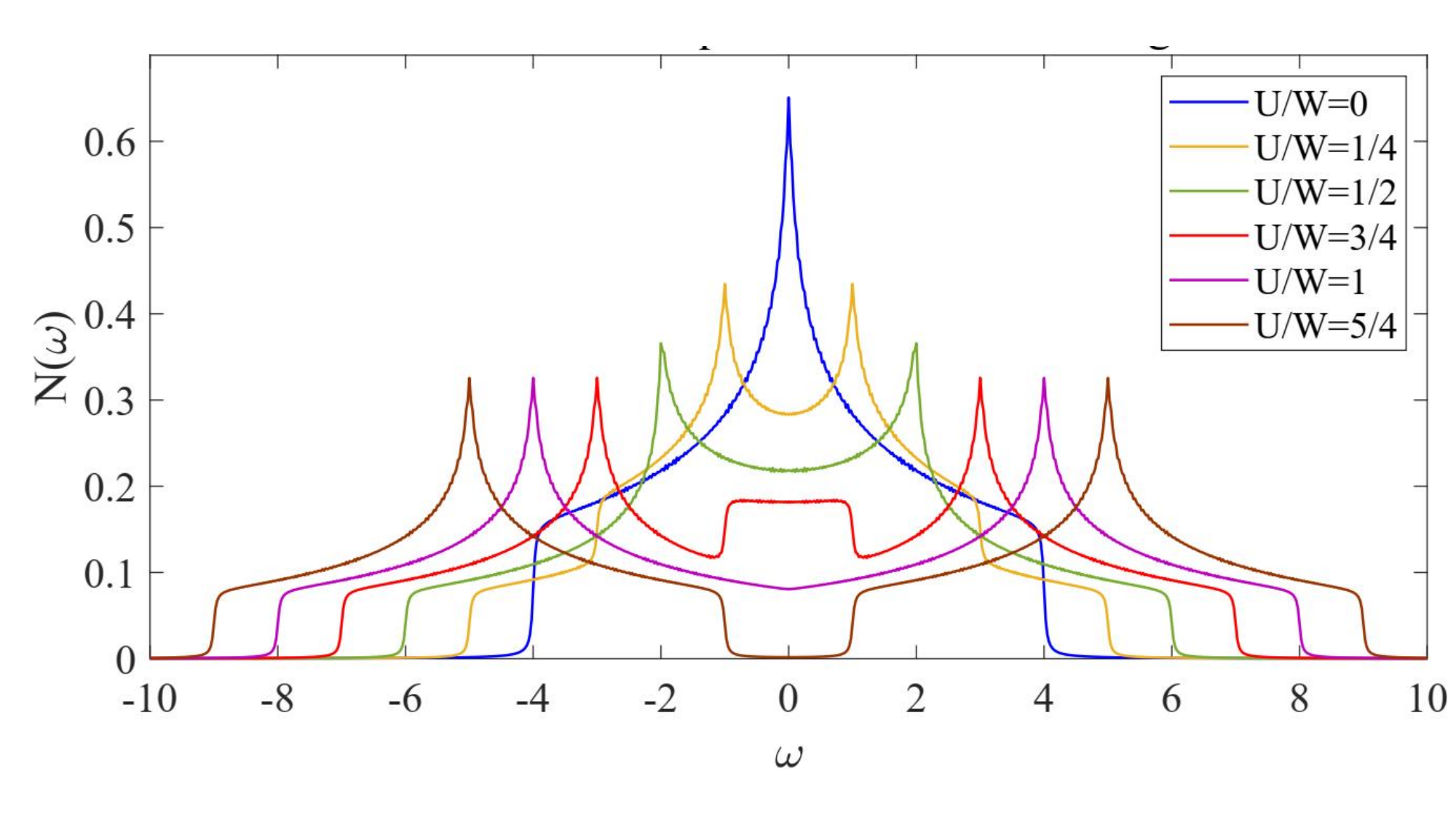}
\caption{\label{fig:HK_dos_square} Density of state of half-filled HK model on square lattice for different interaction $U$.}
\end{center}
\end{figure}

The density of state of electrons is obtained by the single-particle Green function as $N(\omega)=-\frac{1}{\pi N}\sum_{k}\mathrm{Im}G_{\sigma}(k,\omega)$. Fig.~\ref{fig:HK_dos_square} shows the density of state of half-filled square lattice HK model for different interaction $U$. When $U<W$, the density of state near Fermi energy is finite and it decreases with increasing interaction. If $U>W$, the density of state at Fermi energy is zero and the upper and lower Hubbard bands develop, forming the Mott insulator.

\begin{figure}
\begin{center}
\includegraphics[width=0.75\columnwidth]{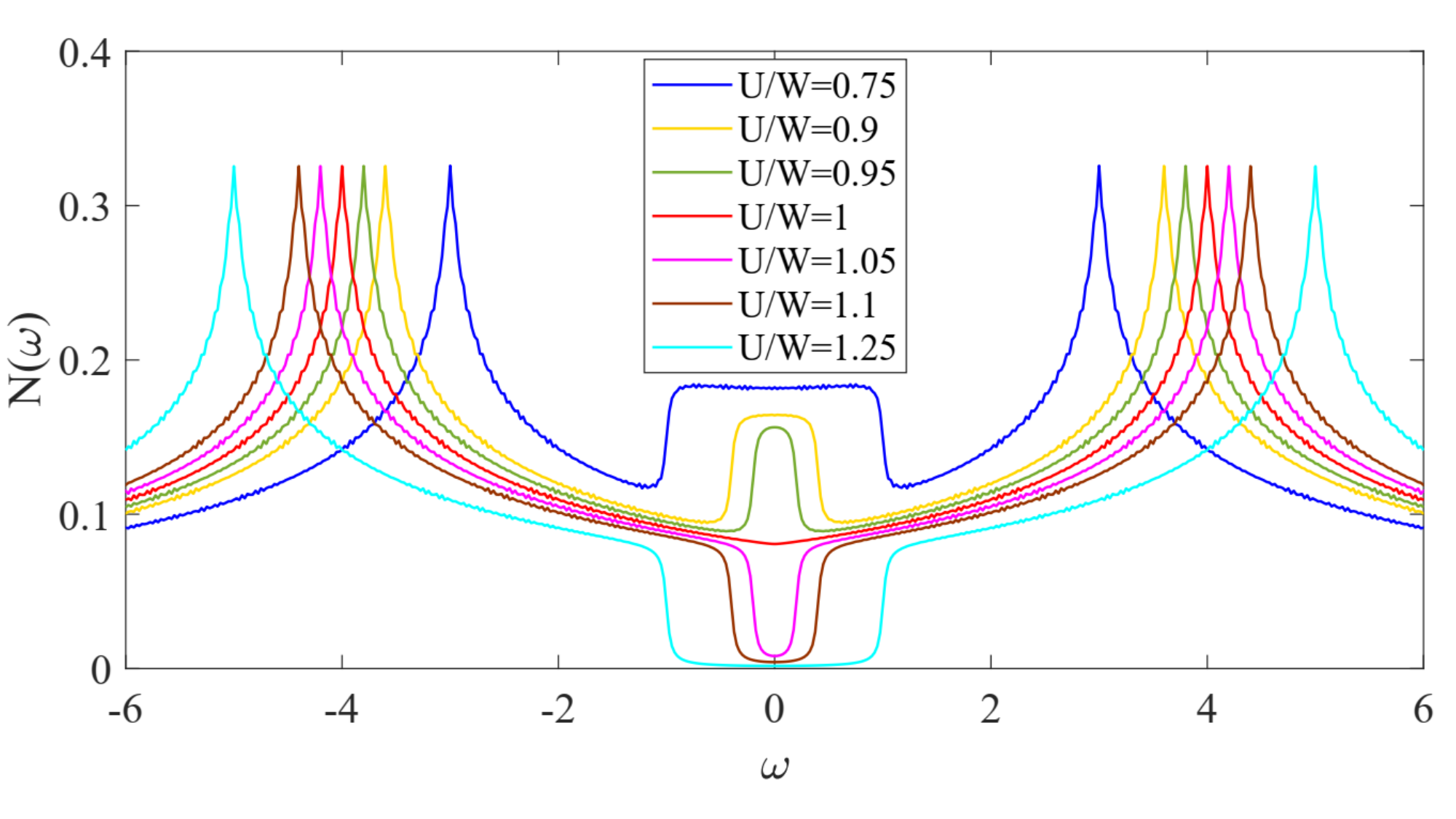}
\caption{\label{fig:HK_dos_square2} Density of state of half-filled HK model on square lattice for different interaction $U$. Here, the quasiparticle-like peak is more visible though it does not come from the Kondo effect.}
\end{center}
\end{figure}

Fig.~\ref{fig:HK_dos_square2} shows the details of density of state near Fermi energy. We find the existence of the quasiparticle peak found in DMFT calculation of Hubbard model. The weight of this quasiparticle peak decreases if approaching Mott transition point. When crossing the transition point, quasiparticle peak disappears and density of state at Fermi energy is zero with the opening of gap.

It should be emphasized that, the finite density of state at Fermi energy in Fig.~\ref{fig:HK_dos_square2} is not the quasiparticle in DMFT. For the latter one, its quasiparticle density of state in the framework of slave-rotor/slave-spin mean-field theory is described by Green function $G=\frac{Z}{\omega-Z\varepsilon_{k}}$ with $Z\neq0$ characterizing the weight of quasiparticle and its nonzero value implies the existence of quasiparticle.\cite{Florens2004,Ruegg2010} In HK model, finite density of state comes from the filling effect of Hubbard bands, (far away from Fermi energy, tuning $\omega$ only involves one band while near Fermi energy involves two bands so the density of state is enhanced) and no Kondo effect and Kondo peak appear, thus it is not the true quasiparticle peak in DMFT.

\subsection{Explanation by exclusion statistics}
In the early time, people have found that the thermodynamics of HK model can be explained by the so-called exclusion statistics or Haldane-Wu statistics. \cite{Haldane1991,YSWu1994,Nayak1994,Byczuk1994,Hatsugai1996,Vitoriano2001a,Vitoriano2001b}
The main point of exclusion statistics is that certain interacting systems may be mapped into non-interacting models with the elementary excitations obeying generalized Pauli principle.

For HK model, we use holon and doulon operator $\hat{h}_{k\uparrow}=\hat{c}_{k\uparrow}(1-\hat{n}_{k\downarrow}),\hat{h}_{k\downarrow}=\hat{c}_{k\uparrow}(1-\hat{n}_{k\uparrow}),\hat{d}_{k}=\hat{c}_{k\uparrow}\hat{c}_{k\uparrow}$, thus the total energy is written as
\begin{eqnarray}
E&=&\langle\hat{H}_{HK}\rangle=\sum_{k}\left[(\varepsilon_{k}-\mu)(\langle\hat{n}_{k\uparrow}\rangle+\langle\hat{n}_{k\downarrow}\rangle)+U\langle\hat{n}_{k\uparrow}\hat{n}_{k\downarrow}\rangle\right]\nonumber\\
&=&\sum_{k}[(\varepsilon_{k}-\mu)(\langle\hat{h}_{k\uparrow}^{\dag}\hat{h}_{k\uparrow}\rangle+\langle\hat{h}_{k\downarrow}^{\dag}\hat{h}_{k\downarrow}\rangle)+2(\varepsilon_{k}-\mu+U/2)\langle\hat{d}_{k}^{\dag}\hat{d}_{k}\rangle]\nonumber
\end{eqnarray}
which implies the system could be seen as a collection of holon and doublon. In the language of
exclusion statistics, the total energy has the following non-interacting form
\begin{equation}
E=\langle \hat{H}_{HK}\rangle=\sum_{k\alpha}\varepsilon_{k\alpha}n_{k\alpha}
\end{equation}
where $n_{k\alpha}$ and $\varepsilon_{k\alpha}$ are distribution function and energy of the particle, which is called excluson with $\alpha=1,2$ and $3$, respectively. Specifically, we have
\begin{eqnarray}
\varepsilon_{k1}=\varepsilon_{k2}=\varepsilon_{k}-\mu,~~\varepsilon_{k3}=2(\varepsilon_{k}-\mu+U/2)
\end{eqnarray}
and
\begin{eqnarray}
n_{k1}=\langle\hat{h}_{k\uparrow}^{\dag}\hat{h}_{k\uparrow}\rangle,~~
n_{k2}=\langle\hat{h}_{k\downarrow}^{\dag}\hat{h}_{k\downarrow}\rangle,~~
n_{k3}=\langle\hat{d}_{k}^{\dag}\hat{d}_{k}\rangle.
\end{eqnarray}
Recall that the anticommutative relations of holon and doublon operators are not zero, and this fact is reflected by introducing the statistical matrix $g_{kk',\alpha\alpha'}$, defined by
\begin{equation}
\Delta D_{k\alpha}=-\sum_{k'\alpha'}g_{kk',\alpha\alpha'}\Delta N_{k'\alpha'}.
\end{equation}
or $D_{k\alpha}=G_{k\alpha}-\sum_{k'\alpha'}g_{kk',\alpha\alpha'}N_{k'\alpha'}$. Here, the variation in the available single-particle states of species $\alpha$ is denoted by $\Delta D_{k\alpha}$, which is caused by
a set of allowed changes on the number of occupied single particle states $\Delta N_{k'\alpha'}$.
$G_{k\alpha}$ is the number of available single-particle states when there is no particle in
the system. Vitoriano et al. find\cite{Vitoriano2001b}
\begin{equation}
g_{kk',\alpha\alpha'}=\delta_{kk'}\left(
                        \begin{array}{ccc}
                          1 & 1 & 1 \\
                          0 & 1 & 1 \\
                          0 & 0 & 1 \\
                        \end{array}
                      \right)
\end{equation}
The meaning of this statistical matrix is understood as follows. Firstly, the $\delta$ function $\delta_{kk'}$ tells us that the holon and doublon with distinct momentum do not affect the occupation of each other. For excluson with the same momentum $k$, e.g. the holon $h_{k\uparrow}$, we have $\Delta D_{k1}=-g_{kk,11}\Delta N_{k1}-g_{kk,12}\Delta N_{k2}-g_{kk,13}\Delta N_{k3}
=-(\Delta N_{k1}+\Delta N_{k2}+\Delta N_{k3})$. $\Delta N_{k1},\Delta N_{k2},\Delta N_{k3}$ are the
changes on the number of occupied states of $h_{k\uparrow}$, $h_{k\downarrow}$ and $d_{k}$. So, if the number of any holon or doublon occupation increases by one, the available single-particle states for $h_{k\uparrow}$ should be reduced by $1$. Now, the Hilbert space of $\hat{H}_{k}$ is formed by $|0\rangle_{k},|\uparrow\rangle_{k},|\downarrow\rangle_{k}$ and $|\uparrow\downarrow\rangle_{k}$.
In empty occupation state $|0\rangle_{k}$, we have $N_{k1}=N_{k2}=N_{k3}=0$. For $|\uparrow\rangle_{k}$, $N_{k1}=1,N_{k2}=N_{k3}=0$ while $N_{k2}=1,N_{k1}=N_{k3}=0$ in $|\downarrow\rangle_{k}$ state. The double occupation state $|\uparrow\downarrow\rangle_{k}$ has $N_{k1}=N_{k2}=0,N_{k3}=1$. Consider $|\uparrow\rangle_{k}$ state and we know it has one holon $\hat{h}_{k\uparrow}$ but no $\hat{h}_{k\downarrow}$ or $\hat{d}_{k}$. A double holon state with two $\hat{h}_{k\uparrow}$ is prohibited due to Pauli principle and this fact is consistent with $\Delta D_{k1}=-(1+0+0)=-1$. (The available single-particle states for $\hat{h}_{k\uparrow}$ is $1$ if other excluson is absent and adding second $\hat{h}_{k\uparrow}$ reduces the number of state by $1$. Thus, there is no single-particle states for $\hat{h}_{k\uparrow}$ to occupy and adding the second $\hat{h}_{k\uparrow}$ particle is prohibited.)

In terms of the statistical matrix, the Wu function $w_{k\alpha}=\frac{D_{k\alpha}}{N_{k\alpha}}$,\cite{YSWu1994} can be found by solving
\begin{equation}
(1+w_{k\alpha})\prod_{k'\alpha'}\left(\frac{w_{k'\alpha'}}{1+w_{k'\alpha'}}\right)^{g_{k'k,\alpha'\alpha}}=e^{\beta \varepsilon_{k\alpha}}.
\end{equation}
In the case of HK model, above equation is simplified as $(1+w_{k\alpha})\prod_{\alpha'=1,2,3}\left(\frac{w_{k\alpha'}}{1+w_{k\alpha'}}\right)^{g_{kk,\alpha'\alpha}}=e^{\beta \varepsilon_{k\alpha}}$. One finds $w_{k1}=e^{\beta \varepsilon_{k1}}$, $w_{k2}=e^{\beta \varepsilon_{k2}}(1+w_{k1}^{-1})$ and $w_{k3}=e^{\beta \varepsilon_{k3}}(1+w_{k1}^{-1})(1+w_{k2}^{-1})$. Then the free energy can be expressed by $w_{k\alpha}$ as
\begin{eqnarray}
\mathcal{F}&=&-T\sum_{k\alpha}\ln(1+w_{k\alpha}^{-1})=-T\sum_{k}\ln\left(1+e^{-\beta \varepsilon_{k1}} +e^{-\beta \varepsilon_{k2}}+e^{-\beta\varepsilon_{k3}}\right)\nonumber\\
&=&-T\sum_{k}\ln\left(1+2e^{-\beta (\varepsilon_{k}-\mu)}+e^{-\beta2(\varepsilon_{k}-\mu+U/2)}\right)\nonumber
\end{eqnarray}
which reproduces the correct free energy of HK model Eq.~\ref{eq:free_energy}.
%The $U=\infty$ limit %has been discussed in Ref.~\cite{Hatsugai1996}, where the authors label each eigenstate with number %of charge $N_{c}$ and number of magnon $N_{s}$ (number of spin-down). ($|0\rangle_{k}$ has %$N_{c}=N_{s}=0$, $|\uparrow\rangle_{k},|\downarrow\rangle_{k}$ have $N_{c}=1,N_{s}=0$ and %$N_{c}=1,N_{s}=1$ but double occupation state is excluded due to $U=\infty$) Then, treat %$N_{c},N_{s}$ as independent variables, it is found that $G_{c}=1, G_{s}=0$ and
%\begin{equation}
%\left(
 % \begin{array}{cc}
 %   g_{cc} & g_{cs} \\
 %   g_{sc} & g_{ss} \\
%  \end{array}
%\right)=\left(
%          \begin{array}{cc}
%            1 & 0 \\
%            -1 & 1 \\
%          \end{array}
%        \right)
%\end{equation}
%for each momentum. Note the nontrivial value $-1$ for
%mutual statistics $g_{sc}$, which means the presence of a charge can create
%a magnon state, though there is no bare available singlemagnon state ($G_{s}=0$) when there is no %charge. In this situation, $w_{kc}$ and $w_{ks}$ obey the equations $w_{ks}=e^{\beta %\varepsilon_{ks}},w_{kc}=e^{\beta \varepsilon_{kc}}(1+w_{ks}^{-1})$.
Before ending this section we note that $w_{k\alpha}$ is related to $n_{k\alpha}$ via $w_{k\alpha}=n_{k\alpha}^{-1}-\sum_{\alpha'}\beta_{kk,\alpha\alpha'}n_{k\alpha'}/n_{k\alpha}$ with $\beta_{kk,\alpha\alpha'}=g_{kk,\alpha\alpha'}G_{k\alpha'}/G_{k\alpha}$. Simple observation gives $G_{k1}=G_{k2}=G_{k3}=1$, then $n_{k3}=\frac{1}{1+w_{k3}}=\langle \hat{n}_{k\uparrow}\hat{n}_{k\downarrow}\rangle=\langle \hat{d}_{k}^{\dag}\hat{d}_{k}\rangle$. Similarly, $n_{k2}=\frac{1-n_{k3}}{1+w_{k3}}=\langle \hat{h}_{k\downarrow}^{\dag}\hat{h}_{k\downarrow}\rangle$ and $n_{k1}=\frac{1-n_{k2}-n_{k3}}{1+w_{k1}}=\langle \hat{h}_{k\uparrow}^{\dag}\hat{h}_{k\uparrow}\rangle$, which completes the non-interacting form $E=\sum_{k\alpha}\varepsilon_{k\alpha}n_{k\alpha}$ of Haldane-Wu statistics. A Boltzmann transport theory for excluson has been established in Ref.~\cite{Bhaduri1996} but we do not know any similar theory for HK-related models.
\subsection{Two-particle correlation}
Unlike the single-particle correlation/Green function, the two-particle correlation function of HK model has not been fully understood. There is controversy about the singular behaviors in charge susceptibility and current-current correlation.\cite{Guerci2024,YMa2024} To avoid confusing the reader, let us consider the impurity-induced Friedel oscillation in HK model.\cite{MZhao2023} Assuming a single (nonmagnetic) impurity located on zero-th site and only electron on this site feels its scattering, the resulting impurity Hamiltonian reads $\hat{H}_{\rm imp}=V\sum_{\sigma}\hat{c}_{0\sigma}^{\dag}\hat{c}_{0\sigma}=\frac{V}{N_{\rm s}} \sum_{k,k',\sigma}\hat{c}_{k\sigma}^{\dag}\hat{c}_{k'\sigma}$ with $V$ being the strength of impurity.

\begin{figure}
\begin{center}
	%\flushleft
	\includegraphics[width=0.75\linewidth]{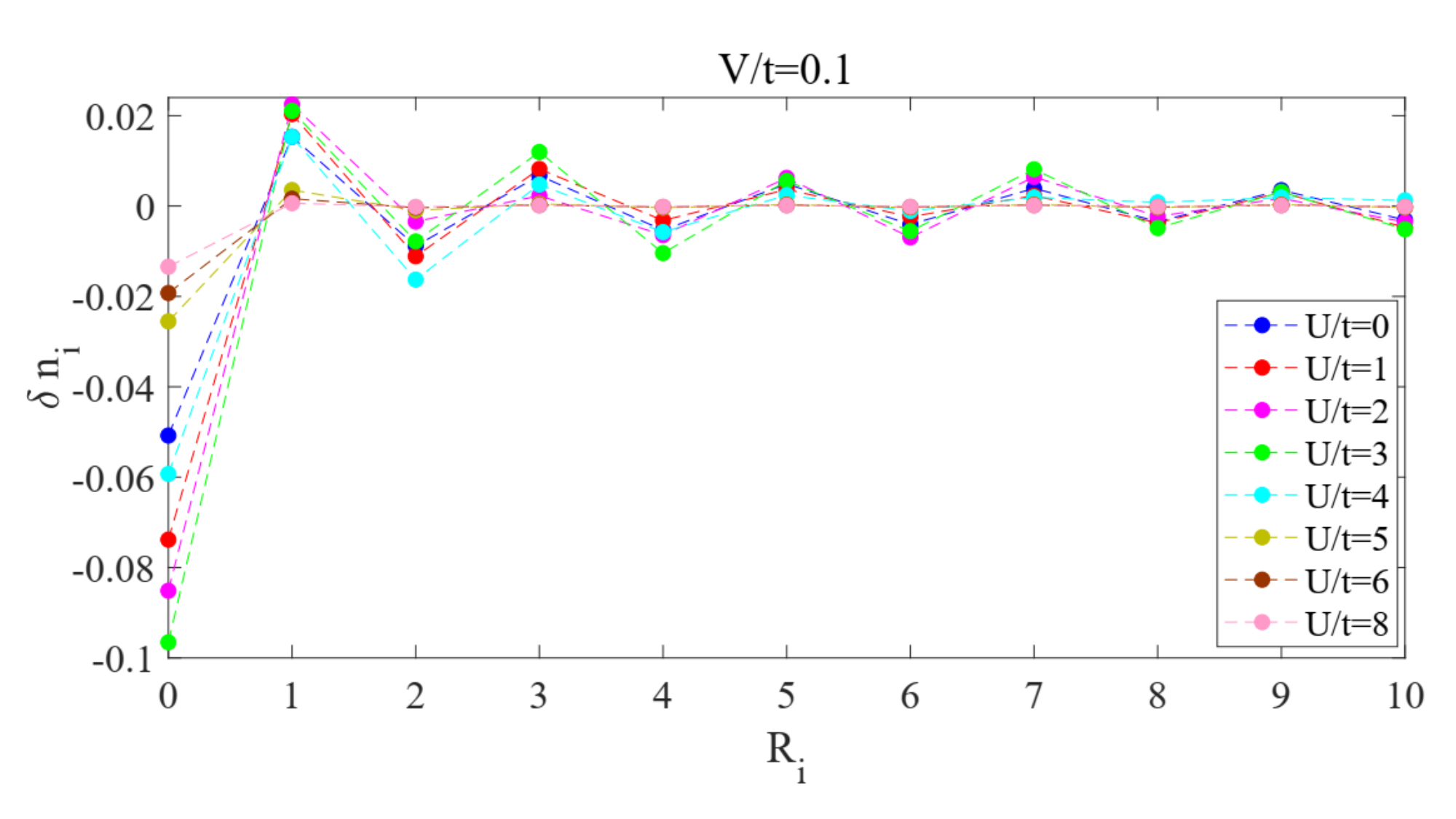}
	\caption{\label{fig:V-01-LRT} $\delta n_{i}$ calculated from the linear-response theory for $U/t=0,1,2,3,4,5,6,8$ with $V/t=0.1$ and $\mu=U/2$.}
\end{center}
\end{figure}

According to linear-response theory,\cite{Coleman2015} the fluctuation of electron density at $i$-site $\delta n_{i}(t)$ can be expressed as
\begin{eqnarray}
	\delta n_{i}(t)&=&\frac{1}{i}\int_{-\infty}^{t} dt'\langle[\hat{n}_{i}(t),\hat{H}_{\rm imp}(t')]\rangle=\frac{1}{i}\int_{-\infty}^{\infty}dt' \theta(t-t')\langle[\hat{n}_{i}(t),\hat{n}_{0}(t')]\rangle V(t')\nonumber\\
	&=&\int_{-\infty}^{\infty}dt' \chi_{c}(R_{i},R_{0},t-t')V(t')\nonumber
\end{eqnarray}
where the charge susceptibility in real space is defined as $\chi_{c}(R_{i},R_{0},t-t') =\frac{1}{i}\theta(t-t')\langle[\hat{n}_{i}(t),\hat{n}_{0}(t')]\rangle$ and its Fourier transformation is $\chi_{c}(q,\omega)$. Considering paramagnetic solution with $n_{k\sigma}=n_{k}$, we find
\begin{eqnarray}
\chi_{c}(q,\omega)&=&-\frac{1}{N_{s}}\sum_{k,\sigma}(1-n_{k})(1-n_{k+q})\frac{f_{F}(\varepsilon_{k}-\mu)-f_{F}(\varepsilon_{k+q}-\mu)}{\omega-\varepsilon_{k+q}+\varepsilon_{k}}\nonumber\\
&-&\frac{1}{N_{s}}\sum_{k,\sigma}(1-n_{k})n_{k+q}\frac{f_{F}(\varepsilon_{k}-\mu)-f_{F}(\varepsilon_{k+q}-\mu+U)}{\omega-\varepsilon_{k+q}-U+\varepsilon_{k}}\nonumber\\
&-&\frac{1}{N_{s}}\sum_{k,\sigma}n_{k}(1-n_{k+q})\frac{f_{F}(\varepsilon_{k}-\mu+U)-f_{F}(\varepsilon_{k+q}-\mu)}{\omega-\varepsilon_{k+q}+\varepsilon_{k}+U}\nonumber\\
&-&\frac{1}{N_{s}}\sum_{k,\sigma}n_{k}n_{k+q}\frac{f_{F}(\varepsilon_{k}-\mu+U)-f_{F}(\varepsilon_{k+q}-\mu+U)}{\omega-\varepsilon_{k+q}+\varepsilon_{k}}.
\end{eqnarray}
It is amusing that this expression can also be derived if one assumes Wick theorem is applicable although we have derived it by exact calculation from equation of motion method in \ref{ap-A}.
The reason is that since HK model is diagonalized in momentum space, one can expand perturbation around HK limit, for correlations such as charge susceptibility, the terms beyond Wick theorem are vanished in the thermodynamics limit.\cite{MZhao2023} Because impurity is static,
$\delta n_{i}=-V \frac{1}{N_{s}} \mathrm{Re}[\sum_{q}\exp(iq(R_{i}-R_{0}))\chi_{c}(q,\omega=0)]$ and site-dependent charge oscillation appears near impurity, as seen in Fig.~\ref{fig:V-01-LRT}. Furthermore, the oscillation is dominated by $q=k_{F1}+k_{F2}$, which underlies the inter-band transition from one quasi-Fermi surface into another one.

In addition, for symmetric half-filled system, the static charge susceptibility at zero temperature is \begin{eqnarray}
\chi_{c}=\lim_{q\rightarrow0}\chi_{c}(q,0)&=&-\frac{1}{4N_{s}}\sum_{k,\sigma}\frac{\partial f_{F}(\varepsilon_{k}-\mu)}{\partial\varepsilon_{k}}-\frac{1}{2N_{s}}\sum_{k,\sigma}\frac{f_{F}(\varepsilon_{k}-\mu+U)-f_{F}(\varepsilon_{k}-\mu)}{U}\nonumber\\
&-&\frac{1}{4N_{s}}\sum_{k,\sigma}\frac{\partial f_{F}(\varepsilon_{k}-\mu+U)}{\partial\varepsilon_{k}}.
\end{eqnarray}
For metallic state with $U<W$, all terms in $\chi_{c}$ are nonzero, thus $\chi_{c}$ is finite. For Mott insulator, the lower band is fully occupied ($f_{F}(\varepsilon_{k}-\mu)=1$) while upper band is empty ($f_{F}(\varepsilon_{k}-\mu+U)=0$), thus the first and third terms are zero but the second term survives as $\chi_{c}=\frac{1}{N_{s}}\sum_{k}\frac{1}{U}=\frac{1}{U}\neq0$. This means the Mott insulator has finite charge susceptibility, a nonphysical result because we require insulator to have $\chi_{c}=0$. This issue is first pointed out by Guerci et al., who have used a two-orbital HK model to illustrate the failure of $\chi_{c}$ to identify Mott insulating state.\cite{Guerci2024} In contrast, the charge susceptibility from thermodynamic relation has $\chi_{c}=\frac{\partial n}{\partial \mu}=N_{0}(\mu)+N_{0}(\mu-U)$, ($N_{0}(\omega)$ is the density of state of non-interacting system). We know $N_{0}(\omega)$ is zero for $\omega>W/2$ and the symmetric half-filled system with $\mu=U/2$ in Mott state must have $N_{0}(\mu)=N_{0}(\mu-U)=0$, therefore the thermodynamic charge susceptibility vanishes in Mott insulating phase as expected. Because the charge susceptibility from linear-response theory or Kubo formula is just the long-wavelength limit of $\chi_{c}(q,0)$, the inconsistency with thermodynamic susceptibility makes Guerci et al. claim that the charge correlation function of HK model is singular in long-wavelength limit.\cite{Guerci2024}

However, the conclusion of Guerci et al. has been challenged by Ma et al., who point out the importance of the non-commutativity of the long-wavelength and thermodynamic limits in the calculation of charge susceptibility and current-current correlation function.\cite{YMa2024} They claim that when the correct limits are taken, the correlation function from Kubo formula can yield physically reasonable results. In spite of these, Ma et al. obtain the charge susceptibility $\chi_{c}=\chi_{dir}+\chi_{cro}$ in long-wavelength and high frequency limit with parabolic non-interacting dispersion. ($\chi_{dir}\sim\frac{q^{2}}{\omega^{2}}$ and $\chi_{cro}\sim\frac{U}{\omega^{2}-U^{2}}$) If three dimensional Coulomb interaction $V(q)\sim 1/q^{2}$ is considered, the RPA charge susceptibility should be $\chi_{c}^{RPA}=\chi_{c}/(1-V(q)\chi_{c})$. Inserting $\chi_{c}$ into $\chi_{c}^{RPA}$, there is a pole around $\omega\sim1/q$, which gives us a plasmon excitation with dispersion $\omega\sim q^{-1}$. Such dispersion is unusual and in contrast to $\omega\sim$ constant in the standard Coulomb electron gas.\cite{Coleman2015} Crucially, the anomalous behavior of $\chi_{c}$ leads to the violation of standard $f$-sum rule, which states that
\begin{eqnarray}
\int_{-\infty}^{\infty}d\omega~\omega \mathrm{Im}\chi_{c}(q,\omega)=-\frac{\pi n}{m}q^{2}.\nonumber
\end{eqnarray}
Although this result is derived with assuming a free-particle dispersion $\varepsilon_{k}=\frac{k^{2}}{2m}$ and momentum-independent interactions, one expects it should be valid generally. The right-hand side of $f$-sum rule is related to commutator $\frac{\pi}{N_{s}}\langle[[\hat{H},\hat{\rho}_{q}],\hat{\rho}_{q}^{\dag}]\rangle$ with charge density operator $\hat{\rho}_{q}=\sum_{k\sigma}\hat{c}_{k\sigma}^{\dag}\hat{c}_{k+q\sigma}$. For HK model with $\hat{H}=\hat{H}_{0}+\hat{H}_{U}$, one finds $\langle[[\hat{H}_{0},\hat{\rho}_{q}],\hat{\rho}_{q}^{\dag}]\rangle=-\frac{q^{2}}{m}N_{e}$ and if the commutator for HK interaction $\hat{H}_{U}$ ($[[\hat{H}_{U},\hat{\rho}_{q}],\hat{\rho}_{q}^{\dag}]$) is zero, the $f$-sum rule is satisfied. However, Ma et al. find
\begin{equation}
\langle[[\hat{H}_{U},\hat{\rho}_{q}],\hat{\rho}_{q}^{\dag}]\rangle=
U\sum_{k}\left[2\langle \hat{n}_{k\uparrow}\hat{n}_{k\downarrow}\rangle-U\langle \hat{n}_{k\uparrow}\hat{n}_{k+q\downarrow}\rangle-U\langle \hat{n}_{k\uparrow}\hat{n}_{k-q\downarrow}\rangle+(\uparrow\leftrightarrow\downarrow)\right]\label{eq:f-sum_com}
\end{equation}
and it reduces into the following form for $q\neq0$
\begin{eqnarray}
\langle[[\hat{H}_{U},\hat{\rho}_{q}],\hat{\rho}_{q}^{\dag}]\rangle=
U\sum_{k}\left[2\langle \hat{n}_{k\uparrow}\hat{n}_{k\downarrow}\rangle-U\langle \hat{n}_{k\uparrow}\rangle\langle\hat{n}_{k+q\downarrow}\rangle-U\langle \hat{n}_{k\uparrow}\rangle\langle\hat{n}_{k-q\downarrow}\rangle+(\uparrow\leftrightarrow\downarrow)\right]\nonumber,
\end{eqnarray}
which is obviously nonzero. If we set $q=0$ in Eq.~\ref{eq:f-sum_com}, it gives $\langle[[\hat{H}_{U},\hat{\rho}_{q}],\hat{\rho}_{q}^{\dag}]\rangle=0$. Therefore, the long-wavelength limit $q\rightarrow0$ is singular for $\chi_{c}$.

\section{Beyond solvable limit}\label{sec:3}
We have seen that the solvable HK model has Mott transition and NFL behaviors. These models are diagonalized in momentum space, which means the quasiparticle has no scattering mechanism and they have infinite lifetime. On the other hand, realistic interaction, particularly the one in correlated electron systems is short-ranged, e.g. the Hubbard interaction and magnetic exchange interaction in $t-J$ or Kondo lattice model. These interactions must mix different momentum and lead to huge Hilbert space to treat. When this situation appears, the solvability of HK-like models is lost, so we think considering the non-HK interaction generally breaks the solvability.

Consider the HK model with Hubbard interaction $\hat{H}_{HKU}=\hat{H}_{HK}+U_{H}\sum_{j}\hat{n}_{j\uparrow}\hat{n}_{j\downarrow}$, where $U_{H}$ is the strength of Hubbard interaction. In momentum space, the above model can be written as,
\begin{equation}
\hat{H}_{HKU}=\sum_{k}\left[(\varepsilon_{k}-\mu)(\hat{n}_{k\uparrow}+\hat{n}_{k\downarrow})+U\hat{n}_{k\uparrow}\hat{n}_{k\downarrow}\right]+\frac{U_{H}}{N_{s}}\sum_{k,k',q}\hat{c}_{k+q\uparrow}^{\dag}\hat{c}_{k'-q\downarrow}^{\dag}\hat{c}_{k'\downarrow}\hat{c}_{k\uparrow}.
\end{equation}
Note that the Hubbard interaction involves transition between different momentum state, thus it breaks the solvability of original HK model. Obviously, solving the above model with Hubbard interaction is a hard problem, therefore certain approximations are needed. The simplest method is the perturbation theory, and other methods including the renormalization group to estimate the instability of HK physics. Besides, if only $d=1$ system is considered, direct calculation in terms of ED is feasible.
\subsection{Perturbation theory}
In principle, we can use the free electron part $\sum_{k\sigma}(\varepsilon_{k}-\mu)\hat{n}_{k\sigma}$ as the starting point of perturbation theory, but note that the ground-state of HK model is rather different from free electrons, thus new starting point is needed. Wang and Yang choose the Hartree-Fock mean-field Hamiltonian as the starting point,\cite{Wang2023} and the HK interaction can be written as
\begin{eqnarray}
\hat{n}_{k\uparrow}\hat{n}_{k\downarrow}&=&
(\langle\hat{n}_{k\uparrow}\rangle+\hat{n}_{k\uparrow}-\langle\hat{n}_{k\uparrow}\rangle)
(\langle\hat{n}_{k\downarrow}\rangle+\hat{n}_{k\downarrow}-\langle\hat{n}_{k\downarrow}\rangle)\nonumber\\
&=&\langle\hat{n}_{k\uparrow}\rangle\langle\hat{n}_{k\downarrow}\rangle+\langle\hat{n}_{k\uparrow}\rangle(\hat{n}_{k\downarrow}-\langle\hat{n}_{k\downarrow}\rangle)\nonumber\\
&+&(\hat{n}_{k\uparrow}-\langle\hat{n}_{k\uparrow}\rangle)\langle\hat{n}_{k\downarrow}\rangle
+(\hat{n}_{k\uparrow}-\langle\hat{n}_{k\uparrow}\rangle)(\hat{n}_{k\downarrow}-\langle\hat{n}_{k\downarrow}\rangle)\nonumber\\
&=&\langle\hat{n}_{k\uparrow}\rangle\hat{n}_{k\downarrow}+\hat{n}_{k\uparrow}\langle\hat{n}_{k\downarrow}\rangle-\langle\hat{n}_{k\uparrow}\rangle\hat{n}_{k\downarrow}-\hat{n}_{k\uparrow}\langle\hat{n}_{k\downarrow}\rangle+\hat{n}_{k\uparrow}\hat{n}_{k\downarrow}.\nonumber
\end{eqnarray}
Now, absorbing the first two terms into the definition of free electron Hamiltonian, and the remaining terms are the interacting ones. So,
\begin{equation}
\hat{H}_{HKU}=\hat{H}_{0}+\hat{H}_{1}+\hat{H}_{2}+\hat{H}_{H}
\end{equation}
\begin{eqnarray}
&&\hat{H}_{0}=\sum_{k\sigma}(\varepsilon_{k}-\mu+U\langle\hat{n}_{k\bar{\sigma}}\rangle)\hat{n}_{k\sigma}\nonumber\\
&&\hat{H}_{1}=-\sum_{k\sigma}U\langle\hat{n}_{k\bar{\sigma}}\rangle\hat{n}_{k\sigma}\nonumber\\
&&\hat{H}_{2}=U\sum_{k}\hat{n}_{k\uparrow}\hat{n}_{k\downarrow}\nonumber\\
&&\hat{H}_{H}=\frac{U_{H}}{N_{s}}\sum_{k,k',q}\hat{c}_{k+q\uparrow}^{\dag}\hat{c}_{k'-q\downarrow}^{\dag}\hat{c}_{k'\downarrow}\hat{c}_{k\uparrow}.
\end{eqnarray}
For $\hat{H}_{0}$, its single-particle Green function is
\begin{equation}
G_{\sigma}^{0}(k,\omega)=\frac{1}{\omega-\varepsilon_{k}+\mu-U\langle\hat{n}_{k\bar{\sigma}}\rangle}
\end{equation}
On the other hand, we know the exact single-particle Green function of HK model is
\begin{equation}
G_{\sigma}(k,\omega)=\frac{1-\langle\hat{n}_{k\bar{\sigma}}\rangle}{\omega-\varepsilon_{k}+\mu}+\frac{\langle\hat{n}_{k\bar{\sigma}}\rangle}{\omega-\varepsilon_{k}+\mu-U}.
\end{equation}
The distribution function at zero temperature is $\langle\hat{n}_{k}\rangle=\sum_{\sigma}\langle\hat{n}_{k\sigma}\rangle=\theta(\mu-\varepsilon_{k})+\theta(\mu-\varepsilon_{k}-U)$, its value is $0,1,2$. We try to show that $G_{\sigma}^{0}(k,\omega)$ at $T=0$ is identical to the exact one $G_{\sigma}(k,\omega)$. This is because, when $\varepsilon_{k}<\mu,\varepsilon_{k}+U<\mu$ or $\varepsilon_{k}+U<\mu$, the distribution function $\langle\hat{n}_{k\sigma}\rangle=1$, at this time
\begin{equation}
G_{\sigma}(k,\omega)=\frac{1}{\omega-\varepsilon_{k}+\mu-U},~~G_{\sigma}^{0}(k,\omega)=\frac{1}{\omega-\varepsilon_{k}+\mu-U}.
\end{equation}
It is seen that these two ones are identical. For $\varepsilon_{k}>\mu,\varepsilon_{k}+U>\mu$, namely $\varepsilon_{k}>\mu$, we have $\langle\hat{n}_{k\sigma}\rangle=0$. The single-particle Green function is
\begin{equation}
G_{\sigma}(k,\omega)=\frac{1}{\omega-\varepsilon_{k}+\mu},~~
G_{\sigma}^{0}(k,\omega)=\frac{1}{\omega-\varepsilon_{k}+\mu}.
\end{equation}
Finally, for $\varepsilon_{k}<\mu,\varepsilon_{k}+U>\mu$, the distribution function is $1$. To lift the spin degeneracy, we assume there exists a small magnetic field along $z$-axis, this means we can set $\langle\hat{n}_{k\uparrow}\rangle=1$,$\langle\hat{n}_{k\downarrow}\rangle=0$, so
\begin{eqnarray}
&&G_{\uparrow}(k,\omega)=\frac{1}{\omega-\varepsilon_{k}+\mu},~~G_{\downarrow}(k,\omega)=\frac{1}{\omega-\varepsilon_{k}+\mu-U}\nonumber\\
&&G_{\uparrow}^{0}(k,\omega)=\frac{1}{\omega-\varepsilon_{k}+\mu},~~G_{\downarrow}^{0}(k,\omega)=\frac{1}{\omega-\varepsilon_{k}+\mu-U}.
\end{eqnarray}
We see that for three different parameters/distribution functions $G_{\sigma}^{0}(k,\omega)$ at $T=0$ is identical to $G_{\sigma}(k,\omega)$ of exact solution. Thus, we think the Hartree-Fock Hamiltonian $\hat{H}_{0}$ can lead to correct Green function at zero temperature, and this is also the reason to use $\hat{H}_{0}$ as the starting point of perturbation theory. The remaining work is to treat $\hat{H}_{1},\hat{H}_{2},\hat{H}_{H}$ as the perturbation and introduce the corresponding Feynman diagrams.
\begin{figure}
\begin{center}
\includegraphics[width=0.65\columnwidth]{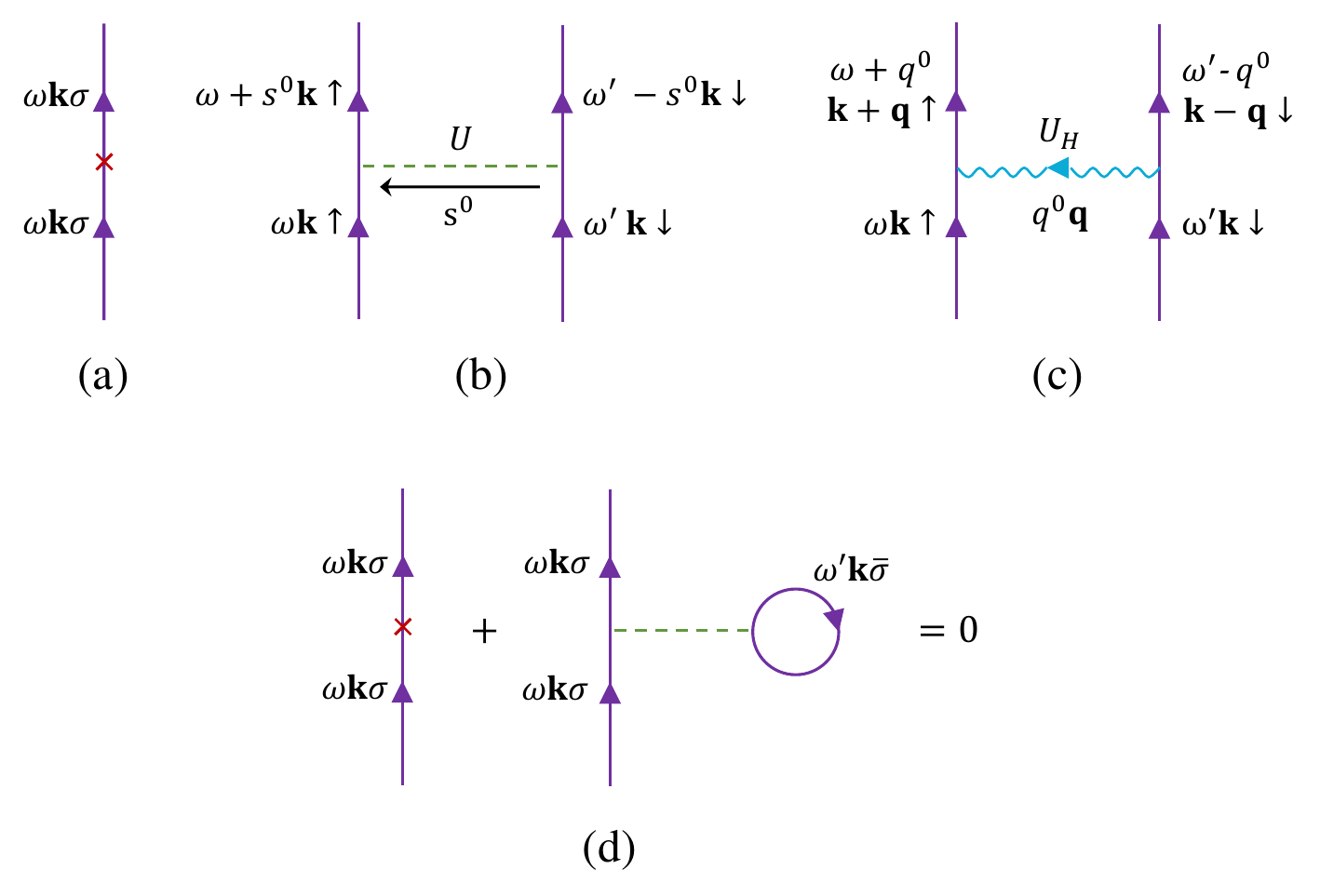}
\caption{\label{fig:HK_Feynman} Interaction vertex induced by HK and Hubbard interaction. (a) $H_{1}$, (b) $H_{2}$, (c) $H_{H}$. The self-energy correction of (a) and Hartree term (b) cancel out, which is expressed in (d).}
\end{center}
\end{figure}

\begin{figure}
\begin{center}
\includegraphics[width=0.75\columnwidth]{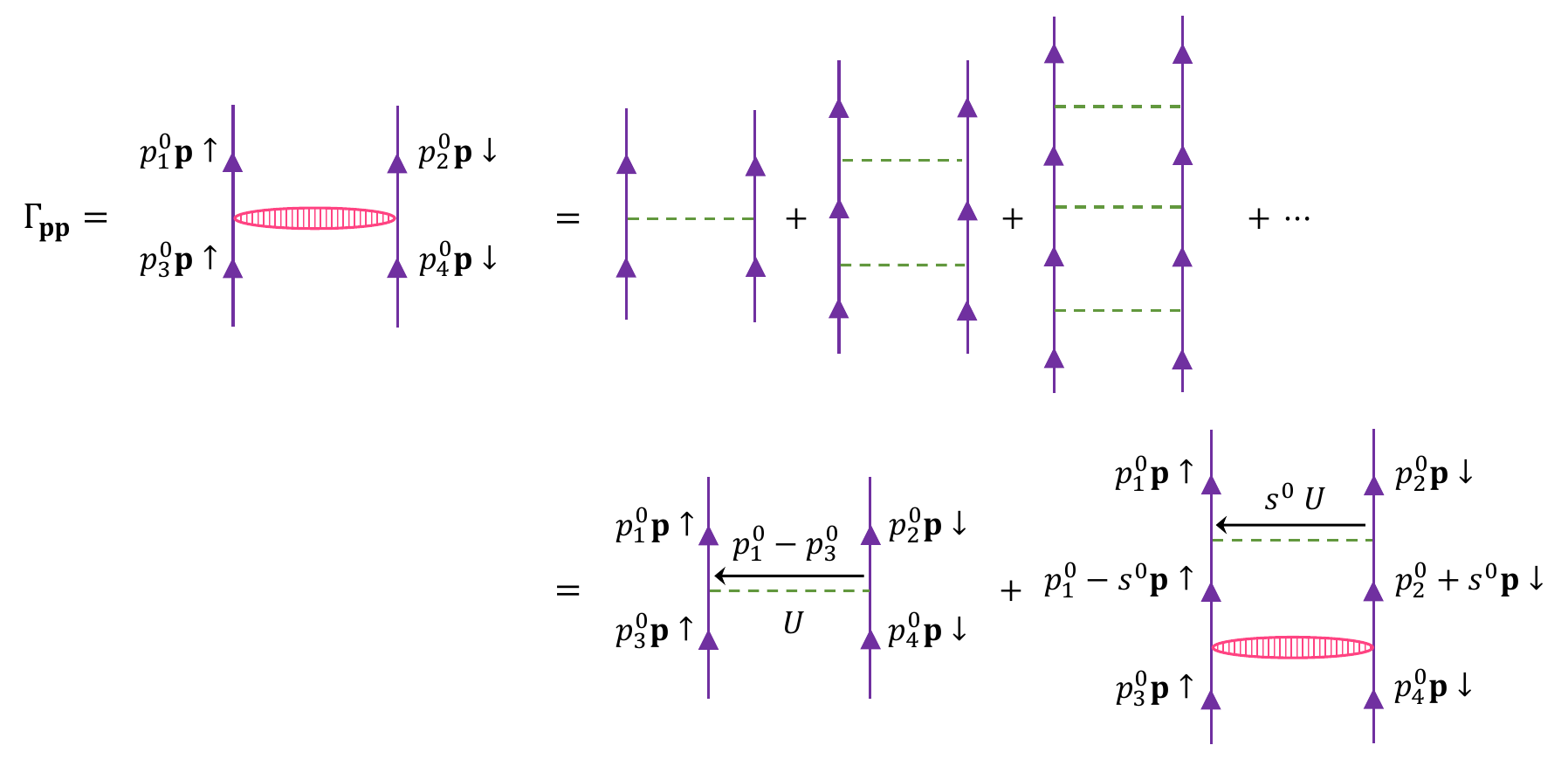}
\caption{\label{fig:HK_Feynman2} Ladder diagrams in the particle-particle channel.}
\end{center}
\end{figure}

\begin{figure}
\begin{center}
\includegraphics[width=0.75\columnwidth]{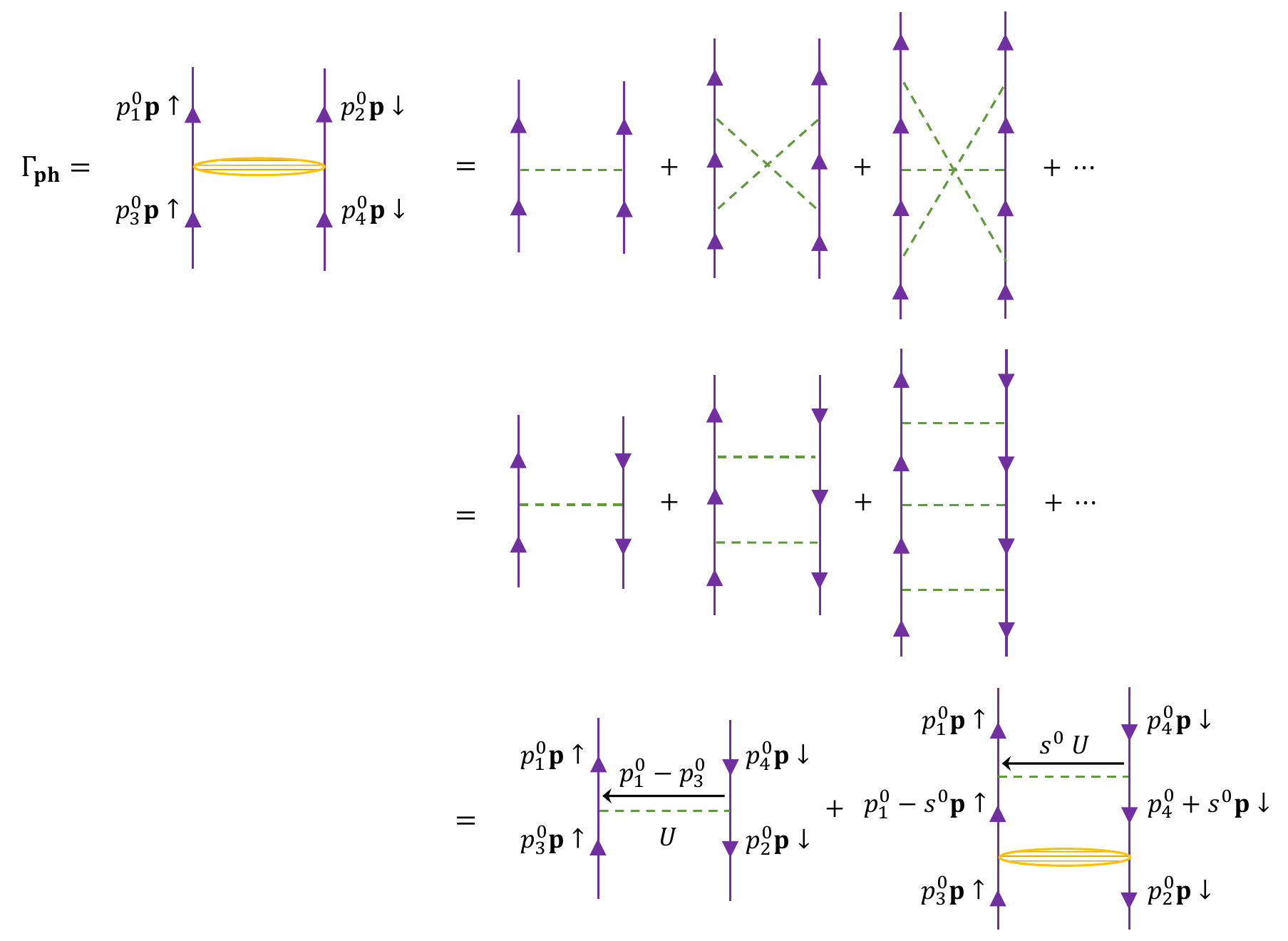}
\caption{\label{fig:HK_Feynman3} Ladder diagrams in the particle-hole channel.}
\end{center}
\end{figure}

Fig.~\ref{fig:HK_Feynman} gives three kinds of vertex Feynman diagram for $\hat{H}_{1},\hat{H}_{2},\hat{H}_{H}$. In (a), momentum and frequency are conserved for $\hat{H}_{1}$, its vertex is $-U\langle \hat{n}_{k\bar{\sigma}}\rangle$. (b) corresponds to the HK interaction whose vertex is $-U$. (c) corresponds to Hubbard interaction with vertex $-U_{H}$. An interesting fact is that the self-energy correction from (a) and (b) cancel out, (Fig.~\ref{fig:HK_Feynman}(d)) and it seems to be valid to any order of perturbation theory,
\begin{eqnarray}
\Sigma_{a}+\Sigma_{bH}&=&-U\langle \hat{n}_{k\bar{\sigma}}\rangle+(-U)(-1)\int\frac{d\omega'}{2\pi}G_{\bar{\sigma}}^{0}(k,i\omega')\nonumber\\
&=&-U\langle \hat{n}_{k\bar{\sigma}}\rangle+U\int\frac{d\omega'}{2\pi}\frac{1}{i\omega'-\varepsilon_{k}+\mu-U\langle \hat{n}_{k\sigma}\rangle}\nonumber\\
&=&-U\langle \hat{n}_{k\bar{\sigma}}\rangle+U\overbrace{\theta(\mu-\varepsilon_{k}-U\langle\hat{n}_{k\sigma}\rangle)}^{=\langle \hat{n}_{k\bar{\sigma}}\rangle}=0.
\end{eqnarray}
Furthermore, Wang and Yang find the contribution of $\hat{H}_{2}$ can be written as the summation of ladder diagram in particle-particle and particle-hole channel. It is easy to check that the contribution of Fock-like exchange diagram vanishes, and up to the $U^{2}$-order, only Hartree-like diagram contributes. At the same time, the one-loop diagram from $\hat{H}_{2}$ gives zero result, so all the ring diagrams do not contribute. Thus, only ladder diagrams contribute, (Fig.~\ref{fig:HK_Feynman2}, Fig.~\ref{fig:HK_Feynman3})
\begin{eqnarray}
\Gamma_{pp}(p_{1}^{0}-p_{2}^{0},p)=\frac{U(1-\langle \hat{n}_{p\sigma}\rangle)(1-\langle \hat{n}_{p\bar{\sigma}}\rangle)}{1-\frac{U}{i(p_{1}^{0}+p_{2}^{0})-(E_{p\sigma}+E_{p\bar{\sigma}})}}
+\frac{U\langle \hat{n}_{p\sigma}\rangle\langle \hat{n}_{p\bar{\sigma}}\rangle}{1+\frac{U}{i(p_{1}^{0}+p_{2}^{0})-(E_{p\sigma}+E_{p\bar{\sigma}})}},\nonumber
\end{eqnarray}
\begin{eqnarray}
\Gamma_{ph}(p_{1}^{0}-p_{4}^{0},p)=\frac{U(1-\langle \hat{n}_{p\sigma}\rangle)\langle \hat{n}_{p\bar{\sigma}}\rangle}{1-\frac{U}{i(p_{1}^{0}-p_{4}^{0})-(E_{p\sigma}-E_{p\bar{\sigma}})}}
+\frac{U\langle \hat{n}_{p\sigma}\rangle(1-\langle \hat{n}_{p\bar{\sigma}}\rangle)}{1+\frac{U}{i(p_{1}^{0}-p_{4}^{0})-(E_{p\sigma}-E_{p\bar{\sigma}})}}.\nonumber
\end{eqnarray}
Here $\Gamma_{pp},\Gamma_{ph}$ denote the vertex function in the particle-particle and particle-hole channel and $E_{p\sigma}=\varepsilon_{p}-\mu+U\langle \hat{n}_{p\bar{\sigma}}\rangle$.
\begin{figure}
\begin{center}
\includegraphics[width=0.55\columnwidth]{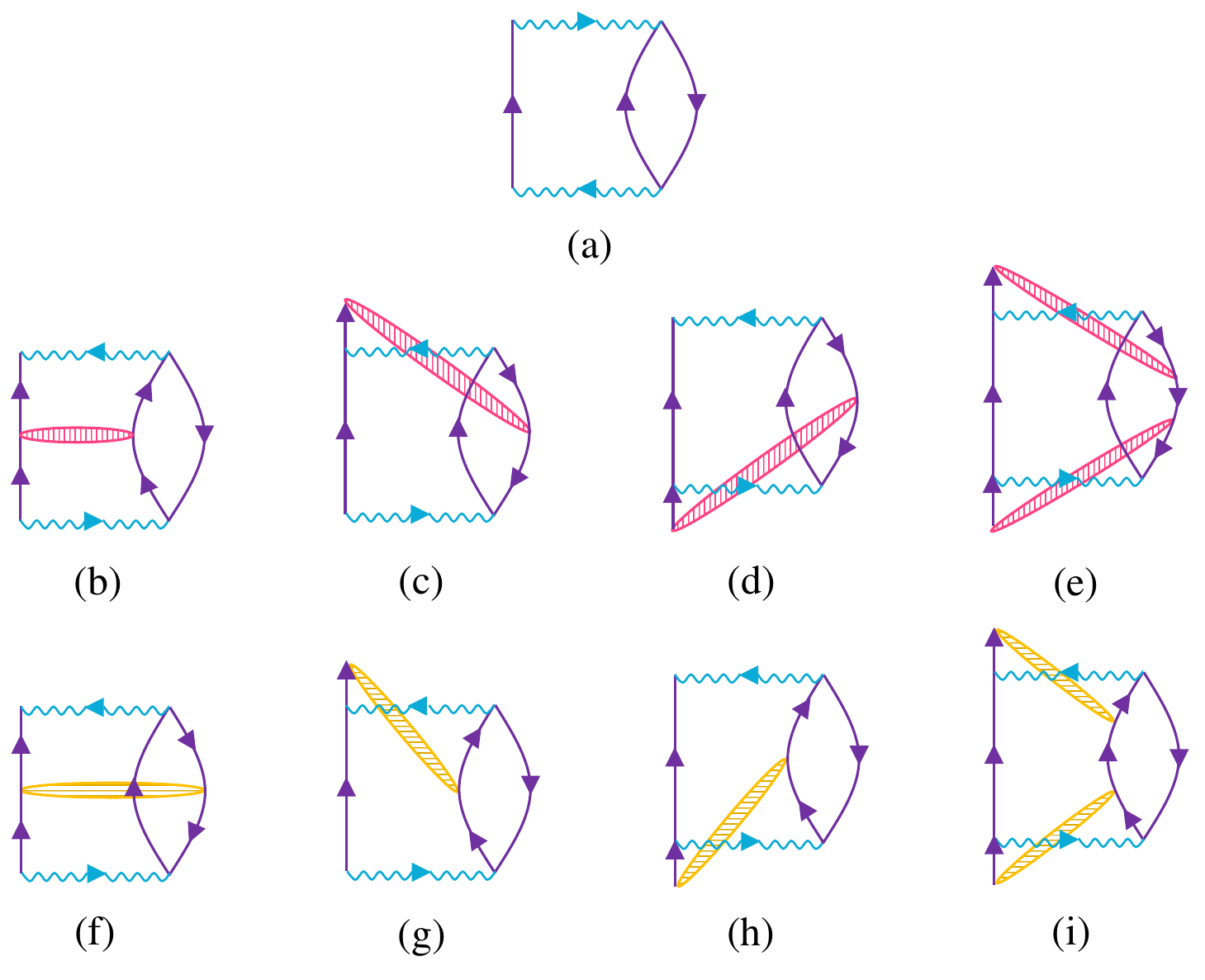}
\caption{\label{fig:HK_Feynman4} Second-order self-energy diagrams with Hubbard interaction $U_{H}$.}
\end{center}
\end{figure}

Using $\Gamma_{pp},\Gamma_{ph}$, one can calculate the self-energy. Fig.~\ref{fig:HK_Feynman4} gives the diagrams of self-energy up to the second order of $U_{H}$. From Fig.~\ref{fig:HK_Feynman4}(a), one finds the imaginary part of self-energy is $\sim -N^{3}(0)U_{H}^{2}\omega^{2}$, and it has a feature of FL. ($N(0)$ is the density of state at Fermi energy) For Fig.~\ref{fig:HK_Feynman4}(b), after a lengthy calculation, the imaginary part of self-energy is $-\pi U_{H}^{2}N^{2}(0)Uf(k)$ which is nonzero at Fermi surface and can be considered as a NFL effect. We recall that a standard FL model has vanished imaginary part of self-energy at Fermi surface and a finite one has been obtained in the example of Kondo impurity problem.\cite{Coleman2015} Here, Kondo screening effect does not work and we have not a clear picture about the origin of this nonvanished self-energy.

\subsection{Renormalization group}
Except for straightforward perturbation calculation, one can follow the treatment of Shankar, who performs an RG calculation for FL,\cite{Shankar1994} and it is useful to discuss the stability of fixed point of HK model. The work of Phillips group concludes that such calculation is able to show the HK fixed point is stable,
certainly, pairing interaction always leads to superconducting instability.\cite{JZhao2023}

The starting point of Phillips group is the path integral formalism of HK model,
\begin{equation}
\mathcal{Z}=\int D\bar{c}Dc e^{-S},~~S=\int d\tau\sum_{k}\left[\sum_{\sigma}\bar{c}_{k\sigma}(\partial_{\tau}+\varepsilon_{k}-\mu)c_{k\sigma}
+U\bar{c}_{k\uparrow}\bar{c}_{k\uparrow}c_{k\downarrow}c_{k\downarrow}\right]
\end{equation}
We know that HK model has pseudo-Fermi surface in its metallic phase. For simple bands like hypercubic lattice, $U<W$ has two but $U>W$ only has one. For simplicity, consider only one pseudo-Fermi surface is active, and it is determined by $\varepsilon_{k}-\mu=0$. Expanding around this Fermi surface leads to $\varepsilon_{k}-\mu\simeq v_{F}(k-k_{L})$, and the action reads
\begin{eqnarray}
S&=&\int d\tau\int \frac{kdkd\theta}{(2\pi)^{2}}\sum_{\sigma}\bar{\psi}_{\sigma}(k,\theta,\tau)(\partial_{\tau}+v_{F}(k-k_{L}))\psi_{\sigma}(k,\theta,\tau)\nonumber\\
&+&\int d\tau\int \frac{kdkd\theta}{(2\pi)^{2}}U\bar{\psi}_{\uparrow}(k,\theta,\tau)\bar{\psi}_{\uparrow}(k,\theta,\tau)\psi_{\downarrow}(k,\theta,\tau)\psi_{\downarrow}(k,\theta,\tau)\nonumber
\end{eqnarray}
Here, we have written fermions around Fermi surface as $\psi$. Furthermore, set $k=k_{L}+\tilde{k}$ and $\tilde{k}\in[-\Lambda,\Lambda]$ with $\Lambda$ as momentum cutoff. With these prescriptions, the effective action describing electron near pseudo-Fermi surface reads
\begin{eqnarray}
S&=&\int d\tau\int \frac{k_{L}d\tilde{k}d\theta}{(2\pi)^{2}}\sum_{\sigma}\bar{\psi}_{\sigma}(\tilde{k},\theta,\tau)(\partial_{\tau}+v_{F}\tilde{k})\psi_{\sigma}(\tilde{k},\theta,\tau)\nonumber\\
&+&\int d\tau\int \frac{k_{L}d\tilde{k}d\theta}{(2\pi)^{2}}U\bar{\psi}_{\uparrow}(\tilde{k},\theta,\tau)\bar{\psi}_{\uparrow}(\tilde{k},\theta,\tau)\psi_{\downarrow}(\tilde{k},\theta,\tau)\psi_{\downarrow}(\tilde{k},\theta,\tau).\nonumber
\end{eqnarray}
\begin{figure}
\begin{center}
\includegraphics[width=0.55\columnwidth]{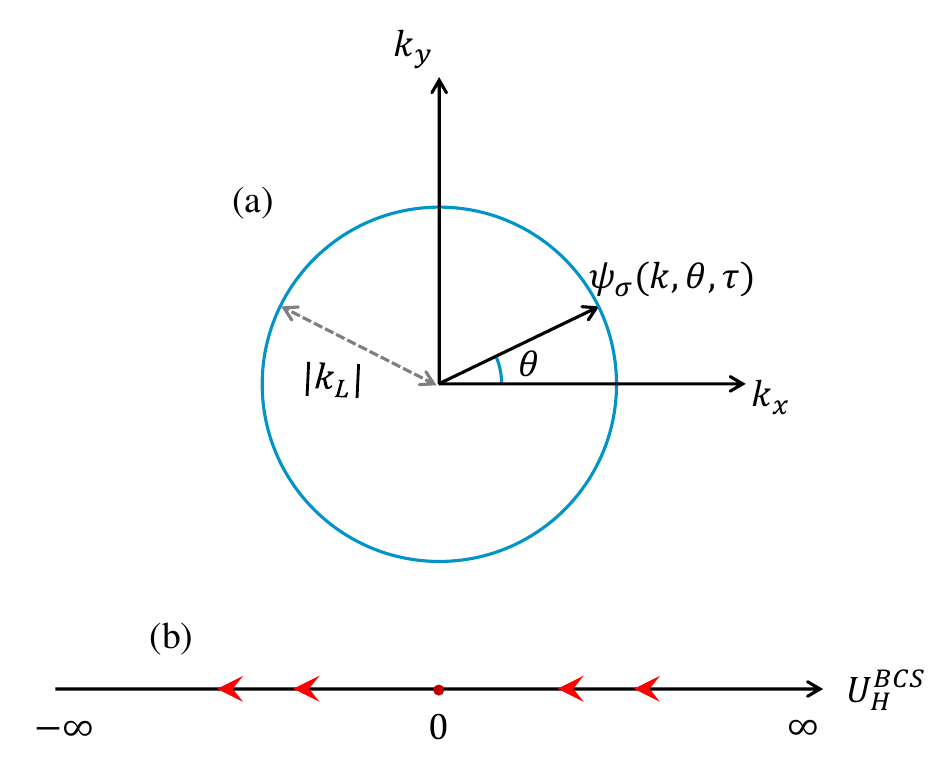}
\caption{\label{fig:RG} (a) Electrons near pseudo-Fermi surface with Fermi wavevector $k_{L}$. (b) The RG flow of HK model with local interaction in BCS-pairing channel.}
\end{center}
\end{figure}

In the standard calculation of RG, we divide the degree of freedom into high energy and low energy part. Here we treat $|\tilde{k}|\in[\Lambda/b,\Lambda]$ as the high energy part while $|\tilde{k}|<\Lambda/b$ is the low energy part. It is seen that high and low energy degree of freedom decouple in $S$, $S=S_{<}+S_{>}$,
\begin{eqnarray}
S_{<}&=&\int d\tau\int_{\Lambda/b} \frac{k_{L}d\tilde{k}d\theta}{(2\pi)^{2}}\sum_{\sigma}\bar{\psi}_{\sigma}(\tilde{k},\theta,\tau)(\partial_{\tau}+v_{F}\tilde{k})\psi_{\sigma<}(\tilde{k},\theta,\tau)\nonumber\\
&+&\int d\tau\int_{\Lambda/b}  \frac{k_{L}d\tilde{k}d\theta}{(2\pi)^{2}}U\bar{\psi}_{\uparrow<}(\tilde{k},\theta,\tau)\bar{\psi}_{\uparrow<}(\tilde{k},\theta,\tau)\psi_{\downarrow<}(\tilde{k},\theta,\tau)\psi_{\downarrow<}(\tilde{k},\theta,\tau).\nonumber
\end{eqnarray}
\begin{eqnarray}
S_{>}&=&\int d\tau\int \frac{k_{L}d\tilde{k}d\theta}{(2\pi)^{2}}\sum_{\sigma}\bar{\psi}_{\sigma>}(\tilde{k},\theta,\tau)(\partial_{\tau}+v_{F}\tilde{k})\psi_{\sigma>}(\tilde{k},\theta,\tau)\nonumber\\
&+&\int d\tau\int \frac{k_{L}d\tilde{k}d\theta}{(2\pi)^{2}}U\bar{\psi}_{\uparrow>}(\tilde{k},\theta,\tau)\bar{\psi}_{\uparrow>}(\tilde{k},\theta,\tau)\psi_{\downarrow>}(\tilde{k},\theta,\tau)\psi_{\downarrow>}(\tilde{k},\theta,\tau).\nonumber
\end{eqnarray}
$\psi_{\sigma<}$ denotes the low energy mode with $|\tilde{k}|<\Lambda/b$ while $\psi_{\sigma>}$ has $|\tilde{k}|\in[\Lambda/b,\Lambda]$ as the high energy mode. Since high and low energy modes decouple, integrating out the high energy mode $\psi_{\sigma>}$ gives the action for low energy mode
\begin{eqnarray}
\mathcal{Z}&=&\int D\bar{\psi}_{<}D\psi_{<}e^{-S_{<}}\int D\bar{\psi}_{>}D\psi_{>}e^{-S_{>}}\propto\int D\bar{\psi}_{<}D\psi_{<}e^{-S_{<}}.
\end{eqnarray}
Then we rescale momentum as $\tilde{k}'=b\tilde{k}$,
\begin{eqnarray}
S_{<}&=&\int d\tau\int_{\Lambda} \frac{k_{L}d\tilde{k}'d\theta}{(2\pi)^{2}}\sum_{\sigma}\bar{\psi}_{\sigma}(\tilde{k}'/b,\theta,\tau)\frac{1}{b}(\partial_{\tau}+v_{F}\tilde{k}'/b)\psi_{\sigma<}(\tilde{k}'/b,\theta,\tau)\nonumber\\
&+&\int d\tau\int_{\Lambda} \frac{k_{L}d\tilde{k}'d\theta}{(2\pi)^{2}}\frac{U}{b}\bar{\psi}_{\uparrow<}(\tilde{k}'/b,\theta,\tau)\bar{\psi}_{\uparrow<}(\tilde{k}'/b,\theta,\tau)\psi_{\downarrow<}(\tilde{k}'/b,\theta,\tau)\psi_{\downarrow<}(\tilde{k}'/b,\theta,\tau).\nonumber
\end{eqnarray}
Obviously, we also make rescaling for $\tau,\psi_{<}$ and $U$ as
\begin{equation}
\tau'=b^{-1}\tau,~~\psi_{<}'(\tilde{k}',\theta,\tau')=b^{-1/2}\psi_{<}(\tilde{k}'/b,\theta,b\tau'),~~U'=b^{2}U.
\end{equation}
So that
\begin{eqnarray}
S_{<}&=&\int d\tau'\int_{\Lambda} \frac{k_{L}d\tilde{k}'d\theta}{(2\pi)^{2}}\sum_{\sigma}\bar{\psi}_{\sigma}'(\tilde{k}',\theta,\tau')(\partial_{\tau'}+v_{F}\tilde{k}')\psi_{\sigma<}'(\tilde{k}',\theta,\tau')\nonumber\\
&+&\int d\tau'\int_{\Lambda} \frac{k_{L}d\tilde{k}'d\theta}{(2\pi)^{2}}U'\bar{\psi}_{\uparrow<}'(\tilde{k}',\theta,\tau')\bar{\psi}_{\uparrow<}'(\tilde{k}',\theta,\tau')\psi_{\downarrow<}'(\tilde{k}',\theta,\tau')\psi_{\downarrow<}'(\tilde{k}',\theta,\tau').\nonumber
\end{eqnarray}
For $U'=b^{2}U$, the HK interaction is highly relevant, whose effect must be included.

Next, considering non-HK interaction, e.g. the local in space interaction, the Hubbard interaction, its action is
\begin{equation}
S_{int}=U_{H}\int d\tau d^{2}\vec{x}\bar{\psi}_{\uparrow}(\vec{x},\tau)\bar{\psi}_{\uparrow}(\vec{x},\tau)\psi_{\downarrow}(\vec{x},\tau)\psi_{\downarrow}(\vec{x},\tau)
\end{equation}
After Fourier transformation, we find
\begin{eqnarray}
S_{int}&=&U_{H}\int d\tau \int\frac{d^{2}\vec{k}_{1}}{(2\pi)^{2}}
\int d^{2}\frac{d^{2}\vec{k}_{2}}{(2\pi)^{2}}
\int d^{2}\frac{d^{2}\vec{k}_{3}}{(2\pi)^{2}}
\int d^{2}\frac{d^{2}\vec{k}_{4}}{(2\pi)^{2}}\nonumber\\
&\times&(2\pi)^{2}\delta(\vec{k}_{1}+\vec{k}_{2}-\vec{k}_{3}-\vec{k}_{4})\bar{\psi}_{\uparrow}(\vec{k}_{1},\tau)\bar{\psi}_{\uparrow}(\vec{k}_{2},\tau)\psi_{\downarrow}(\vec{k}_{3},\tau)\psi_{\downarrow}(\vec{k}_{4},\tau)\nonumber.
\end{eqnarray}
Then, expanding near Fermi surface gives
\begin{eqnarray}
\vec{k}_{1}=(k_{L}+\tilde{k}_{1})\vec{n}_{1}(\theta_{1}),~\vec{k}_{2}=(k_{L}+\tilde{k}_{2})\vec{n}_{2}(\theta_{2}),
~\vec{k}_{3}=(k_{L}+\tilde{k}_{3})\vec{n}_{3}(\theta_{3}),~\vec{k}_{4}=(k_{L}+\tilde{k}_{4})\vec{n}_{4}(\theta_{4}).\nonumber
\end{eqnarray}
Note that $\vec{n}_{i}(\theta_{i})$ is the unit vector for $\theta_{i}$, $|\tilde{k}_{i}|<\Lambda$. Inserting this result into $S_{int}$ leads to
\begin{eqnarray}
S_{int}&=&U_{H}\int d\tau \int\frac{d\theta_{1}k_{L}d\tilde{k}_{1}}{(2\pi)^{2}}
\int\frac{d\theta_{2}k_{L}d\tilde{k}_{2}}{(2\pi)^{2}}
\int\frac{d\theta_{3}k_{L}d\tilde{k}_{3}}{(2\pi)^{2}}
\int\frac{d\theta_{4}k_{L}d\tilde{k}_{4}}{(2\pi)^{2}}
\nonumber\\
&\times&(2\pi)^{2}\delta(k_{L}(\vec{n}_{1}(\theta_{1})+\vec{n}_{2}(\theta_{2})-\vec{n}_{3}(\theta_{3})-\vec{n}_{4}(\theta_{4}))\nonumber\\
&+&\tilde{k}_{1}\vec{n}_{1}(\theta_{1})+\tilde{k}_{2}\vec{n}_{2}(\theta_{2})-\tilde{k}_{3}\vec{n}_{3}(\theta_{3})-\tilde{k}_{4}\vec{n}_{4}(\theta_{4}))\nonumber\\
&\times&\bar{\psi}_{\uparrow}(\tilde{k}_{1},\theta_{1},\tau)\bar{\psi}_{\uparrow}(\tilde{k}_{2},\theta_{2},\tau)\psi_{\downarrow}(\tilde{k}_{3},\theta_{3},\tau)\psi_{\downarrow}(\tilde{k}_{4},\theta_{4},\tau)\nonumber.
\end{eqnarray}
To satisfy the $\delta$ function, the part with order $k_{L}$ can be zero, and it gives three kinds of conditions. The first case is $\theta_{1}=\theta_{4}=\theta,\theta_{2}=\theta_{3}=\theta'$, which corresponds to forward scattering. The second one is $\theta_{1}=\theta_{3}=\theta,\theta_{2}=\theta_{3}=\theta'$ and this is called exchange scattering. The last one is $\vec{n}_{1}=-\vec{n}_{2}$,$\vec{n}_{3}=-\vec{n}_{4}$, ($\theta_{1}=-\theta_{2}=\theta,\theta_{4}=-\theta_{3}=\theta'$) and this describes the BCS pairing scattering. The corresponding action is
\begin{eqnarray}
S_{int}^{1}&=&U_{H}\int d\tau \int d\theta \int d\theta'\int\frac{k_{L}d\tilde{k}_{1}}{(2\pi)^{2}}
\int\frac{k_{L}d\tilde{k}_{2}}{(2\pi)^{2}}
\int\frac{k_{L}d\tilde{k}_{3}}{(2\pi)^{2}}
\int\frac{k_{L}d\tilde{k}_{4}}{(2\pi)^{2}}
\nonumber\\
&\times&(2\pi)^{2}\delta\left(\tilde{k}_{1}\vec{n}(\theta)+\tilde{k}_{2}\vec{n}(\theta')-\tilde{k}_{3}\vec{n}(\theta')-\tilde{k}_{4}\vec{n}(\theta)\right)\nonumber\\
&\times&\bar{\psi}_{\uparrow}(\tilde{k}_{1},\theta,\tau)\bar{\psi}_{\uparrow}(\tilde{k}_{2},\theta',\tau)\psi_{\downarrow}(\tilde{k}_{3},\theta',\tau)\psi_{\downarrow}(\tilde{k}_{4},\theta,\tau)\nonumber.
\end{eqnarray}
\begin{eqnarray}
S_{int}^{2}&=&U_{H}\int d\tau \int d\theta \int d\theta'\int\frac{k_{L}d\tilde{k}_{1}}{(2\pi)^{2}}
\int\frac{k_{L}d\tilde{k}_{2}}{(2\pi)^{2}}
\int\frac{k_{L}d\tilde{k}_{3}}{(2\pi)^{2}}
\int\frac{k_{L}d\tilde{k}_{4}}{(2\pi)^{2}}
\nonumber\\
&\times&(2\pi)^{2}\delta\left(\tilde{k}_{1}\vec{n}(\theta)+\tilde{k}_{2}\vec{n}(\theta')-\tilde{k}_{3}\vec{n}(\theta)-\tilde{k}_{4}\vec{n}(\theta')\right)\nonumber\\
&\times&\bar{\psi}_{\uparrow}(\tilde{k}_{1},\theta,\tau)\bar{\psi}_{\uparrow}(\tilde{k}_{2},\theta',\tau)\psi_{\downarrow}(\tilde{k}_{3},\theta,\tau)\psi_{\downarrow}(\tilde{k}_{4},\theta',\tau)\nonumber.
\end{eqnarray}
\begin{eqnarray}
S_{int}^{3}&=&U_{H}\int d\tau \int d\theta \int d\theta'\int\frac{k_{L}d\tilde{k}_{1}}{(2\pi)^{2}}
\int\frac{k_{L}d\tilde{k}_{2}}{(2\pi)^{2}}
\int\frac{k_{L}d\tilde{k}_{3}}{(2\pi)^{2}}
\int\frac{k_{L}d\tilde{k}_{4}}{(2\pi)^{2}}
\nonumber\\
&\times&(2\pi)^{2}\delta\left(\tilde{k}_{1}\vec{n}(\theta)-\tilde{k}_{2}\vec{n}(\theta)+\tilde{k}_{3}\vec{n}(\theta')-\tilde{k}_{4}\vec{n}(\theta')\right)\nonumber\\
&\times&\bar{\psi}_{\uparrow}(\tilde{k}_{1},\theta,\tau)\bar{\psi}_{\uparrow}(\tilde{k}_{2},-\theta,\tau)\psi_{\downarrow}(\tilde{k}_{3},-\theta',\tau)\psi_{\downarrow}(\tilde{k}_{4},\theta',\tau)\nonumber.
\end{eqnarray}
In the tree-level, or just assuming high and low energy modes decouple, and using the rescaling rule $\tilde{k}'=b\tilde{k},\tau'=b^{-1}\tau,~~\psi_{<}'(\tilde{k}',\theta,\tau')=b^{-1/2}\psi_{<}(\tilde{k}'/b,\theta,b\tau')$ and $U'=b^{2}U$, we find the rescaled $U_{H}'$ satisfies,
\begin{equation}
U_{H}'=b^{2}b^{-4}b^{\frac{1}{2}4}U_{H}=U_{H}.
\end{equation}
where the following relation for $\delta$ function is used,
\begin{equation}
\delta\left(\tilde{k}_{i}'\vec{n}(\theta)/b+...\right)=b\delta\left(\tilde{k}_{i}'\vec{n}(\theta)+...\right).
\end{equation}
It is seen that the rescaling does not change $U_{H}$, which means the interaction is marginal. To confirm it is a true behavior, further calculation is needed.

Just like the calculation of FL performed by Shankar, but also note that the single-particle Green function of HK model depends on its distribution function, we should consider the regime with distribution function $\langle\hat{n}_{k}\rangle=0,1,2$. Phillips group finds that the forward and exchange scattering do not have correction up to the one-loop diagram level, which means they are still marginal. In other words, if these interactions are weak, they must be weak after the RG flow, and it confirms the perturbative nature of them. In contrast, the BCS pairing scattering is not marginal after considering the one-loop diagrams, and its corresponding $\beta$ function is (we use $U_{H}^{BCS}$ to emphasize $U_{H}$ is in the BCS pairing scattering channel)
\begin{equation}
\beta(U_{H}^{BCS})=\frac{dU_{H}^{BCS}}{d\ln b}\sim-\frac{1}{v_{F}}(U_{H}^{BCS})^{2}.
\end{equation}
This equation states that, if the original $U_{H}^{BCS}>0$, it will flow to $U_{H}^{BCS}=0$, this means the repulsive interaction is a irrelevant perturbation in BCS paring scattering channel, and it can be treated perturbatively. If original $U_{H}^{BCS}<0$, this is a different situation, and $U_{H}^{BCS}$ will flow to strong coupling case with $U_{H}^{BCS}\rightarrow-\infty$, which emphasizes that the system will be dominated by the attractive interaction and the resulting states will be the superconducting pairing states discussed in Ref.~\cite{Phillips2020,Zhu2021,Li2022,JZhao2022}. In essence, those works just add BCS-like pairing interaction $\hat{H}_{BCS}=\sum_{kk'}V_{kk'}\hat{c}^{\dag}_{k\uparrow}\hat{c}^{\dag}_{-k\downarrow}\hat{c}_{-k'\downarrow}\hat{c}_{k'\uparrow}$ into the HK Hamiltonian $\hat{H}_{HK}$. (Pairing model with finite Cooper-pair center-of-mass momentum has been investigated in Ref.~\cite{Froldi2024}, where the pair-density-wave fluctuations emerges.) Next, with the assumption of separable pairing interaction $V_{kk'}=-\frac{g}{N_{s}}\gamma_{k}\gamma_{k'}$ ($\gamma_{k}$ is pairing function), a mean-field approximation on $\hat{H}_{BCS}$ results in the following Hamiltonian
\begin{equation}
\hat{H}=\hat{H}_{HK}+\sum_{k}\Delta_{k}(\hat{c}^{\dag}_{k\uparrow}\hat{c}^{\dag}_{-k\downarrow}+\hat{c}_{-k\downarrow}\hat{c}_{k\uparrow})+\frac{N_{s}}{g}\Delta^{2}
\end{equation}
where $\Delta=-\frac{g}{N_{s}}\sum_{k}\gamma_{k'}\langle \hat{c}_{-k'\downarrow}\hat{c}_{k'\uparrow}\rangle$ acts as the superconducting energy gap and $\Delta_{k}=\Delta\gamma_{k}$. If we choose $\gamma_{k}=1$, the pairing term will lead to $s$-wave superconductivity. Although the solvability of $\hat{H}_{HK}$ is broken by pairing interaction, combining Hamiltonian with $k$ and $-k$ sector still admits an exact solution. Solving the resultant $16\times16$-matrix with the definition of $\Delta$ can tell us whether the system is superconducting. If yes, physical observable such as specific heat, susceptibility, optical conductivity density of state and dynamical spin correlation are readily to be calculated following the standard treatment.\cite{Xiang2022}

After all, we find weakly repulsive local interaction cannot change the basic behavior of HK model but attractive interaction can lead to superconducting instability.

\subsection{Exact diagonalization}
In the previous discussion on perturbation theory and RG, we treat the perturbation around HK model, e.g. Hubbard model, as perturbative, and its nature is perturbative. When these perturbations become larger, calculation based on perturbation theory may fail, thus we need some non-perturbative techniques to study non-HK interaction. In one spatial dimension with small lattice size, ED is the method of choice.\cite{Skolimowski2024}

Consider the $d=1$ HK Hamiltonian in real space $\hat{H}_{HK}=-t\sum_{j\sigma}(\hat{c}_{j\sigma}^{\dag}\hat{c}_{j+1\sigma}+h.c.)-\mu\sum_{j\sigma}\hat{c}_{j\sigma}^{\dag}\hat{c}_{j\sigma}+\frac{U}{N_{s}}\sum_{j_{1},j_{2},j_{3},j_{4}}\delta_{j_{1}+j_{3}=j_{2}+j_{4}}\hat{c}_{j_{1}\uparrow}^{\dag}\hat{c}_{j_{2}\uparrow}\hat{c}_{j_{3}\downarrow}^{\dag}\hat{c}_{j_{4}\downarrow}$. For periodic system, the constraint of $\delta$ function can be written as $\mathrm{mod}(j_{1}+j_{3}-j_{2}-j_{4},N_{s})=0$, which means the position of center of mass can have any multiple of system size. For system with size $N_{s}$, the number of terms of HK interaction in PBC is $N_{s}^{3}$, while for OBC, we must have $j_{1}+j_{3}-j_{2}-j_{4}=0$, which leads to smaller number of term. For example, for $N_{s}=10$, the number of terms in PBC is $1000$ while it is $670$ in OBC. For $N_{s}=100$, PBC has $1000000$ and OBC has $666700$, so the number of terms in OBC and PBC has the ratio $1/3$ when the size of system is large enough.

The simplest case is just two sites and PBC gives $8$ terms while OBC gives $6$. We can write down all terms in HK interaction. For OBC,
\begin{eqnarray}
\hat{H}_{U}&=&\frac{U}{2}\left(\hat{c}_{1\uparrow}^{\dag}\hat{c}_{1\uparrow}\hat{c}_{1\downarrow}^{\dag}\hat{c}_{1\downarrow}+
\hat{c}_{1\uparrow}^{\dag}\hat{c}_{1\uparrow}\hat{c}_{2\downarrow}^{\dag}\hat{c}_{2\downarrow}+
\hat{c}_{1\uparrow}^{\dag}\hat{c}_{2\uparrow}\hat{c}_{2\downarrow}^{\dag}\hat{c}_{1\downarrow}+
\hat{c}_{2\uparrow}^{\dag}\hat{c}_{1\uparrow}\hat{c}_{1\downarrow}^{\dag}\hat{c}_{2\downarrow}+
\hat{c}_{2\uparrow}^{\dag}\hat{c}_{2\uparrow}\hat{c}_{1\downarrow}^{\dag}\hat{c}_{1\downarrow}+
\hat{c}_{2\uparrow}^{\dag}\hat{c}_{2\uparrow}\hat{c}_{2\downarrow}^{\dag}\hat{c}_{2\downarrow}\right)\nonumber\\
&=&\frac{U}{2}\left(\hat{n}_{1\uparrow}\hat{n}_{1\downarrow}+\hat{n}_{2\uparrow}\hat{n}_{2\downarrow}+\hat{n}_{1\uparrow}\hat{n}_{2\downarrow}+\hat{n}_{2\uparrow}\hat{n}_{1\downarrow}+
-\hat{c}_{1\uparrow}^{\dag}\hat{c}_{1\downarrow}\hat{c}_{2\downarrow}^{\dag}\hat{c}_{2\uparrow}
-\hat{c}_{2\uparrow}^{\dag}\hat{c}_{2\downarrow}\hat{c}_{1\downarrow}^{\dag}\hat{c}_{1\uparrow}
\right)\nonumber\\
&=&\frac{U}{2}\left(\hat{n}_{1\uparrow}\hat{n}_{1\downarrow}+\hat{n}_{2\uparrow}\hat{n}_{2\downarrow}+\hat{n}_{1\uparrow}\hat{n}_{2\downarrow}+\hat{n}_{2\uparrow}\hat{n}_{1\downarrow}
-\hat{S}_{1}^{+}\hat{S}_{2}^{-}-\hat{S}_{2}^{+}\hat{S}_{1}^{-}\right)\nonumber\\
&=&\frac{U}{2}(\hat{n}_{1\uparrow}\hat{n}_{1\downarrow}+\hat{n}_{2\uparrow}\hat{n}_{2\downarrow})+\frac{U}{4}\hat{n}_{1}\hat{n}_{2}
-U\hat{\vec{S}}_{1}\cdot\hat{\vec{S}}_{2}.\nonumber
\end{eqnarray}
It corresponds to on-site Hubbard interaction, nearest-neighbor charge-charge interaction and Hund-like exchange interaction. When $U>0$, the Hund interaction tends to ferromagnetic correlation.

For PBC, there are two extra terms,
\begin{eqnarray}
\frac{U}{2}\left(\hat{c}_{1\uparrow}^{\dag}\hat{c}_{2\uparrow}\hat{c}_{1\downarrow}^{\dag}\hat{c}_{2\downarrow}+
\hat{c}_{2\uparrow}^{\dag}\hat{c}_{1\uparrow}\hat{c}_{2\downarrow}^{\dag}\hat{c}_{1\downarrow}\right)=\frac{U}{2}\left(\hat{c}_{1\uparrow}^{\dag}\hat{c}_{1\downarrow}^{\dag}\hat{c}_{2\downarrow}\hat{c}_{2\uparrow}+
\hat{c}_{2\uparrow}^{\dag}\hat{c}_{2\downarrow}^{\dag}\hat{c}_{1\downarrow}\hat{c}_{1\uparrow}\right).\nonumber
\end{eqnarray}
These two terms can be seen as the pair-hopping term, which means it transfer a double occupation from one site to another. When interaction is large, both the Hubbard interaction and Hund interaction tend to single occupation on each site while the charge-charge and pair-hopping term favor the double or empty occupation, so these two kinds of interaction are competing. Note that the pair-hopping term disappears in OBC, thus the Hubbard and Hund interaction dominate, and the latter ones make the ferromagnetic correlation dominate.
The work of Skolimowski shows that the ferromagnetic trend is valid for larger size in OBC, in other words, HK system with OBC and large interaction tends to have ferromagnetic spin correlation.\cite{Skolimowski2024}
\begin{figure}
\begin{center}
\includegraphics[width=0.95\columnwidth]{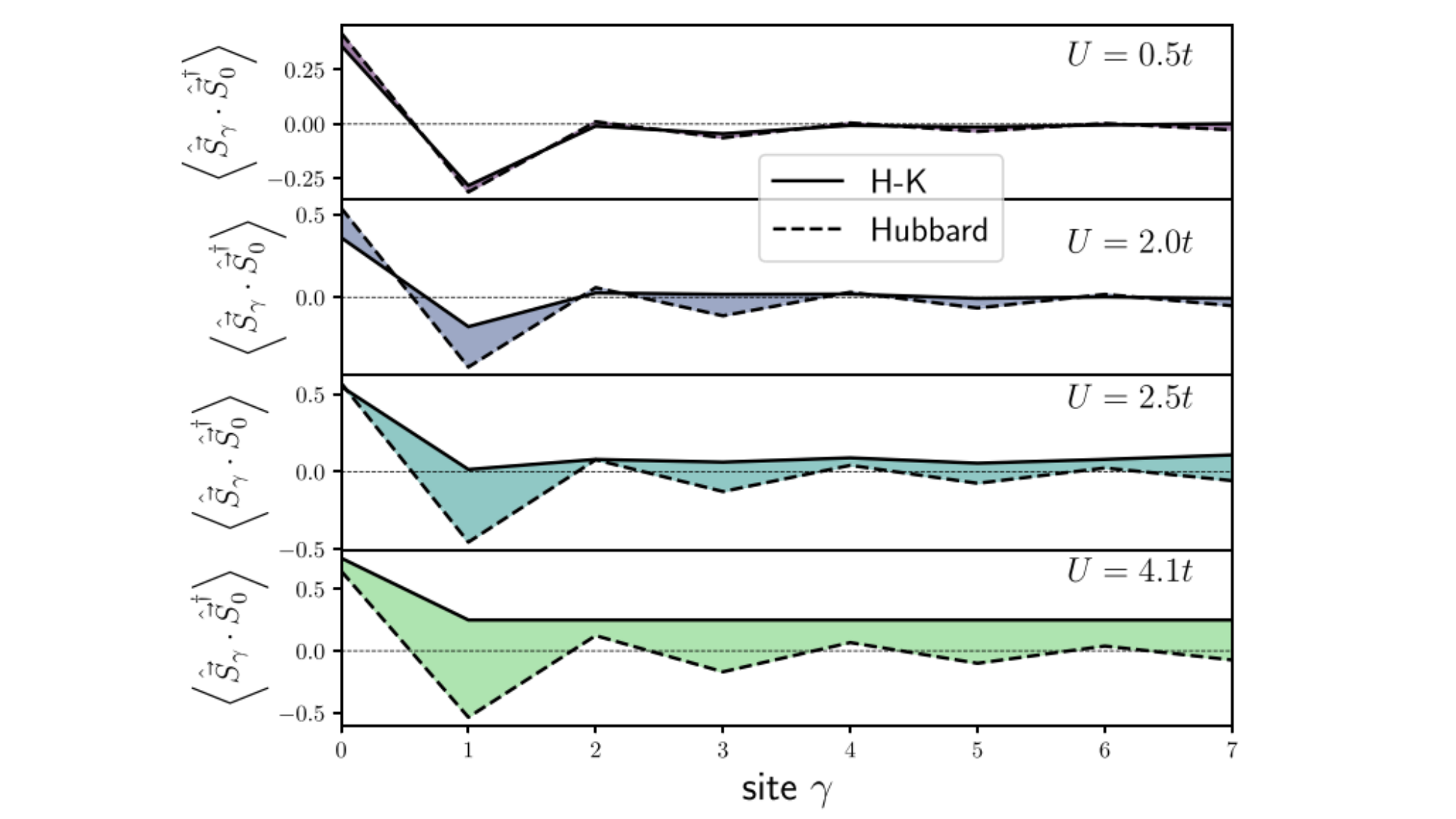}
\caption{\label{fig:HK_ED1} Spin correlation function $\langle \hat{\vec{S}}_{\gamma}\cdot \hat{\vec{S}}_{0}\rangle$ for half-filled HK and Hubbard model with OBC. (adapted from Ref.~\cite{Skolimowski2024})}
\end{center}
\end{figure}

Fig.~\ref{fig:HK_ED1} shows the comparison of spin correlation function $\langle \hat{\vec{S}}_{\gamma}\cdot \hat{\vec{S}}_{0}\rangle$ in symmetric half-filled HK and Hubbard model. ($8$-site and OBC) Here $\hat{\vec{S}}_{0}$ denotes the first site while $\vec{S}_{\gamma}$ denotes the site with distance $\gamma$. The expression of spin correlation function is
\begin{eqnarray}
\langle \hat{\vec{S}}_{\gamma}\cdot \hat{\vec{S}}_{0}\rangle
=\frac{1}{2}\langle \hat{c}_{\gamma\uparrow}^{\dag}\hat{c}_{\gamma\downarrow}\hat{c}_{0\downarrow}^{\dag}\hat{c}_{0\uparrow} +\hat{c}_{\gamma\downarrow}^{\dag}\hat{c}_{\gamma\uparrow}\hat{c}_{0\uparrow}^{\dag}\hat{c}_{0\downarrow}\rangle+\frac{1}{4}\langle \hat{n}_{\gamma\uparrow}\hat{n}_{0\uparrow}+\hat{n}_{\gamma\downarrow}\hat{n}_{0\downarrow}-\hat{n}_{\gamma\uparrow}\hat{n}_{0\downarrow}-\hat{n}_{\gamma\downarrow}\hat{n}_{0\uparrow}\rangle\nonumber
\end{eqnarray}
It is clear that, for HK model, weak interaction leads to antiferromagnetic correlation while strong interaction has weak antiferromagnetic but strong ferromagnetic correlation. When $U>W=4t$ the ferromagnetic correlation does not decay and just acts as a long-ranged order. In contrast, the spin correlation in Hubbard model is always antiferromagnetic and not significant ferromagnetic correlation is observed.
We should emphasize that whether the ferromagnetic correlation can develop into long-ranged order
in HK model needs to be examined in large size system.

To discuss the stability of HK physics, we can inspect the system with Hubbard interaction. Particularly, because the strong coupling system has dominating ferromagnetic correlation, the evolution of spin correlation versus the Hubbard interaction can provide the criteria of stability. In fact, if choosing $U=4.1t$, one finds the spin correlation is still ferromagnetic when Hubbard interaction $U_{H}$ has been changed from zero to $100t$, and the ground-state energy is just the linear function of $U_{H}$, which means the system with Hubbard interaction does not destabilize the fixed point of HK model.

Another interesting feature of real space calculation is that when adding $s$-wave pairing term $\Delta\sum_{j}(\hat{c}_{j\uparrow}^{\dag}\hat{c}_{j\downarrow}^{\dag}+\hat{c}_{j\downarrow}\hat{c}_{j\uparrow})$, the resultant superconductor with OBC can be topologically nontrivial with fermion zero-modes on its edge.\cite{Zhu2021} Because of the $s$-wave pairing, these zero-modes are not Majorana zero-mode in the $p$-wave superconducting chain.\cite{Kitaev2001,Kallin2016} The bulk-boundary correspondence works for this system and Zhu et al identify the ground-state fermion parity as a many-body topological invariant for PBC. Considering the finding of Skolimowski, which shows strong ferromagnetic correlation develops in OBC, the fermion zero-mode found by Zhu et al seems to be stabilized by such ferromagnetic correlation.\cite{Skolimowski2024} Currently, ED calculation for $d\geqslant2$ HK model do not exist, so we are unable to claim the ferromagnetic correlation always leads to boundary fermion zero-modes.

\section{Extension of HK model}\label{sec:4}
\subsection{Fermi arc}
A modified HK model proposed by Yang is able to generate Fermi arc, which does not form closed Fermi surface and violate the Luttinger theorem.\cite{Yang2021} Yang's model has the following Hamiltonian
\begin{eqnarray}
\hat{H}=\sum_{k}[\xi_{k}(\hat{n}_{k\uparrow}+\hat{n}_{k\downarrow})+U_{k}\hat{n}_{k\uparrow}\hat{n}_{k\downarrow}]
\end{eqnarray}
where $\xi_{k}=-2t_{x}\cos k_{x}-2t_{y}\cos k_{y}-\mu$ and a momentum-dependent HK interaction is chosen as $U_{k}=U-2T_{x}\cos k_{x}-2T_{y}\cos k_{y}$. If $U_{k}$ is either positive or negative definite, the solution
is qualitatively very similar to HK model. But, when $U_{k}$
changes sign in momentum space, the surface with $U_{k}=0$ will intersect with the noninteracting Fermi surface $\varepsilon_{k}=0$, and the ground state supports Fermi arcs.
\begin{figure}
\begin{center}
\includegraphics[width=0.85\columnwidth]{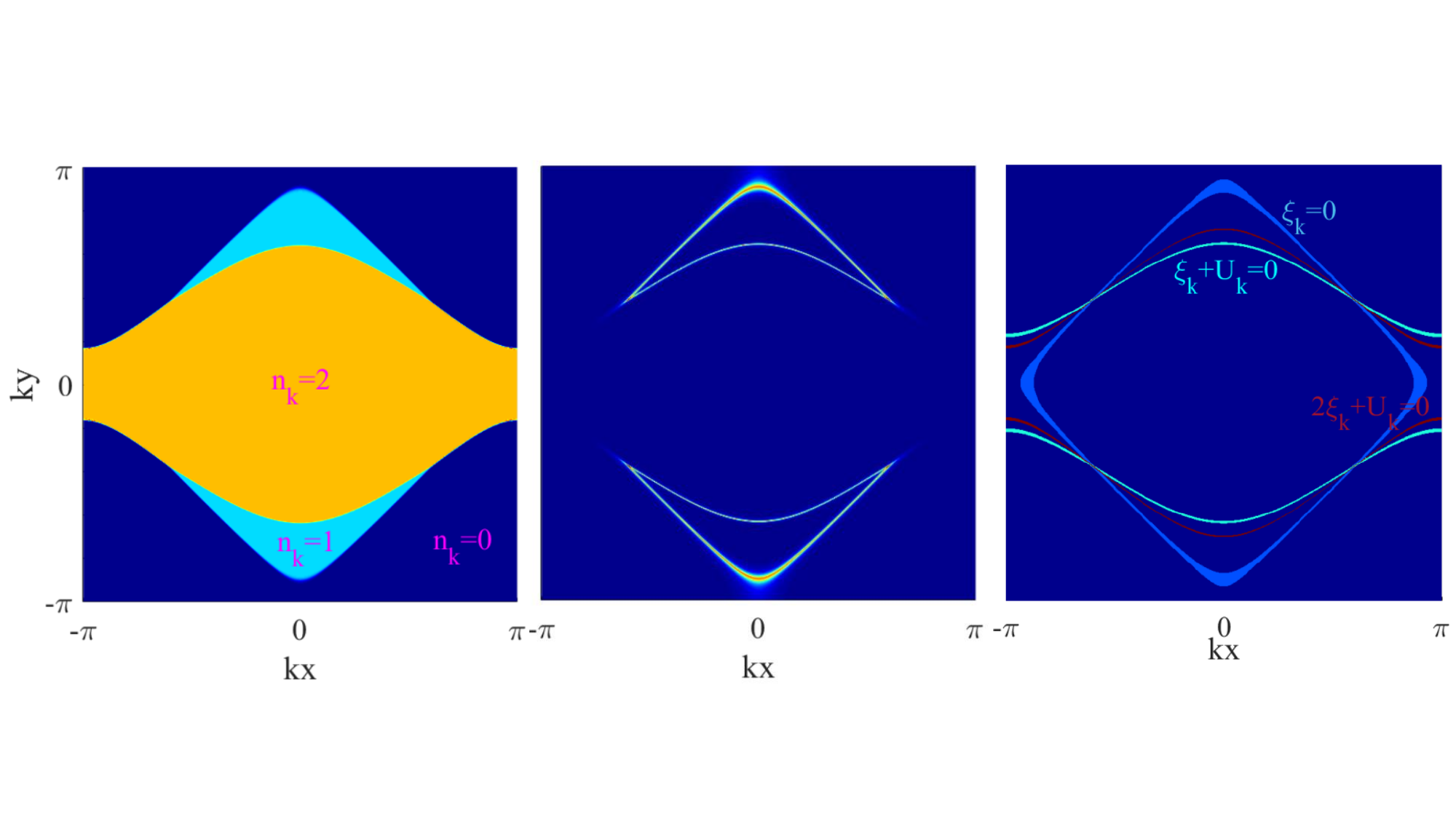}
\caption{\label{fig:Fermi_arc} (Left) Electron distribution function $n_{k}$; (Middle) Spectral function $A(k,0)$; (Right) $\xi_{k}=0$, $\xi_{k}+U_{k}=0$ and $2\xi_{k}+U_{k}=0$. ($t_{x}=t_{y}=1$, $\mu=-0.1,U=0.6$, $T_{x}=1,T_{y}=0.2$)}
\end{center}
\end{figure}

The single-particle Green function of $\hat{H}$ is
\begin{eqnarray}
G(k,\omega)=\frac{1-n_{k}/2}{\omega-\xi_{k}+i0^{+}}+\frac{n_{k}/2}{\omega-\xi_{k}-U_{k}+i0^{+}}\nonumber
\end{eqnarray}
and the electron distribution function $n_{k}=\sum_{\sigma}\langle \hat{n}_{k\sigma}\rangle$ is found to be
\begin{eqnarray}
n_{k}=\left\{
        \begin{array}{ll}
          0, & \hbox{$\xi_{k}>0,2\xi_{k}+U_{k}>0$;} \\
          1, & \hbox{$\xi_{k}<0,2\xi_{k}+U_{k}>0$;} \\
          2, & \hbox{$\xi_{k}<0,2\xi_{k}+U_{k}<0$.}
        \end{array}
      \right.\nonumber
\end{eqnarray}
Thus, boundaries with $\delta n_{k}=1$ or $\delta n_{k}=2$ are determined by $\xi_{k}=0,\xi_{k}+U_{k}=0$ and $2\xi_{k}+U_{k}=0$. All these lines intersect at $U_{k}=0$.

As shown in Fig.~\ref{fig:Fermi_arc}, the zero-frequency spectral function $A(k,0)=-\frac{1}{\pi}\mathrm{Im} G(k,\omega=0)$ indeed exhibits Fermi arc-like structure when $U_{k}$ changes sign in momentum space. ($U_{k}=0.6-2\cos k_{x}-0.4\cos k_{y}$) The Fermi arcs are located at momentums satisfying $\delta n_{k}=1$ while the Fermi surface with $\delta n_{k}=2$ has been gapped. To understand the gapped Fermi surface, we recall that the regime with $n_{k}=2$ has
$G(k,\omega)=\frac{1}{\omega-\xi_{k}-U_{k}+i0^{+}}$ and $2\xi_{k}+U_{k}=0$ in the line with $\delta n_{k}=2$. Inserting $2\xi_{k}+U_{k}=0$ into $G(k,\omega)$ gives
$A(k,0)=-\frac{1}{\pi}\mathrm{Im}G(k,0)=-\frac{1}{\pi}\mathrm{Im}\frac{1}{\xi_{k}+i0^{+}}=\delta(\xi_{k})$, which must vanish since $\xi_{k}$ cannot be zero in this $\delta n_{k}=2$ line.

Inspired by the finding of Fermi arc, Yang considers how the system responds to a perpendicular magnetic field, which gives rise to magnetic quantum oscillations used to map out Fermi surface and effective mass. Ignoring Zeeman energy, Yang argues that the effect of orbital magnetic field can be included by the semiclassical equation of motion $\hbar\frac{d\vec{k}}{dt}=-q_{\alpha} \vec{v}_{\vec{k}}^{\alpha}\times \vec{B}$, accompanied by $\vec{v}_{\vec{k}}^{\alpha}=\nabla_{k}E_{k\alpha}$. Here, $q_{1}=q_{2}=e,q_{3}=2e$ $E_{k1}=\xi_{k}$ represents excitation of spin-up fermions, spin-down fermions has $E_{k2}=\xi_{k}+U_{k}$ and the singlet fermion pairs (doublon) have dispersion $E_{k3}=2\xi_{k}+U_{k}$. Quantizing the motion of each excitation leads to Landau level and observable will exhibit periodic oscillation with period $|\vec{B}|^{-1}$.\cite{Shoenberg1984} However, as discussed in Sec.~\ref{sec:2}, the true quasiparticle of HK system has to be the holon and doublon obeying the exclusion statistics. A phenomenological Boltzmann kinetic equation for particle with exclusion statistics has been established in Ref.~\cite{Bhaduri1996} but its microscopic justification is still lacking. Thus, the usefulness of the semiclassical equation of motion is questioned and the related magnetic quantum oscillation may not exist.

Instead of a momentum-dependent HK interaction in Yang's construction, Worm et al. consider a HK interaction involving a non-zero momentum transfer,\cite{Worm2024}
\begin{equation}
\hat{H}=\sum_{k}[\xi_{k}(\hat{n}_{k\uparrow}+\hat{n}_{k\downarrow})+U\hat{n}_{k\uparrow}\hat{n}_{k+Q\downarrow}]
\end{equation}
with $Q=(\pi,\pi)$ being the antiferromagnetic characteristic wavevector on square lattice. The construction of above model is motivated by strong antiferromagnetic spin-fluctuation in Hubbard model near half-filling, e.g. the widely-used antiferromagnetic mean-field theory leads to reconstructed disconnected small Fermi pockets.\cite{Sachdev2023}

To visualize Fermi arc structure, the corresponding single-particle Green function can be readily found in terms of equation of motion method,
\begin{eqnarray}
G(k,\omega)=\frac{1-n_{k+Q}}{\omega-\xi_{k}+i0^{+}}+\frac{n_{k+Q}}{\omega-\xi_{k}-U+i0^{+}}\nonumber
\end{eqnarray}
and $n_{k}=\sum_{\sigma}\langle \hat{c}_{k\sigma}^{\dag}\hat{c}_{k\sigma}\rangle/2$ satisfies the equations
\begin{eqnarray}
&&n_{k}=(1-n_{k+Q})f_{F}(\xi_{k})+n_{k+Q}f_{F}(\xi_{k}+U)\nonumber\\
&&n_{k+Q}=(1-n_{k})f_{F}(\xi_{k+Q})+n_{k}f_{F}(\xi_{k+Q}+U).\nonumber
\end{eqnarray}
Solving above two equations gives rise to $n_{k},n_{k+Q}$ and the zero-frequency spectral function indeed produces the expected Fermi arcs located at nodal or antinodal direction while hole or electron-like Fermi surface has also been found if $|\mu|\gg U$. When comparing with numerical calculation from dynamical vertex approximation ($D\Gamma A$) on Hubbard model,\cite{Toschi2007} Worm et al. state that both HK-type model
and the $D\Gamma A$ solution of Hubbard model yield the same evolution of Fermi and Luttinger arcs. At the same time, they note that the Fermi arcs computed in $D\Gamma A$ do not end at the boundary of antiferromagnetic BZ, but are more gradually smeared out due to the momentum evolution of $D\Gamma A$ self-energy.

\subsection{Quantum oscillation}
Although Yang argues HK model can have magnetic oscillation, this issue has not been considered seriously until the works of Ref.~\cite{Zhong2023,Leeb2023}. We know that physical observable such as magnetization and resistivity in metals exhibit periodic oscillations under external magnetic fields.\cite{Shoenberg1984} Such magnetic quantum oscillation (QO) is generally believed to result from the oscillation of electron's density of state at Fermi energy when Landau level periodically crosses Fermi energy. The underlying microscopic description is captured by the famous Lifshitz-Kosevich (LK) formula, which is still valid if the electron-electron interaction effect does not alter the FL nature of system under consideration.\cite{Luttinger1961,Wasserman1996} However, a natural question arises when the system is not described by FL and the concept of quasiparticle breaks down.\cite{Chakravarty2011} Inspired by the NFL state in solvable HK model, Zhong, Leeb and Knolle have investigated the magnetic QO on square lattice and in continuum limit.

On square lattice with uniform magnetic field along $z$-axis, the Hamiltonian is given by
\begin{eqnarray}
\hat{H}&=&-t\sum_{j\sigma}e^{i2\pi wj_{y}}\hat{c}_{j\sigma}^{\dag}\hat{c}_{j+x,\sigma}
-t\sum_{j\sigma}\hat{c}_{j\sigma}^{\dag}\hat{c}_{j+y,\sigma}+h.c.\nonumber\\
&-&\mu\sum_{j\sigma}\hat{c}_{j\sigma}^{\dag}\hat{c}_{j\sigma}+\frac{U}{N_{s}}\sum_{j_{1},j_{2},j_{3},j_{4}}\delta_{j_{1}+j_{3}=j_{2}+j_{4}}
\hat{c}_{j_{1}\uparrow}^{\dag}\hat{c}_{j_{2}\uparrow}\hat{c}_{j_{3}\downarrow}^{\dag}\hat{c}_{j_{4}\downarrow}.\label{eq8}
\end{eqnarray}
where Landau gauge $\vec{A}=(-By,0,0)$ is assumed and $w=\frac{Ba^{2}}{h/e}$. ($a$ is lattice constant)
With Luttinger's approximation,\cite{Luttinger1961} Hofstadter butterfly exists in all phases of the ground-state of HK model whatever they are NFL or Mott insulator. By examining the magnetic-field-dependent density of state, magnetization and particle's density, NFL states indeed show QO and their zero-temperature behaviors are captured by an LK-like formula, which reflects the existence of two-Fermi-surface structure of the (non-Landau) quasiparticle in NFL. To go beyond Luttinger's approximation, Leeb and Knolle expand electron operators in terms of Landau level wavefunction, in continuum limit. Their model reads as
\begin{eqnarray}
\hat{H}=\sum_{l,k_{x},\sigma}\varepsilon_{l}\hat{c}_{l,k_{x},\sigma}^{\dag}\hat{c}_{l,k_{x},\sigma}
+\frac{Ul_{B}}{L}\sum_{k_{x},l_{1},l_{2},l_{3},l_{4}}V_{l_{1},l_{2},l_{3},l_{4}}
\hat{c}_{l_{1},k_{x},\uparrow}^{\dag}\hat{c}_{l_{2},k_{x},\uparrow}\hat{c}_{l_{3},k_{x},\downarrow}^{\dag}\hat{c}_{l_{4},k_{x},\downarrow},\nonumber
\end{eqnarray}
which is diagonal in $k_{x}$ due to the conservation of center of mass. The index $l$ denotes Landau level and its energy is $\varepsilon_{l}=\omega_{c}(l+1/2)$. $l_{B}=L\sqrt{2\pi/w}$ is the magnetic length. The interaction matrix-element $V_{l_{1},l_{2},l_{3},l_{4}}$ has complicated expression and interested readers can consult appendix C of
Ref.~\cite{Leeb2023}. Remarkably, for sufficiently high Landau levels, one finds a Landau-level-HK Hamiltonian $\hat{H}=\sum_{l,k_{x}}[\omega_{c}(l+1/2)(\hat{n}_{l,k_{x},\uparrow}+\hat{n}_{l,k_{x},\downarrow})
+U'\hat{n}_{l,k_{x},\uparrow}\hat{n}_{l,k_{x},\downarrow}]$. Thus, Landau levels with $\varepsilon_{l}<\mu-U'$ are doubly occupied, Landau levels with $\varepsilon_{l}<\mu$ are singly occupied, and higher energetic Landau levels are
not occupied. The occupation edges at $\mu$ and $\mu-U'$ lead to QO with frequencies $\frac{\mu}{\omega_{c}}$ and $\frac{\mu-U'}{\omega_{c}}$. At sufficiently strong magnetic fields, there are jumps in the magnetization and these jumps are aperiodic, thus QO becomes aperiodic, breaking Onsager's relation valid in usual QO theories.
\section{Summary and Perspective}
We have given an introduction to the solvable HK model, whose thermodynamics and spectral properties are discussed in detail. The presented working knowledge is expected to be useful for graduate students or researchers interested in HK-related phenomena, which has sharpened our understanding on NFL, interacting topological phase and unconventional superconductivity.

The infinite-ranged interaction renders the solvability of HK Hamiltonian but we all know it is not realistic, and the existing condensed matter systems do not provide realization of such radical interaction. If we make the HK interaction short-ranged (cutoff the interaction into nearest-neighbor and next-nearest-neighbor sites) but preserve the motion of center of mass, the resulting model should be the dipolar Hubbard model,\cite{Lake2023} which can be realized in strongly tilted optical lattices.\cite{Guardado-Sanchez2020} However, the physics of dipolar Hubbard model involves with dipolar bound states, rather than individual electron, thus it is rather different from HK-like model. We feel the mild cutoff of HK interaction is not a good choice and more long-ranged terms have to be included. Research in this direction is highly-desirable. On the other hand, digital quantum simulation by existing quantum processor may realize the HK interaction if all qubits are explained in momentum space.\cite{Fauseweh2024,XZhang2022,RShen2023,TChen2023}
In the perspective of RG, the unrealistic HK interaction stabilizes the Mott insulator and related Luttinger-theorem-violating NFL, therefore the physics obtained from analysis on HK Hamiltonian is realistic and the HK-related systems will always stimulate fascinating study in near future.

\ack
We thank Jiangfan Wang and Yu Li for their discussion on related issues and proofreading on this manuscript.

\appendix
%\section*{Appendix}
%\setcounter{section}{1}
\section{Derivation of $\chi_{c}$}\label{ap-A}
To begin, let us consider $\chi_{c}(R_{i},R_{j},t)=i\theta(t)\langle[\hat{n}_{i}(t),\hat{n}_{j}]\rangle$, which can be rewritten as
\begin{eqnarray}
\chi_{c}(R_{i},R_{j},t)=\frac{i}{N_{s}^{2}}\sum_{k_{1},k_{2},k_{3},k_{4}}\sum_{\sigma,\sigma'}e^{-i(k_{1}-k_{2})R_{i}}e^{-i(k_{3}-k_{4})R_{j}}\theta(t)\langle[\hat{c}_{k_{1}\sigma}^{\dag}(t)\hat{c}_{k_{2}\sigma}(t),\hat{c}_{k_{3}\sigma'}^{\dag}\hat{c}_{k_{4}\sigma'}]\rangle.\nonumber
\end{eqnarray}
For a translation-invariant system, we should have $\chi_{c}(R_{i},R_{j},t)=\frac{1}{N_{s}}\sum_{q}e^{iq(R_{i}-R_{j})}\chi(q,t)$, where
\begin{eqnarray}
\chi_{c}(q,t)&=&\frac{i}{N_{s}}\sum_{k_{1},k_{3}}\sum_{\sigma,\sigma'}\theta(t)\langle[\hat{c}_{k_{1}\sigma}^{\dag}(t)\hat{c}_{k_{1}+q\sigma}(t),\hat{c}_{k_{3}\sigma'}^{\dag}\hat{c}_{k_{3}-q\sigma'}]\rangle\nonumber\\
&=&\frac{i}{N_{s}}\theta(t)\langle[\overbrace{\sum_{k_{1}\sigma}\hat{c}_{k_{1}\sigma}^{\dag}(t)\hat{c}_{k_{1}+q\sigma}(t)}^{\hat{\rho}_{q}(t)},\overbrace{\sum_{k_{3}\sigma'}\hat{c}_{k_{3}\sigma'}^{\dag}\hat{c}_{k_{3}-q\sigma'}}^{\hat{\rho}_{-q}}]\rangle\nonumber\\
&=&\frac{i}{N_{s}}\theta(t)\langle[\hat{\rho}_{q}(t),\hat{\rho}_{-q}]\rangle
\end{eqnarray}

Now, let us define retarded Green's function
\begin{eqnarray}
\langle\langle\hat{c}_{k\sigma}^{\dag}\hat{c}_{k+q\sigma}|\hat{c}_{k'\sigma'}^{\dag}\hat{c}_{k'-q\sigma'}\rangle\rangle_{\omega}\equiv-i\int_{-\infty}^{\infty} dte^{i\omega t}\theta(t)\langle[\hat{c}_{k\sigma}^{\dag}(t)\hat{c}_{k+q,\sigma}(t),\hat{c}_{k'\sigma'}^{\dag}\hat{c}_{k'-q,\sigma'}]\rangle\nonumber
\end{eqnarray}
which is related to charge susceptibility via
\begin{equation}
\chi_{c}(q,\omega)=-\frac{1}{N_{s}}\sum_{k\sigma}\sum_{k'\sigma'}\langle\langle\hat{c}_{k\sigma}^{\dag}\hat{c}_{k+q\sigma}|\hat{c}_{k'\sigma'}^{\dag}\hat{c}_{k'-q\sigma'}\rangle\rangle_{\omega}
\end{equation}

According to the general formalism of equation of motion for the retarded Green's function, we find
\begin{eqnarray}
\omega\langle\langle\hat{c}_{k\sigma}^{\dag}\hat{c}_{k+q\sigma}|\hat{c}_{k'\sigma'}^{\dag}\hat{c}_{k'-q,\sigma'}\rangle\rangle_{\omega}
=\langle[\hat{c}_{k\sigma}^{\dag}\hat{c}_{k+q\sigma},\hat{c}_{k'\sigma'}^{\dag}\hat{c}_{k'-q\sigma'}]\rangle+\langle\langle[\hat{c}_{k\sigma}^{\dag}\hat{c}_{k+q\sigma},\hat{H}]|\hat{c}_{k'\sigma'}^{\dag}\hat{c}_{k'-q,\sigma'}\rangle\rangle_{\omega}\nonumber
\end{eqnarray}
Here, $[\hat{c}_{k\sigma}^{\dag}\hat{c}_{k+q\sigma},\hat{c}_{k'\sigma'}^{\dag}\hat{c}_{k'-q\sigma'}]=\delta_{\sigma,\sigma'}\delta_{k',k+q}
(\hat{c}_{k\sigma}^{\dag}\hat{c}_{k\sigma}-\hat{c}_{k+q,\sigma}^{\dag}\hat{c}_{k+q,\sigma})$.

For $[\hat{c}_{k\sigma}^{\dag}\hat{c}_{k+q,\sigma},\hat{H}]$, utilizing $[\hat{c}_{k\sigma},H]=(\varepsilon_{k}-\mu)\hat{c}_{k\sigma}+U\hat{c}_{k\sigma}\hat{n}_{k\bar{\sigma}}$, one finds
\begin{equation}
[\hat{c}_{k\sigma}^{\dag}\hat{c}_{k+q,\sigma},\hat{H}]=(\varepsilon_{k+q}-\varepsilon_{k})\hat{c}_{k\sigma}^{\dag}\hat{c}_{k+q,\sigma}+U\hat{c}_{k\sigma}^{\dag}\hat{c}_{k+q,\sigma}(\hat{n}_{k+q,\bar{\sigma}}-\hat{n}_{k\bar{\sigma}}).
\end{equation}
So,
\begin{eqnarray}
  (\omega-\varepsilon_{k+q}+\varepsilon_{k})\langle\langle \hat{c}_{k\sigma}^{\dag}\hat{c}_{k+q,\sigma}|\hat{c}_{k'\sigma'}^{\dag}\hat{c}_{k'-q,\sigma'}\rangle\rangle_{\omega}&=&\delta_{\sigma,\sigma'}\delta_{k',k+q}
(n_{k\sigma}-n_{k+q,\sigma})\nonumber\\
  &+&U\langle\langle\hat{c}_{k\sigma}^{\dag}\hat{c}_{k+q,\sigma}(\hat{n}_{k+q,\bar{\sigma}}-\hat{n}_{k\bar{\sigma}})|\hat{c}_{k'\sigma'}^{\dag}\hat{c}_{k'-q,\sigma'}\rangle\rangle_{\omega}.\nonumber
\end{eqnarray}
To proceed, we need
\begin{eqnarray}
[\hat{c}_{k\sigma}^{\dag}\hat{c}_{k+q,\sigma}\hat{n}_{k+q,\bar{\sigma}},\hat{H}]=(\varepsilon_{k+q}-\varepsilon_{k}+U)\hat{c}_{k\sigma}^{\dag}\hat{c}_{k+q,\sigma}\hat{n}_{k+q,\bar{\sigma}}-U\hat{c}_{k\sigma}^{\dag}\hat{c}_{k+q,\sigma}\hat{n}_{k\bar{\sigma}}\hat{n}_{k+q,\bar{\sigma}},\nonumber
\end{eqnarray}
\begin{eqnarray}
[\hat{c}_{k\sigma}^{\dag}\hat{c}_{k+q,\sigma}\hat{n}_{k,\bar{\sigma}},\hat{H}]=(\varepsilon_{k+q}-\varepsilon_{k}-U)\hat{c}_{k\sigma}^{\dag}\hat{c}_{k+q,\sigma}\hat{n}_{k,\bar{\sigma}}+U\hat{c}_{k\sigma}^{\dag}\hat{c}_{k+q,\sigma}\hat{n}_{k\bar{\sigma}}\hat{n}_{k+q,\bar{\sigma}}.\nonumber
\end{eqnarray}
Thus,
\begin{eqnarray}
  (\omega-\varepsilon_{k+q}+\varepsilon_{k}-U)\langle\langle \hat{c}_{k\sigma}^{\dag}\hat{c}_{k+q,\sigma}\hat{n}_{k+q,\bar{\sigma}}|\hat{c}_{k'\sigma'}^{\dag}\hat{c}_{k'-q,\sigma'}\rangle\rangle_{\omega}&=&\langle[\hat{c}_{k\sigma}^{\dag}\hat{c}_{k+q,\sigma}\hat{n}_{k+q,\bar{\sigma}},\hat{c}_{k'\sigma'}^{\dag}\hat{c}_{k'-q,\sigma'}]\rangle\nonumber\\
  &-&U\langle\langle\hat{c}_{k\sigma}^{\dag}\hat{c}_{k+q,\sigma}\hat{n}_{k\bar{\sigma}}\hat{n}_{k+q,\bar{\sigma}}|\hat{c}_{k'\sigma'}^{\dag}\hat{c}_{k'-q,\sigma'}\rangle\rangle_{\omega}.\nonumber
\end{eqnarray}
\begin{eqnarray}
  (\omega-\varepsilon_{k+q}+\varepsilon_{k}+U)\langle\langle \hat{c}_{k\sigma}^{\dag}\hat{c}_{k,\sigma}\hat{n}_{k+q,\bar{\sigma}}|\hat{c}_{k'\sigma'}^{\dag}\hat{c}_{k'-q,\sigma'}\rangle\rangle_{\omega}&=&\langle[\hat{c}_{k\sigma}^{\dag}\hat{c}_{k+q,\sigma}\hat{n}_{k,\bar{\sigma}},\hat{c}_{k'\sigma'}^{\dag}\hat{c}_{k'-q,\sigma'}]\rangle\nonumber\\
  &+&U\langle\langle\hat{c}_{k\sigma}^{\dag}\hat{c}_{k+q,\sigma}\hat{n}_{k\bar{\sigma}}\hat{n}_{k+q,\bar{\sigma}}|\hat{c}_{k'\sigma'}^{\dag}\hat{c}_{k'-q,\sigma'}\rangle\rangle_{\omega}.\nonumber
\end{eqnarray}
\begin{eqnarray}
\langle[\hat{c}_{k\sigma}^{\dag}\hat{c}_{k+q,\sigma}\hat{n}_{k+q,\bar{\sigma}},\hat{c}_{k'\sigma'}^{\dag}\hat{c}_{k'-q,\sigma'}]\rangle
&=&\delta_{\sigma,\sigma'}\delta_{k',k+q}
(n_{k\sigma}n_{k+q,\bar{\sigma}}-\langle\hat{n}_{k+q,\sigma}\hat{n}_{k+q,\bar{\sigma}}\rangle)\nonumber\\
&=&\delta_{\sigma,\sigma'}\delta_{k',k+q}
(n_{k\sigma}n_{k+q,\bar{\sigma}}-n_{k+q,\bar{\sigma}}f_{F}(\varepsilon_{k+q}-\mu+U))\nonumber\\
&=&\delta_{\sigma,\sigma'}\delta_{k',k+q}n_{k+q,\bar{\sigma}}
(n_{k\sigma}-f_{F}(\varepsilon_{k+q}-\mu+U))\nonumber\\
&\equiv&\delta_{\sigma,\sigma'}\delta_{k',k+q}A_{1}(k,q,\sigma),\nonumber
\end{eqnarray}
\begin{equation}
A_{1}(k,q,\sigma)=n_{k+q,\bar{\sigma}}(n_{k\sigma}-f_{F}(\varepsilon_{k+q}-\mu+U))
\end{equation}
\begin{eqnarray}
\langle[\hat{c}_{k\sigma}^{\dag}\hat{c}_{k+q,\sigma}\hat{n}_{k\bar{\sigma}},\hat{c}_{k'\sigma'}^{\dag}\hat{c}_{k'-q,\sigma'}]\rangle&=&
\delta_{\sigma,\sigma'}\delta_{k',k+q}(\langle\hat{n}_{k\sigma}\hat{n}_{k,\bar{\sigma}}\rangle-n_{k+q,\sigma}n_{k\bar{\sigma}})\nonumber\\
&=&\delta_{\sigma,\sigma'}\delta_{k',k+q}(n_{k\bar{\sigma}}f_{F}(\varepsilon_{k}-\mu+U)-n_{k+q,\sigma}n_{k\bar{\sigma}})\nonumber\\
&=&\delta_{\sigma,\sigma'}\delta_{k',k+q}n_{k\bar{\sigma}}(f_{F}(\varepsilon_{k}-\mu+U)-n_{k+q,\sigma})\nonumber\\
&\equiv&\delta_{\sigma,\sigma'}\delta_{k',k+q}A_{2}(k,q,\sigma).\nonumber
\end{eqnarray}
\begin{equation}
A_{2}(k,q,\sigma)=n_{k\bar{\sigma}}(f_{F}(\varepsilon_{k}-\mu+U)-n_{k+q,\sigma})
\end{equation}
One must calculate $\langle\langle\hat{c}_{k\sigma}^{\dag}\hat{c}_{k+q,\sigma}\hat{n}_{k\bar{\sigma}}\hat{n}_{k+q,\bar{\sigma}}|\hat{c}_{k'\sigma'}^{\dag}\hat{c}_{k'-q,\sigma'}\rangle\rangle_{\omega}$, and finds
\begin{eqnarray}
[\hat{c}_{k\sigma}^{\dag}\hat{c}_{k+q,\sigma}\hat{n}_{k\bar{\sigma}}\hat{n}_{k+q,\bar{\sigma}},\hat{H}]&=&[\hat{c}_{k\sigma}^{\dag}\hat{c}_{k+q,\sigma},\hat{H}]\hat{n}_{k\bar{\sigma}}\hat{n}_{k+q,\bar{\sigma}}\nonumber\\
&=&(\varepsilon_{k+q}-\varepsilon_{k})\hat{c}_{k\sigma}^{\dag}\hat{c}_{k+q,\sigma}\hat{n}_{k\bar{\sigma}}\hat{n}_{k+q,\bar{\sigma}}+U\hat{c}_{k\sigma}^{\dag}\hat{c}_{k+q,\sigma}(\hat{n}_{k+q,\bar{\sigma}}-\hat{n}_{k\bar{\sigma}})\hat{n}_{k\bar{\sigma}}\hat{n}_{k+q,\bar{\sigma}}\nonumber\\
&=&(\varepsilon_{k+q}-\varepsilon_{k})\hat{c}_{k\sigma}^{\dag}\hat{c}_{k+q,\sigma}\hat{n}_{k\bar{\sigma}}\hat{n}_{k+q,\bar{\sigma}},\nonumber
\end{eqnarray}
which fortunately implies the closeness of equation of motion and we obtain
\begin{eqnarray}
&&(\omega-\varepsilon_{k+q}+\varepsilon_{k})\langle\langle\hat{c}_{k\sigma}^{\dag}\hat{c}_{k+q,\sigma}\hat{n}_{k\bar{\sigma}}\hat{n}_{k+q,\bar{\sigma}}|\hat{c}_{k'\sigma'}^{\dag}\hat{c}_{k'-q,\sigma'}\rangle\rangle_{\omega}
=\langle[\hat{c}_{k\sigma}^{\dag}\hat{c}_{k+q,\sigma}\hat{n}_{k\bar{\sigma}}\hat{n}_{k+q,\bar{\sigma}},\hat{c}_{k'\sigma'}^{\dag}\hat{c}_{k'-q,\sigma'}]\rangle\nonumber\\
&=&\delta_{\sigma,\sigma'}\delta_{k',k+q}(\langle \hat{n}_{k\sigma}\hat{n}_{k\bar{\sigma}}\rangle n_{k+q,\bar{\sigma}}-\langle \hat{n}_{k+q,\sigma}\hat{n}_{k+q,\bar{\sigma}}\rangle n_{k\bar{\sigma}})\nonumber\\
&=&\delta_{\sigma,\sigma'}\delta_{k',k+q}(n_{k\bar{\sigma}}f_{F}(\varepsilon_{k}-\mu+U)n_{k+q,\bar{\sigma}}-n_{k+q,\bar{\sigma}}f_{F}(\varepsilon_{k+q}-\mu+U) n_{k\bar{\sigma}})\nonumber\\
&=&\delta_{\sigma,\sigma'}\delta_{k',k+q}n_{k\bar{\sigma}}n_{k+q,\bar{\sigma}}(f_{F}(\varepsilon_{k}-\mu+U)-f_{F}(\varepsilon_{k+q}-\mu+U))\equiv\delta_{\sigma,\sigma'}\delta_{k',k+q}A_{3}(k,q,\sigma)\nonumber
\end{eqnarray}
\begin{equation}
A_{3}(k,q,\sigma)=n_{k\bar{\sigma}}n_{k+q,\bar{\sigma}}(f_{F}(\varepsilon_{k}-\mu+U)-f_{F}(\varepsilon_{k+q}-\mu+U))
\end{equation}
The above equations imply
\begin{eqnarray}
\langle\langle \hat{c}_{k\sigma}^{\dag}\hat{c}_{k+q,\sigma}\hat{n}_{k+q,\bar{\sigma}}|\hat{c}_{k'\sigma'}^{\dag}\hat{c}_{k'-q,\sigma'}\rangle\rangle_{\omega}&=&\frac{\delta_{\sigma,\sigma'}\delta_{k',k+q}(A_{1}(k,q,\sigma)-A_{3}(k,q,\sigma))}{\omega-\varepsilon_{k+q}\varepsilon_{k}-U}\nonumber\\
&+&\frac{\delta_{\sigma,\sigma'}\delta_{k',k+q}A_{3}(k,q,\sigma)}{\omega-\varepsilon_{k+q}+\varepsilon_{k}}.
\end{eqnarray}

\begin{eqnarray}
\langle\langle \hat{c}_{k\sigma}^{\dag}\hat{c}_{k+q,\sigma}\hat{n}_{k,\bar{\sigma}}|\hat{c}_{k'\sigma'}^{\dag}\hat{c}_{k'-q,\sigma'}\rangle\rangle_{\omega}&=&\frac{\delta_{\sigma,\sigma'}\delta_{k',k+q}(A_{2}(k,q,\sigma)-A_{3}(k,q,\sigma))}{\omega-\varepsilon_{k+q}+\varepsilon_{k}+U}\nonumber\\
&+&\frac{\delta_{\sigma,\sigma'}\delta_{k',k+q}A_{3}(k,q,\sigma)}{\omega-\varepsilon_{k+q}+\varepsilon_{k}}
\end{eqnarray}
Finally, we obtain
\begin{eqnarray}
\langle\langle \hat{c}_{k\sigma}^{\dag}\hat{c}_{k+q,\sigma}|\hat{c}_{k'\sigma'}^{\dag}\hat{c}_{k'-q,\sigma'}\rangle\rangle_{\omega}&=&
  \delta_{\sigma,\sigma'}\delta_{k',k+q}
\frac{n_{k\sigma}-n_{k+q,\sigma}-A_{1}(k,q,\sigma)-A_{2}(k,q,\sigma)+2A_{3}(k,q,\sigma)}{\omega-\varepsilon_{k+q}+\varepsilon_{k}}\nonumber\\
&+&\delta_{\sigma,\sigma'}\delta_{k',k+q}\frac{A_{1}(k,q,\sigma)-A_{3}(k,q,\sigma)}{\omega-\varepsilon_{k+q}+\varepsilon_{k}-U}\nonumber\\
&+&\delta_{\sigma,\sigma'}\delta_{k',k+q}\frac{A_{2}(k,q,\sigma)-A_{3}(k,q,\sigma)}{\omega-\varepsilon_{k+q}+\varepsilon_{k}+U}.
\end{eqnarray}
Inserting the formula into $\chi_{c}(q,\omega)$, it is found that
\begin{eqnarray}
\chi_{c}(q,\omega)&=&
-\frac{1}{N_{s}}\sum_{k\sigma}\frac{n_{k\sigma}-n_{k+q,\sigma}-A_{1}(k,q,\sigma)-A_{2}(k,q,\sigma)+2A_{3}(k,q,\sigma)}{\omega-\varepsilon_{k+q}+\varepsilon_{k}}\nonumber\\
&-&\frac{1}{N_{s}}\sum_{k\sigma}\frac{A_{1}(k,q,\sigma)-A_{3}(k,q,\sigma)}{\omega-\varepsilon_{k+q}+\varepsilon_{k}-U}-\frac{1}{N_{s}}\sum_{k\sigma}\frac{A_{2}(k,q,\sigma)-A_{3}(k,q,\sigma)}{\omega-\varepsilon_{k+q}+\varepsilon_{k}+U}\nonumber
\end{eqnarray}
with
\begin{eqnarray}
&&A_{1}(k,q,\sigma)=n_{k+q,\bar{\sigma}}(n_{k\sigma}-f_{F}(\varepsilon_{k+q}-\mu+U))
\nonumber\\
&&A_{2}(k,q,\sigma)=n_{k\bar{\sigma}}(f_{F}(\varepsilon_{k}-\mu+U)-n_{k+q,\sigma})
\nonumber\\
&&A_{3}(k,q,\sigma)=n_{k\bar{\sigma}}n_{k+q,\bar{\sigma}}(f_{F}(\varepsilon_{k}-\mu+U)-f_{F}(\varepsilon_{k+q}-\mu+U)).
\nonumber
\end{eqnarray}
We have checked that $\chi_{c}(q,\omega)$ can be written as the following form in the paramagnetic state ($n_{k\uparrow}=n_{k\downarrow}=n_{k}$)
\begin{eqnarray}
\chi_{c}(q,\omega)&=&-\frac{1}{N_{s}}\sum_{k,\sigma}(1-n_{k})(1-n_{k+q})\frac{f_{F}(\varepsilon_{k}-\mu)-f_{F}(\varepsilon_{k+q}-\mu)}{\omega-\varepsilon_{k+q}+\varepsilon_{k}}\nonumber\\
&-&\frac{1}{N_{s}}\sum_{k,\sigma}(1-n_{k})n_{k+q}\frac{f_{F}(\varepsilon_{k}-\mu)-f_{F}(\varepsilon_{k+q}-\mu+U)}{\omega-\varepsilon_{k+q}-U+\varepsilon_{k}}\nonumber\\
&-&\frac{1}{N_{s}}\sum_{k,\sigma}n_{k}(1-n_{k+q})\frac{f_{F}(\varepsilon_{k}-\mu+U)-f_{F}(\varepsilon_{k+q}-\mu)}{\omega-\varepsilon_{k+q}+\varepsilon_{k}+U}\nonumber\\
&-&\frac{1}{N_{s}}\sum_{k,\sigma}n_{k}n_{k+q}\frac{f_{F}(\varepsilon_{k}-\mu+U)-f_{F}(\varepsilon_{k+q}-\mu+U)}{\omega-\varepsilon_{k+q}+\varepsilon_{k}}.
\end{eqnarray}

\end{document}